%% file: ms.tex
\documentclass[12pt]{article}
\usepackage[letterpaper,margin=1.25in]{geometry}
\usepackage{amsmath,amssymb,amsthm,mathtools,bm,mathrsfs}
\usepackage{graphicx,booktabs,threeparttable,array,longtable,float}
\graphicspath{{figures/}{generated/}}
\usepackage{flafter}
\usepackage{setspace}
\usepackage{enumitem}
\usepackage{microtype}
\usepackage{xcolor}
\usepackage[colorlinks=true,allcolors=blue]{hyperref}
\newcommand{\oaref}[1]{\ref{#1}}
\usepackage[authoryear,round]{natbib}
\newtheorem{theorem}{Theorem}
\newtheorem{lemma}{Lemma}
\newtheorem{corollary}{Corollary}
\newtheorem{proposition}{Proposition}
\theoremstyle{definition}
\newtheorem{assumption}{Assumption}
\newtheorem{definition}{Definition}
\theoremstyle{remark}
\newtheorem{remark}{Remark}

\newcommand{\RR}{\mathbb R}
\newcommand{\Pp}{\mathbb P}
\newcommand{\EE}{\mathbb E}
\newcommand{\cR}{\mathcal R}
\newcommand{\cP}{\mathcal P}
\newcommand{\ThetaMR}{\Theta^{MR}}
\newcommand{\Rstar}{\mathcal R^\star}
\newcommand{\PR}{\mathrm{PR}}
\newcommand{\RA}{\mathrm{RA}}

\DeclareMathOperator{\dist}{dist}

\title{\textbf{Inference After Model Refutation:}\\Uniform Post-Relaxation Inference for Partially Identified Models}
\author{Ulrich Hounyo\\[0.35em]
\normalsize Department of Economics, University at Albany, State University of New York\\[-0.05em]
\normalsize \texttt{khounyo@albany.edu}}
\date{This version: \today}

\makeatletter
\renewcommand{\maketitle}{%
  \begin{center}
    {\LARGE \@title\par}
    \vspace{1.1em}
    {\large \@author\par}
    \vspace{0.8em}
    {\normalsize \@date\par}
  \end{center}
  \vspace{0.4em}}
\makeatother

\begin{document}

\maketitle
\vspace{-2.5em}

\begin{abstract}
When data reject a model, the researcher must decide which identifying assumptions to give up. If several equally preferred repairs restore compatibility, the repair itself is unknown, and inference under any one of them can exclude structural values supported by the others. I propose Post-Relaxation Projection (PR), which carries one simultaneous confidence event through every repair that may be minimal. For a rectangular band on the repair-specific restrictions, PR contains exactly the post-relaxation targets compatible with that band. I also derive limits on learning the repair, its exact perturbation radius, oracle-equivalent adaptation, and versions for instrumental variables without a first-stage relevance bound. In cross-country data on institutions and income, the data do not settle which historical instrument to drop, yet every admissible minimum repair implies a positive effect, and the sign survives sizable direct effects of the instruments on income.
\end{abstract}

\begingroup
\endgroup

\section{Introduction}\label{sec:intro}

Empirical work rarely stops when a maintained model is rejected. The usual response is to weaken it: drop an exclusion restriction, remove a block of moment inequalities, or relax a shape or behavioral restriction. This step is innocuous when one relaxation is clearly the right one. It is not innocuous when several relaxations are judged equally costly on substantive grounds, each is just enough to restore compatibility with the data, and they support different values of the structural parameter. Choosing among them then changes the object being estimated, not only the estimator. In this paper each repair is a candidate relaxation written down before the sample is examined, the repairs are ranked by a scientific order fixed in advance, and the post-refutation target is built from every minimum feasible repair.

Let $\theta$ denote the structural parameter, taking values in a compact set $\Theta$, and let $R$ index a finite menu of repairs fixed before observing the sample. Repair $R$ has ex ante scientific cost $c(R)$ and identified set $\Theta_R(P)\subseteq\Theta$ under the population distribution $P$. Define
\[
c^\star(P)=\min_{R:\Theta_R(P)\neq\varnothing}c(R),\qquad
\Rstar(P)=\{R:\Theta_R(P)\neq\varnothing,\ c(R)=c^\star(P)\}.
\]
When several repairs attain the minimum feasible cost, singling out one of them adds a restriction that the repair rule does not contain. The post-relaxation target is therefore
\begin{equation}\label{eq:target-intro}
\boxed{\ThetaMR(P)=\bigcup_{R\in\Rstar(P)}\Theta_R(P).}
\end{equation}
The cost can encode equal deletion costs, unequal substantive priorities, or any other ranking chosen in advance, and Section \ref{subsec:partial-order} replaces it with any finite strict partial order. With two candidate instruments, for instance, if both equal-cost one-deletion repairs are feasible, both repair-specific identified sets belong to the target. In every case the target is conditional on the candidate restrictions and their ranking: the data determine which repairs are compatible with the population, not what a relaxation should cost.

The difficulty is that $\Rstar(P)$ is a discrete population object that has to be learned from the same data used for inference. Two populations that cannot be told apart at the usual local rates may have different minimum repairs and therefore different targets. Selecting a repair and then doing inference can fail even if the second step is exact, because the selection step may delete a whole component of the target. Throughout, fix $0<\alpha<1$; $\Phi$ is the standard-normal cdf, $z_{1-\alpha}=\Phi^{-1}(1-\alpha)$, and $\Pp_P$ denotes probability under $P$.

A scalar example shows the problem. Two repairs have the same ex ante cost. The first always supports $[1,2]$. The second supports $[m,0]$ when $m\le0$ and is infeasible when $m>0$. At $m=0$ both repairs are population-minimum, and the target, which I write $\ThetaMR(m)$ in this example, is $\ThetaMR(0)=\{0\}\cup[1,2]$. Let $X_i\sim N(m,1)$, $\bar X_n=n^{-1}\sum_{i=1}^nX_i$, and suppose the researcher reports a single repair: the second when $\bar X_n\le0$ and the first otherwise. Write $\Pp_0$ for probability at $m=0$, and let $C_n^{\mathrm{hard}}$ be the exact population identified set of the selected repair. There is no inferential error within the selected repair, yet
\begin{equation}\label{eq:half}
\boxed{\Pp_0\{0\in C_n^{\mathrm{hard}}\}=\tfrac12,\qquad \Pp_0\{\ThetaMR(0)\subseteq C_n^{\mathrm{hard}}\}=0.}
\end{equation}
The first probability concerns the single value $0$, the second the whole target. Selection loses $0$ half of the time and, because the target is disconnected, never reports all of it. I call a repair boundary \emph{target-changing} when crossing it adds values to $\ThetaMR(P)$ or removes them. With several unresolved decisions, the probability of retaining a structural value is governed by a Gaussian carrier region (Section \ref{sec:uncertainty}).

\paragraph{Information limits.}
Two lower bounds describe what any procedure can achieve. A two-point argument bounds the minimax risk of estimating the minimum-repair correspondence. A local asymptotic normality (LAN) argument bounds how quickly an honest procedure can discard a value that belongs to the target at the boundary but drops out under a contiguous alternative. Let $C_n$ be a confidence set and $P_{n,h}\{\cdot\}$ probability under the $n$-sample local law indexed by $h$, with $h=0$ the boundary law. If $\theta^\dagger$ is such a value in the experiment of Assumption \ref{ass:R}, with Fisher information $I$, then
\begin{equation}\label{eq:lower-bound-intro}
\liminf_n P_{n,h}\{\theta^\dagger\in C_n\}
\ge \Phi\!\left(z_{1-\alpha}-|h|\sqrt I\right).
\end{equation}
When the Gaussian membership region is convex, a simple exposed-face test attains this bound (Section \ref{sec:frontier}). Finite unions of IV repairs need not be convex, and for them the bound can be far from attainable.

\paragraph{Uniform inference without learning the repair.}
PR has no repair-selection step. Suppose repair $R$ has the primitive representation
\[
\Theta_R(P)=\{\theta\in\Theta:b_{Rj}(\theta,P)\le0,\ j=1,\ldots,K_R\},
\]
with $K_R<\infty$ restrictions. A single simultaneous confidence event covers all these functions for all repairs. PR keeps every repair that, on this event, could still be population-minimum and takes the union of the corresponding structural outer sets. (``Projection'' refers to carrying the event through the repair rule, not to a Euclidean projection.) On the event, every population-minimum repair and every value it supports are retained, so uniform coverage of the entire target needs neither consistent estimation of the optimal repair, nor of its active set, nor a derivative of the nonsmooth feasibility map.

When the feasibility bounds are computed from the same rectangular primitive band, the PR set is exactly the union of all post-relaxation targets compatible with that band (Theorem \ref{thm:bandwise-complete}). No smaller set built from the realized band can contain the target for every compatible configuration, and the conclusions PR certifies are exactly those the band implies. This is a statement about the information in the band, not about statistical efficiency, and it can fail when repairs share primitives and that information is exploited (Remark \ref{rem:shared-primitives}). Section \ref{subsec:partial-order} also gives the exact sup-norm radius within which the minimal-repair correspondence cannot change.

PR sits between two familiar alternatives. Hard selection is too narrow near a target-changing boundary, as the example shows. The union over every repair in the menu is too wide, because it keeps costly relaxations that a cheaper feasible repair has made unnecessary; the same is true of taking the union of Anderson--Rubin or sensitivity sets computed repair by repair. Sensitivity analysis along a chosen path of relaxations does not account for sampling uncertainty about which discrete relaxation is minimal. PR keeps exactly the branches that the band cannot rule out as population-minimum. Section \ref{subsec:benchmarks} compares these procedures in the application.

Coverage of the entire target also delivers conclusions shared by all minimum repairs. If the PR set lies inside a prespecified sign, threshold, or exclusion region, I say that the claim is \emph{certified}. Any number of such claims can be read from the same event without a multiplicity correction, but near a target-changing boundary certifying a new claim is subject to the same local information limit as retaining a value.

\paragraph{Resolution and adaptation.}
Broad protection is unnecessary once the repair configuration is well separated. Let $\rho_n$ be the stochastic resolution of the feasibility-index bounds and $\Delta(P)$ the distance to the nearest feasibility or minimality boundary. If $\Delta(P_n)/\rho_n\to\infty$, PR recovers the minimum repairs with probability approaching one, and a version that certifies them on a small independent pilot sample is first-order equivalent to oracle inference that knows them.

\paragraph{Weak first stages.}
In the IV model, uniformity need not rely on ratios of reduced-form coefficients. Let $\gamma_j$ and $\pi_j$ be the covariances of instrument $j$ with the outcome and with the endogenous regressor, after partialling out the controls. Pushing a simultaneous region for $(\gamma_j,\pi_j)_j$ through the minimum-repair map with a single common witness gives a set that is uniformly valid without any lower bound on $|\pi_j|$, including at $\pi_j=0$. When the first stages are nonzero but possibly weak, inverting Anderson--Rubin statistics instrument by instrument \citep{AndersonRubin1949,Fieller1954} gives a sharper set. Simulations calibrated to the application show why this matters: with the weakest first stage at $F\approx10$, as in the data, the regular ratio band covers only about $89\%$ of the time, whereas the score inversion stays at its nominal level.

\paragraph{Institutions application.}
The application revisits the cross-country regressions of Rodrik, Subramanian, and Trebbi \citeyearpar{RodrikSubramanianTrebbi2002NBER,RodrikSubramanianTrebbi2004}, hereafter RST, and asks whether the positive effect of institutions on income survives uncertainty about which historical instrument should be abandoned. The tolerance $r$ bounds how far each instrument may depart from exact validity, in units of the structural coefficient. In the 80-country sample, where settler mortality and two language-based instruments are all available, the data cannot separate the seven repairs that keep at least one instrument. Even so, every band-compatible minimum-repair target is positive for every $r\in[0,0.60]$ and every strictly positive additive deletion-cost vector, with or without a first-stage relevance bound. In RST's 140-country sample only the two language instruments are available and Hansen's test rejects their joint validity ($p<0.01$). Dropping one or the other gives intervals that barely overlap, and the conventional estimate under the rejected model excludes part of each. PR keeps both repairs: its lower end, about $0.75$, holds whichever instrument is dropped, while its upper end moves by about one unit of $\beta$. With equal costs PR coincides in both samples with the union of the one-instrument intervals from the same band. The theory's contribution there is to show that this union, rather than a selected repair or the uninformative structural bounds, is the honest report; the minimum-repair rule narrows it once the data certify a cheaper repair, which at the same estimates would take about three times as many countries and would then lift the lower bound at $r=0.42$ from $0.46$, the value for the union at that precision, to $0.87$.

The tolerance has a direct economic reading: it bounds the direct effect of an instrument on income relative to its first stage. At the estimated first stages, the sign in the 80-country sample withstands direct effects of $0.20$ to $0.26$ log point of income per standard deviation of each instrument, all at once; for the English-language share this is two-thirds of its reduced-form association with income. Carrying the first-stage uncertainty through cuts the guaranteed protection to about $0.13$ log point for the English-language share and to $0.04$--$0.05$ for the other two instruments, whose first stages the data cannot bound well away from zero. The evidence for the sign therefore rests mainly on the language instrument with the strongest first stage. Across both samples, a move across the interquartile range of the rule-of-law index raises log income by at least about half of the interquartile income gap when the instruments are held to exact validity; whether institutions account for all of the gap depends on which historical identifying argument one trusts. All of these statements are conditional on the prespecified repair menu and on relaxing as little as necessary.

\paragraph{Relation to existing work.}
The target builds on work on falsification and minimal relaxation. \citet{MastenPoirier2021} study the falsification frontier of continuous relaxations, and \citet{LiKedagniMourifie2024} take unions of identified sets across inclusion-minimal discrete relaxations; in the finite, compact setting of this paper, partial-order PR provides inference for that union. \citet{ApfelWindmeijer2024} extend the falsification-adaptive set to instruments that may violate either exclusion or conditional exogeneity. Sensitivity analysis studies how conclusions move along specified relaxations \citep{ConleyHansenRossi2012,MastenPoirier2026}; here sampling uncertainty also decides which discrete relaxation enters the target. The problem differs from set inference under a maintained model and from local misspecification \citep{ChernozhukovHongTamer2007,AndrewsSoares2010,BugniCanayGuggenberger2012,AndrewsKwon2024}. Constructions that maximize over a confidence set for a nuisance parameter are closer in spirit \citep{BergerBoos1994,RomanoShaikhWolf2014,McCloskey2017}, but there the unresolved nuisance does not change the target. The failure of select-then-infer is related to the post-model-selection problem of \citet{LeebPotscher2005}; the difference is that selection here changes the estimand, not only the sampling distribution of the estimator. Finally, \citet{Bei2025} studies inference on union bounds whose index set is known; in the present problem the index set is the minimum-repair correspondence and must itself be learned.

\paragraph{Outline.}
Section \ref{sec:population} describes the population geometry of minimum repair, Section \ref{sec:uncertainty} the local limits, Section \ref{sec:pr} the PR procedure, Section \ref{sec:frontier} the Gaussian benchmark, and Section \ref{sec:ra} resolution-adaptive inference. Section \ref{sec:iv} specializes to instrumental variables, Section \ref{sec:mc} reports simulations, and Section \ref{sec:application} the application. Proofs, the IV calibration theory, and additional empirical material are in the Appendix.

\section{Population Minimum Repair}\label{sec:population}

Let $(\cP,d_{\cP})$ be a metric space of distributions, let $(\Theta,d_{\Theta})$ be a nonempty compact metric parameter space, and let $\cR$ be a finite collection of admissible repairs. Each repair $R\in\cR$ has fixed ex ante cost $c(R)$ and identified set $\Theta_R(P)\subseteq\Theta$. For $\theta\in\Theta$ and $A,B\subseteq\Theta$, write $\dist\{\theta,B\}=\inf_{b\in B}d_\Theta(\theta,b)$, $\dist(A,B)=\inf_{a\in A,b\in B}d_\Theta(a,b)$, and $e(A,B)=\sup_{a\in A}\dist\{a,B\}$ for the directed excess of $A$ over $B$, with $\dist\{\theta,\varnothing\}=\dist(A,\varnothing)=+\infty$ and $e(\varnothing,B)=0$. For compact sets, possibly empty, $d_H(A,B)=\max\{e(A,B),e(B,A)\}$ is the extended Hausdorff distance; it is the ordinary one for nonempty sets, and $d_H(A,\varnothing)=+\infty$ for $A\ne\varnothing$.

\begin{assumption}[Population repair geometry]\label{ass:P}
For every $P\in\cP$: (i) at least one repair is feasible; (ii) each repair has a continuous feasibility index $q_R:\cP\to\RR$ satisfying $\Theta_R(P)\neq\varnothing$ if and only if $q_R(P)\le0$; and (iii) the correspondence $P\mapsto\Theta_R(P)$ has sequentially closed graph. The repair collection is finite, costs are fixed ex ante, and $\Theta$ is compact.
\end{assumption}

When each repair has the primitive representation of Section \ref{subsec:primitive}, continuity of the primitive functions in $\theta$ together with their uniform convergence along convergent sequences of distributions implies part (iii), and with the canonical index of that section also part (ii) (Appendix, Lemma \oaref{lem:primitive-graph-app}). PR coverage itself requires neither.

Define
\[
\mathcal F(P)=\{R\in\cR:\Theta_R(P)\neq\varnothing\},\qquad
c^\star(P)=\min_{R\in\mathcal F(P)}c(R),
\]
\[
\Rstar(P)=\{R\in\mathcal F(P):c(R)=c^\star(P)\},\qquad
\ThetaMR(P)=\bigcup_{R\in\Rstar(P)}\Theta_R(P).
\]
Because $\cR$ is finite and at least one repair is feasible (Assumption \ref{ass:P}(i)), both $c^\star(P)$ and $\Rstar(P)$ are well defined. Under Assumption \ref{ass:P}(iii), each repair-specific set is closed in compact $\Theta$; hence $\ThetaMR(P)$ is nonempty and compact, so the Hausdorff distances used below are well defined.

If repairs are deletion sets and cost rises strictly with inclusion, as it does under positive additive deletion costs, every cost-minimum feasible repair is inclusion-minimal, but not conversely: two inclusion-minimal repairs may carry different costs. The cost-minimum target is the union over inclusion-minimal relaxations only when those share the minimum cost.

The finite-repair formulation is not specific to IV: it covers any finite collection of exclusion, moment, shape, sign, behavioral, or measurement restrictions that can be relaxed in prespecified blocks, provided each repaired model has the primitive representation of Section \ref{sec:pr}.

\begin{theorem}[Population geometry of minimum repair]\label{thm:geometry}
Under Assumption \ref{ass:P}:
\begin{enumerate}[label=(\roman*)]
\item $c^\star(P)$ is lower semicontinuous.
\item If some $R\in\Rstar(P_0)$ satisfies $q_R(P_0)<0$, then locally $c^\star(P)=c^\star(P_0)$ and $\Rstar(P)\subseteq\Rstar(P_0)$.
\item If $q_R(P_0)<0$ for every $R\in\Rstar(P_0)$, then locally $\Rstar(P)=\Rstar(P_0)$.
\item Let $P_n\to P_0$ and define $\mathscr E=\{R:R\in\Rstar(P_n)\text{ infinitely often}\}$. If $\theta^\dagger\in\ThetaMR(P_0)$ but $\theta^\dagger\notin\Theta_R(P_0)$ for every $R\in\mathscr E$, then there exist $\delta>0$ and $N<\infty$ such that
\[
\dist\{\theta^\dagger,\ThetaMR(P_n)\}\ge\delta\qquad\forall n\ge N.
\]
Consequently,
\[
\liminf_{n\to\infty}d_H\{\ThetaMR(P_n),\ThetaMR(P_0)\}\ge\delta>0.
\]
\end{enumerate}
\end{theorem}

Parts (i)--(iii) say that strict feasibility stabilizes both the minimum cost and the repair correspondence. Part (iv) describes the mechanism by which the target can jump; it does not classify all discontinuities. Its proof uses finiteness of the menu to find a repair that is minimal infinitely often and graph closedness to pass membership to the limit. Compactness of $\Theta$ makes the repair-specific sets and their finite union compact, and repairability keeps the target nonempty.

\begin{definition}[Target-changing repair boundary]\label{def:target-changing}
A distribution $P_0$ is a target-changing repair boundary if there exist a sequence $P_n\to P_0$, a value $\theta^\dagger\in\ThetaMR(P_0)$, and $\delta>0$ such that
\[
\dist\{\theta^\dagger,\ThetaMR(P_n)\}\ge \delta
\]
for all sufficiently large $n$, and the jump is not generated by an ordinary discontinuity of a repair-specific set that remains feasible. Formally, for every repair $R$ with $\Theta_R(P_0)\neq\varnothing$ and every subsequence $n_k$ on which $\Theta_R(P_{n_k})\neq\varnothing$,
\[
d_H\{\Theta_R(P_{n_k}),\Theta_R(P_0)\}\to0.
\]
The extra continuity clause isolates changes attributable to the feasibility or minimality status of a repair rather than to a discontinuity inside a fixed feasible repaired model. Theorem \ref{thm:geometry}(iv) supplies the separation; the continuity clause is what justifies calling the jump a repair boundary.
\end{definition}

Two feasible repairs of equal cost both belong to $\Rstar(P)$, however negative their feasibility indices; $q_R$ records feasibility and is not a secondary ranking. The nonregular event is therefore a repair becoming feasible or infeasible.

\section{Repair Uncertainty and Local Adaptation}\label{sec:uncertainty}

Three criteria recur below and should be kept apart. For an estimator $\widehat\cR_n$ of the minimum-repair correspondence, repair recovery concerns $\Pp_P\{\widehat\cR_n=\Rstar(P)\}$, entire-target coverage concerns $\Pp_P\{\ThetaMR(P)\subseteq C_n\}$, and structural-value coverage concerns $\Pp_P\{\theta\in C_n\}$ for $\theta\in\ThetaMR(P)$. PR targets entire-target coverage, the Gaussian adaptation problem studies structural-value coverage, and learning the repair correspondence is a classification problem. Confidence procedures are measurable random closed subsets of $\Theta$, or measurable acceptance rules in the Gaussian limit experiment, so all coverage events below are well defined.

\subsection{Two-point repair classification}

Let $P_{n,0}$ and $P_{n,1}$ denote two sequences of local distributions whose population-minimum repair correspondences differ. For an estimator $\widehat\cR_n$, let
\[
e_{n,j}=P_{n,j}^n\{\widehat\cR_n\neq\Rstar(P_{n,j})\},\qquad j=0,1.
\]
For probability laws $P$ and $Q$, write $\|P-Q\|_{TV}=\sup_A|P(A)-Q(A)|$ for their total-variation distance.

\begin{theorem}[Repair classification]\label{thm:classification}
The minimax risk of estimating the population-minimum repair correspondence satisfies
\begin{equation}\label{eq:tv-bound}
\inf_{\widehat\cR_n}\max\{e_{n,0},e_{n,1}\}
\ge \frac{1-\|P_{n,0}^n-P_{n,1}^n\|_{TV}}{2}.
\end{equation}
If, in addition, $\|P_{n,0}^n-P_{n,1}^n\|_{TV}\to2\Phi(d_G/2)-1$, the total-variation distance between $N(0,\Sigma_G)$ and $N(h,\Sigma_G)$ for some shift $h$ and covariance $\Sigma_G$, where $d_G^2=h'\Sigma_G^+h$ for $h$ in the range of $\Sigma_G$ ($\Sigma_G^+$ the Moore--Penrose inverse) and $d_G=+\infty$ otherwise, then
\begin{equation}\label{eq:gaussian-bound}
\liminf_n\inf_{\widehat\cR_n}\max\{e_{n,0},e_{n,1}\}\ge\Phi(-d_G/2).
\end{equation}
\end{theorem}

The classification bound becomes a coverage statement when a structural value can be carried only by particular selected repair collections. For a repair collection $\mathcal D\subseteq\cR$, write
\[
\Theta_{\mathcal D}(P)=\bigcup_{R\in\mathcal D}\Theta_R(P).
\]

\begin{corollary}[Oracle-assisted select-then-infer failure]\label{cor:carrier}
Let $P_n\in\cP$ be a sequence and $\theta_n^\dagger\in\ThetaMR(P_n)$; all stochastic orders and probabilities in this corollary are under the $n$-sample law $P_n^n$. Let $\widehat{\mathcal D}_n\subseteq\cR$ denote the repair collection selected by a hard rule, and define
\[
\mathfrak G_n=\{\mathcal D\subseteq\cR:\theta_n^\dagger\in\Theta_{\mathcal D}(P_n)\}.
\]
Suppose $C_n^{\mathrm{hard}}$ and $\Theta_{\widehat{\mathcal D}_n}(P_n)$ are nonempty compact sets with probability tending to one and
\begin{equation}\label{eq:selection-faithful}
d_H\{C_n^{\mathrm{hard}},\Theta_{\widehat{\mathcal D}_n}(P_n)\}=o_{P_n}(1).
\end{equation}
Suppose also that, for some $\delta>0$,
\begin{equation}\label{eq:carrier-separation}
\inf_{\mathcal D\notin\mathfrak G_n}
\dist\{\theta_n^\dagger,\Theta_{\mathcal D}(P_n)\}\ge\delta
\end{equation}
eventually, where the infimum is over collections with nonempty $\Theta_{\mathcal D}(P_n)$. Then
\begin{equation}\label{eq:carrier-bound}
P_n^n\{\theta_n^\dagger\in C_n^{\mathrm{hard}}\}
\le P_n^n\{\widehat{\mathcal D}_n\in\mathfrak G_n\}+o(1).
\end{equation}
Because $\theta_n^\dagger\in\ThetaMR(P_n)$, the same right-hand side also bounds
$P_n^n\{\ThetaMR(P_n)\subseteq C_n^{\mathrm{hard}}\}$.
If, in an isolated-boundary experiment, the asymptotic carrier event is $\{Z\le0\}$ with $Z\sim N(0,1)$ and all other repair decisions are separated, the bound in \eqref{eq:carrier-bound} converges to $1/2$. More generally, with several unresolved Gaussian decisions it converges to the probability of the corresponding carrier region, which equals $2^{-k}$ only when the region is the negative orthant of $k$ independent standard normal decisions.
\end{corollary}

The corollary separates repair-selection failure from within-repair inference error. A procedure can estimate one repair label accurately and still miss the post-relaxation target when another same-cost minimum repair carries structural values absent from the selected target.

\subsection{The local adaptation problem}

Suppose a value $\theta^\dagger$ belongs to the boundary target and disappears under a contiguous alternative. The confidence procedure faces a one-sided testing problem: exclusion of $\theta^\dagger$ is a rejection of the boundary state.

\begin{assumption}[Regular LAN boundary]\label{ass:R}
Along a scalar local submodel $\{P_{n,h}:h\in\mathbb R\}\subset\cP$, for every fixed $h\in\mathbb R$, let $P_{n,h}^n$ denote the $n$-sample law and abbreviate probability under it by $P_{n,h}\{\cdot\}$. Use $o_{P_{n,0}}(1)$ for convergence in probability under $P_{n,0}^n$. Then
\[
\log\frac{dP_{n,h}^n}{dP_{n,0}^n}
=h\zeta_n-\frac12h^2 I+o_{P_{n,0}}(1),
\qquad \zeta_n\Rightarrow N(0,I),
\]
with $0<I<\infty$.
\end{assumption}

\begin{theorem}[Local adaptation lower bound]\label{thm:adapt-lb}
Suppose Assumption \ref{ass:R} holds and $\theta^\dagger\in\ThetaMR(P_{n,0})$ for all sufficiently large $n$. If
\[
\liminf_n P_{n,0}\{\ThetaMR(P_{n,0})\subseteq C_n\}\ge1-\alpha,
\]
then for every fixed $h$,
\begin{equation}\label{eq:adapt-lb}
\liminf_n P_{n,h}\{\theta^\dagger\in C_n\}
\ge \Phi\!\left[z_{1-\alpha}-|h|\sqrt I\right].
\end{equation}
If, additionally, for some $\delta>0$, $\dist\{\theta^\dagger,\ThetaMR(P_{n,h})\}\ge\delta$ eventually, then
\[
\liminf_n P_{n,h}\{e(C_n,\ThetaMR(P_{n,h}))\ge\delta\}
\ge \Phi\!\left[z_{1-\alpha}-|h|\sqrt I\right].
\]
\end{theorem}

The proof is the usual LAN power-envelope argument; what the repair structure adds is the local paths along which the target changes through feasibility or minimality status while every repair-specific set stays well behaved (Definition \ref{def:target-changing}). The scalar example of the introduction is such a path with $I=1$. Appendix \oaref{oa:iv-lan} gives an IV version with four instruments and equal deletion costs: two pairs of instruments are both feasible at the boundary, and a local move of one coefficient makes the second pair infeasible, so a component of the target at distance $10r$ from the rest disappears. If that coefficient is estimated with asymptotic variance $\sigma^2$, independently of the others, then $I=\sigma^{-2}$, and any honest procedure keeps the vanished component with probability at least $\Phi(z_{1-\alpha}-|h|/\sigma)$.

\section{Uniform Post-Relaxation Projection}\label{sec:pr}

The construction retains the finite repair menu and fixed ex ante costs and assumes \emph{repairability}: for every $P\in\cP$, at least one $R\in\cR$ satisfies $\Theta_R(P)\neq\varnothing$. Then $c^\star(P)$, $\Rstar(P)$, and $\ThetaMR(P)$ are all well defined. PR itself does not need continuity of $P\mapsto q_R(P)$ or sequentially closed repair graphs; those conditions were used only for the population stability result.

\subsection{Primitive representation}\label{subsec:primitive}

For each repair $R$, suppose
\begin{equation}\label{eq:primitive-rep}
\Theta_R(P)=\{\theta\in\Theta:b_{Rj}(\theta,P)\le0,\ j=1,\ldots,K_R\},
\end{equation}
where $K_R<\infty$, $\Theta$ is compact, and $\theta\mapsto b_{Rj}(\theta,P)$ is continuous. If a repair leaves no substantive restriction, take $K_R=1$ and set $b_{R1}(\theta,P)\equiv-1$.

For a continuous array $f=(f_1,\ldots,f_{K_R})$ on $\Theta$, define $b_R(P)=(b_{Rj}(\cdot,P))_{j=1}^{K_R}$ and the canonical min--max map and its induced feasibility index by
\begin{equation}\label{eq:Q}
Q_R(f)=\min_{\theta\in\Theta}\max_{1\le j\le K_R} f_j(\theta),
\qquad
q_R^Q(P)=Q_R\{b_R(P)\}.
\end{equation}
By compactness and continuity, $q_R^Q(P)\le0$ if and only if $\Theta_R(P)\neq\varnothing$, and $Q_R$ is coordinatewise monotone and one-Lipschitz in the sup norm, $|Q_R(f)-Q_R(g)|\le\|f-g\|_\infty$.

PR does not require this particular scale. Any scalar $q_R(P)$ with
\begin{equation}\label{eq:q-sign}
q_R(P)\le0\quad\Longleftrightarrow\quad\Theta_R(P)\neq\varnothing
\end{equation}
is allowed; in the IV application of Section \ref{sec:iv}, $q_R=L_R-U_R$ is twice the canonical index when the structural bounds do not bind.

\subsection{A simultaneous primitive and feasibility band}

\begin{assumption}[Common primitive-and-feasibility honesty]\label{ass:E}
For every $R,j$, the random endpoints $\underline b_{Rj,n}(\theta)$ and $\overline b_{Rj,n}(\theta)$ are jointly measurable in the data and $\theta$, continuous in $\theta$ almost surely, and satisfy $\underline b_{Rj,n}(\theta)\le\overline b_{Rj,n}(\theta)$ for all $\theta$ almost surely. For every $R$, the scalar endpoints $\underline q_{R,n}$ and $\overline q_{R,n}$ are measurable and satisfy $\underline q_{R,n}\le\overline q_{R,n}$ almost surely. Define
\begin{equation}\label{eq:E-event}
\mathcal E_n(P)=\left\{
\begin{array}{l}
\underline b_{Rj,n}(\theta)\le b_{Rj}(\theta,P)\le\overline b_{Rj,n}(\theta)\quad\forall R,j,\theta,\\
\underline q_{R,n}\le q_R(P)\le\overline q_{R,n}\quad\forall R
\end{array}
\right\}.
\end{equation}
Assume
\begin{equation}\label{eq:E}
\liminf_n\inf_{P\in\cP}\Pp_P\{\mathcal E_n(P)\}\ge1-\alpha.
\end{equation}
\end{assumption}

Here $\mathcal E_n(P)$ is one joint event over all primitive and feasibility bounds; marginal coordinatewise coverage is not enough.

If $q_R=q_R^Q$, one valid construction is
\begin{equation}\label{eq:q-bounds-canonical}
\underline q_{R,n}=Q_R(\underline b_{R,n}),
\qquad
\overline q_{R,n}=Q_R(\overline b_{R,n}),
\end{equation}
because $Q_R$ is monotone. Assumption \ref{ass:E} also permits sharper scalar bounds obtained directly from contrasts or exact algebra, as in the IV implementation below.

Repair $R$ is \emph{plausibly population-minimum} when its own feasibility remains possible and no cheaper repair is known to be feasible:
\begin{equation}\label{eq:plausible}
\widehat{\mathfrak R}_n=
\left\{R:\underline q_{R,n}\le0,\ \overline q_{S,n}>0\ \forall S:c(S)<c(R)\right\}.
\end{equation}
The strict inequality for cheaper repairs matters: if $\overline q_{S,n}=0$, cheaper repair $S$ is feasible throughout the coordinatewise box and every more costly repair is ruled out.

For each repair define the structural outer set
\begin{equation}\label{eq:C_R}
C_{n,R}=\{\theta\in\Theta:\underline b_{Rj,n}(\theta)\le0\ \forall j\},
\end{equation}
and let
\begin{equation}\label{eq:PR}
C_n^{\PR}=\bigcup_{R\in\widehat{\mathfrak R}_n}C_{n,R}.
\end{equation}
The feasibility sandwich decides which branches remain possible, and the lower primitive band supplies the structural outer set on each retained branch; a retained branch with an empty outer set contributes nothing.

For the canonical construction, let $\mathfrak B_n$ be the set of continuous primitive arrays $f=(f_{Rj})$ satisfying $\underline b_{Rj,n}\le f_{Rj}\le\overline b_{Rj,n}$ pointwise. Define $\Theta_R(f)=\{\theta:f_{Rj}(\theta)\le0\ \forall j\}$ and $q_R(f)=Q_R(f_R)$, let $\mathfrak B_n^{\mathrm{rep}}$ contain the arrays with at least one feasible repair, and define $T(f)$ by the same minimum-cost union rule as $\ThetaMR(P)$.

\begin{theorem}[Bandwise completeness of canonical PR]\label{thm:bandwise-complete}
Suppose the scalar feasibility bounds are canonical as in \eqref{eq:q-bounds-canonical}. For every realization of the primitive rectangle,
\begin{equation}\label{eq:bandwise-complete}
C_n^{\PR}
=
\bigcup_{f\in\mathfrak B_n^{\mathrm{rep}}}T(f).
\end{equation}
Hence any $D\subseteq\Theta$ satisfying $T(f)\subseteq D$ for every $f\in\mathfrak B_n^{\mathrm{rep}}$ must contain $C_n^{\PR}$. Moreover, for every claim region $A\subseteq\Theta$,
\begin{equation}\label{eq:bandwise-certificate}
C_n^{\PR}\subseteq A
\quad\Longleftrightarrow\quad
T(f)\subseteq A\quad\forall f\in\mathfrak B_n^{\mathrm{rep}}.
\end{equation}
The same statements hold for any fixed finite strict partial order after replacing minimum-cost repairs by $\prec$-minimal feasible repairs.
\end{theorem}

The result is conditional on the realized rectangle: canonical PR discards a value only when every primitive configuration permitted by the band excludes it. A smaller honest set must therefore use information that the rectangle does not record, such as restrictions linking the primitives of different repairs or a nonrectangular joint region. Separately supplied sharper feasibility bounds still deliver the coverage in Theorem \ref{thm:PR}, but the equality in \eqref{eq:bandwise-complete} then requires a joint-compatibility condition.

\begin{remark}[Shared primitives]\label{rem:shared-primitives}
The set $\mathfrak B_n$ lets the primitive bounds of different repairs move independently. When two repairs share a primitive function, as two IV repairs that retain the same instrument do, the honest information set is smaller, and canonical PR can be strictly larger than the union of targets compatible with it. For example, let $\Theta=[-1,2]$, $r=0.1$, let $R_0$ retain both instruments (cost $0$) and $R_1$ only the first (cost $1$), and let the band be $\beta_1\in[0,1]$, $\beta_2\in[0.10,0.15]$. Repair by repair, $R_0$ cannot be shown feasible, so $R_1$ is retained and contributes $[-0.1,1.1]$. If the same $\beta_1$ must enter both repairs, a value $\theta<0$ requires $\beta_1<0.1$, hence $|\beta_1-\beta_2|\le0.15<2r$; then $R_0$ is feasible and its target $[\beta_2-r,\beta_1+r]$ lies in $[0,\infty)$. Hence $[-0.1,0)$ belongs to $C_n^{\PR}$ but to no target compatible with the shared band, and the gap persists if the menu also contains the repair that keeps only the second instrument. Theorem \ref{thm:bandwise-complete} is therefore a sharpness result for the rectangular information set. Exploiting shared primitives requires a joint projection of the kind used in Section \ref{subsec:weak-iv}. In the simulation of Appendix \oaref{oa:shared}, built on this configuration, the joint projection is shorter than PR in at most a fifth of the samples, and its average length is within one percent of PR's.
\end{remark}

For repair recovery, measure uncertainty in the scale that actually governs the repair rule:
\begin{equation}\label{eq:eta}
\eta_n(P)=\max_{R\in\cR}\max\left\{|\underline q_{R,n}-q_R(P)|,|\overline q_{R,n}-q_R(P)|\right\}.
\end{equation}

\begin{assumption}[Statistical resolution rate]\label{ass:C}
There exists a deterministic sequence $\rho_n\downarrow0$ such that
\begin{equation}\label{eq:C-uniform}
\lim_{M\to\infty}\limsup_{n\to\infty}\sup_{P\in\cP}
\Pp_P\{\eta_n(P)>M\rho_n\}=0.
\end{equation}
\end{assumption}

Define the population repair-resolution margin
\begin{align}\label{eq:Delta}
\Delta(P)=\min\Big(&\{-q_R(P):R\in\Rstar(P)\}\nonumber\\
&\cup\{q_S(P):c(S)<c^\star(P)\}\nonumber\\
&\cup\{q_S(P):c(S)=c^\star(P),\ S\notin\Rstar(P)\}\Big),
\end{align}
with empty groups omitted. In particular, if a true minimum repair lies on its feasibility boundary, then $\Delta(P)=0$. Since $\Delta$ and $\eta_n$ use the same scale, the recovery condition is unchanged by a common positive rescaling of all feasibility indices. Rescaling repairs by different factors does change $\Delta(P)/\rho_n$; with indices on different scales, the condition should be read repair by repair, each margin against the resolution of its own index.

Under Assumption \ref{ass:E}, $C_n^{\PR}$ is a measurable random closed set (Appendix, Lemma \oaref{lem:PR-measurable}).

\paragraph{The procedure in four steps.} For readers who want the computation before the theory:
\begin{enumerate}[leftmargin=*,itemsep=1pt]
\item List the repairs $R\in\cR$, their costs (or a partial order), and for each repair the restrictions $b_{Rj}(\theta,P)\le0$, all fixed before seeing the data.
\item Compute one simultaneous $(1-\alpha)$ band for all primitive functions and all feasibility indices, giving $\underline b_{Rj,n}\le\overline b_{Rj,n}$ and $\underline q_{R,n}\le\overline q_{R,n}$.
\item Keep repair $R$ if $\underline q_{R,n}\le0$ and $\overline q_{S,n}>0$ for every cheaper $S$; this is the set $\widehat{\mathfrak R}_n$ in \eqref{eq:plausible}.
\item Report $C_n^{\PR}=\bigcup_{R\in\widehat{\mathfrak R}_n}\{\theta:\underline b_{Rj,n}(\theta)\le0\ \forall j\}$, and read off any conclusion that holds on the whole set.
\end{enumerate}
All the statistical work is in step 2.

\begin{theorem}[Uniform PR and repair recovery]\label{thm:PR}
Under the primitive repair setup and Assumption \ref{ass:E}, on $\mathcal E_n(P)$,
\[
\Rstar(P)\subseteq\widehat{\mathfrak R}_n
\qquad\text{and}\qquad
\ThetaMR(P)\subseteq C_n^{\PR}.
\]
Consequently,
\begin{equation}\label{eq:PR-coverage}
\liminf_n\inf_{P\in\cP}\Pp_P\{\ThetaMR(P)\subseteq C_n^{\PR}\}\ge1-\alpha.
\end{equation}
If, in addition, Assumption \ref{ass:C} holds and a sequence $P_n\in\cP$ satisfies
\begin{equation}\label{eq:repair-recovery-rate}
\Delta(P_n)/\rho_n\to\infty,
\end{equation}
then
\begin{equation}\label{eq:repair-recovery}
\Pp_{P_n}\{\widehat{\mathfrak R}_n=\Rstar(P_n)\}\to1.
\end{equation}
\end{theorem}

For recovery, a costlier false repair is blocked by a cheaper true minimum repair, whereas a same-cost infeasible repair is excluded when its own lower feasibility endpoint becomes positive. The latter is the third group in $\Delta(P)$ in \eqref{eq:Delta}.

Coverage and repair recovery answer different questions. Coverage \eqref{eq:PR-coverage} holds at a fixed level and does not require learning the repair; recovery \eqref{eq:repair-recovery} describes what happens under separation. In a regular fixed-dimensional problem $\rho_n\asymp n^{-1/2}$, so the repair correspondence becomes learnable only when the feasibility margins dominate root-$n$ uncertainty.

The inclusions in Theorem \ref{thm:PR} are deterministic on $\mathcal E_n(P)$, so PR inherits exactly the coverage of the simultaneous band: a finite-sample band gives finite-sample coverage, and no further resampling is needed once the band is available. Section \ref{sec:iv} verifies the band uniformly for the IV class used here.

Theorem \ref{thm:bandwise-complete} compares PR only with sets that use the rectangular band. It is not a statement about the shortest honest set among all procedures, and PR need not attain the bound of Section \ref{sec:frontier}.

\subsection{Partial orders and the stability radius of the repair map}\label{subsec:partial-order}

Scalar costs are convenient, but the projection argument does not need them. Let $\prec$ be a fixed strict partial order on the finite repair menu, chosen before observing the sample, where $S\prec R$ means that repair $S$ is scientifically preferred to repair $R$. The ex-ante qualification is essential: if the order were estimated from the same sample, uncertainty about the ranking would create another target-selection problem not covered by the procedure below. Define the population-minimal feasible repairs and their structural target by
\begin{equation}\label{eq:partial-order-target}
\Rstar_{\prec}(P)
=
\{R:q_R(P)\le0,\ q_S(P)>0\ \forall S\prec R\},
\qquad
\Theta^{MR}_{\prec}(P)
=
\bigcup_{R\in\Rstar_{\prec}(P)}\Theta_R(P).
\end{equation}
When $S\prec R$ means $c(S)<c(R)$, \eqref{eq:partial-order-target} reduces exactly to the scalar cost-minimum target. When repairs are deletion sets and $S\prec R$ means $S\subsetneq R$, it instead retains every inclusion-minimal feasible deletion.

Define
\begin{equation}\label{eq:partial-order-plausible}
\widehat{\mathfrak R}_{n,\prec}
=
\{R:\underline q_{R,n}\le0,\ \overline q_{S,n}>0\ \forall S\prec R\},
\qquad
C^{\PR}_{n,\prec}
=
\bigcup_{R\in\widehat{\mathfrak R}_{n,\prec}}C_{n,R}.
\end{equation}
This is again an existential projection of the coordinatewise feasibility-index box: $R$ is retained exactly when the box contains some vector for which $R$ is $\prec$-minimal feasible.
For recovery, let
\[
a_R(P)=\min_{S\prec R}q_S(P),
\qquad \min\varnothing=+\infty,
\]
and define the repair-specific classification margin
\begin{equation}\label{eq:partial-order-margin}
\delta_R^\prec(P)
=
\begin{cases}
\min\{-q_R(P),\,a_R(P)\},&R\in\Rstar_\prec(P),\\[2pt]
\max\{q_R(P),\,-a_R(P)\},&R\notin\Rstar_\prec(P),
\end{cases}
\qquad
\Delta_\prec(P)=\min_{R\in\cR}\delta_R^\prec(P),
\end{equation}
where the minimum over an empty set of predecessors is $+\infty$, so that $a_R(P)=+\infty$ and $-a_R(P)=-\infty$ for a repair with no predecessor. A zero value records an unresolved feasibility or dominance boundary.

To make the meaning of this margin precise, for a deterministic vector $v=(v_R)_{R\in\cR}$ define
\[
\mathcal M_\prec(v)=\{R:v_R\le0,\ v_S>0\ \forall S\prec R\},
\]
and extend $a_R$, $\delta_R^\prec$, and $\Delta_\prec$ by replacing $q(P)$ with $v$.

\begin{proposition}[Exact minimality-stability radius]\label{prop:partial-order-radius}
For every $v\in\RR^{|\cR|}$ and every repair $R$,
\begin{equation}\label{eq:repair-status-radius}
\delta_R^\prec(v)
=
\inf\left\{\|w-v\|_\infty:
\mathbf 1\{R\in\mathcal M_\prec(w)\}
\neq
\mathbf 1\{R\in\mathcal M_\prec(v)\}
\right\}.
\end{equation}
Consequently,
\begin{equation}\label{eq:repair-map-radius}
\Delta_\prec(v)
=
\inf\{\|w-v\|_\infty:\mathcal M_\prec(w)\neq\mathcal M_\prec(v)\}.
\end{equation}
Thus $\mathcal M_\prec(w)=\mathcal M_\prec(v)$ whenever $\|w-v\|_\infty<\Delta_\prec(v)$, and for every $\varepsilon>0$ some $w$ with $\|w-v\|_\infty<\Delta_\prec(v)+\varepsilon$ has $\mathcal M_\prec(w)\neq\mathcal M_\prec(v)$. The infima need not be attained, because feasibility is a weak inequality and predecessor infeasibility a strict one. If lower and upper feasibility-index endpoints satisfy
$\max_R\max\{|\underline q_R-v_R|,|\overline q_R-v_R|\}<\Delta_\prec(v)$,
then the interval rule
$\{R:\underline q_R\le0,\ \overline q_S>0\ \forall S\prec R\}$
equals $\mathcal M_\prec(v)$. Conversely, no deterministic radius strictly larger than $\Delta_\prec(v)$ can guarantee this equality uniformly over endpoint vectors. For the cost order $S\prec R$ iff $c(S)<c(R)$, whenever the model is repairable, $\Delta_\prec\{q(P)\}=\Delta(P)$.
\end{proposition}

For a minimal $R$, status changes when $v_R$ crosses above zero or a predecessor crosses from $v_S>0$ to $v_S\le0$, giving distances $-v_R$ and $a_R(v)$. For a nonminimal $R$, entry requires correcting its own infeasibility, predecessor feasibility, or both, giving $\max\{v_R,-a_R(v)\}$. Taking the finite minimum gives \eqref{eq:repair-map-radius}. If $v_R=0$ for a minimal repair, the radius is zero; using $v_S\ge0$ for predecessor infeasibility would be wrong because equality is feasible.

Replacing ``cheaper'' by ``preceding'' in the proof of Theorem \ref{thm:PR} shows that, on the same event $\mathcal E_n(P)$, $C^{\PR}_{n,\prec}$ contains $\Theta^{MR}_\prec(P)$, and that $\widehat{\mathfrak R}_{n,\prec}=\Rstar_\prec(P_n)$ with probability tending to one when $\Delta_\prec(P_n)/\rho_n\to\infty$ (Appendix, Corollary \oaref{cor:partial-order}). With deletion-set repairs ordered by strict inclusion, this gives inference on the union over inclusion-minimal relaxations studied by \citet{LiKedagniMourifie2024}, in the finite, compact setting of this paper. When the ranking itself is unsettled, one can fix a finite family of orders in advance and take unions: coverage, bandwise completeness, and recovery carry over, and the least separated order governs recovery (Appendix, Proposition \oaref{thm:order-robust}). The local limits of Section \ref{sec:uncertainty} concern target membership and therefore apply to any of these targets.

\subsection{Certified conclusions}\label{subsec:certificates}

\begin{corollary}[Uniform certificates for repair-robust conclusions]\label{cor:robust-conclusions}
Let $T(P)$ denote $\ThetaMR(P)$, $\Theta^{MR}_\prec(P)$, or the order-robust target of Appendix \oaref{oa:partial-order}, and let $C_n\subseteq\Theta$ be a measurable random closed set satisfying
\[
\liminf_n\inf_{P\in\cP}\Pp_P\{T(P)\subseteq C_n\}\ge1-\alpha.
\]
Fix continuous functionals $g_1,\ldots,g_L:\Theta\to\RR$ and use the conventions $\inf_\varnothing g_\ell=+\infty$ and $\sup_\varnothing g_\ell=-\infty$. Then, with asymptotic probability at least $1-\alpha$ uniformly in $P$, $[\inf_{C_n}g_\ell,\sup_{C_n}g_\ell]\supseteq[\inf_{T(P)}g_\ell,\sup_{T(P)}g_\ell]$ for every $\ell$ simultaneously. Fix claim regions $A_1,\ldots,A_K\subseteq\Theta$ for which $\{C_n\subseteq A_k\}$ is measurable. Then
\begin{equation}\label{eq:robust-claim-fwer}
\limsup_n\sup_{P\in\cP}\Pp_P\{\exists k:C_n\subseteq A_k,\ T(P)\not\subseteq A_k\}\le\alpha.
\end{equation}
Thus one entire-target event protects all prespecified universal conclusions read deterministically from $C_n$; no additional multiplicity correction is needed. If a claim is not certified, the procedure draws no conclusion about it.

The same logic applies to any fixed finite sensitivity family---for example, alternative cost rankings or tolerances---when one common event of probability at least $1-\alpha$ implies target containment for every reported set, as it does for deterministic PR projections of that event. This does not justify choosing the sensitivity family after inspecting the data.
\end{corollary}

When containment in $C_n$ is all that is used, this rule cannot be improved, and near a target-changing boundary a newly true claim is certified with probability at most $\Phi(|h|\sqrt I-z_{1-\alpha})$ along the experiment of Assumption \ref{ass:R}; at $\alpha=0.05$ and $|h|\sqrt I=1$ this is about $0.26$ (Appendix, Corollary \oaref{cor:certificate-resolution}).

\section{The Statistical Price of Repair Uncertainty}\label{sec:frontier}

Fix a candidate structural value $\theta$. In a regular local experiment, let
\[
Y\sim N(h,\Sigma),\qquad \Sigma\succ0.
\]
Write $P_u=N(u,\Sigma)$ for the Gaussian law with mean $u$. Let $K_\theta^0$ be the set of local means under which $\theta$ remains in the post-relaxation target, and define
\begin{equation}\label{eq:K-closure}
K_\theta=\operatorname{cl}(K_\theta^0).
\end{equation}
Taking the closure is without loss: $u\mapsto P_u\{Y\in G\}$ is continuous for every measurable acceptance region $G$, so honesty over $K_\theta^0$ and over $K_\theta$ are equivalent. Define
\begin{equation}\label{eq:mahalanobis}
d_\Sigma(h,K_\theta)=\inf_{u\in K_\theta}\sqrt{(h-u)'\Sigma^{-1}(h-u)}.
\end{equation}
An acceptance region $G$ is honest for $\theta$ if $P_u\{Y\in G\}\ge1-\alpha$ for every $u\in K_\theta$. For $h\notin K_\theta$, define the optimal retention probability
\[
\mathcal A_\theta^\star(h)=
\inf_{G:\,\inf_{u\in K_\theta}P_u\{Y\in G\}\ge1-\alpha}
P_h\{Y\in G\},
\]
where the infimum is over measurable acceptance regions. Thus $1-\mathcal A_\theta^\star(h)$ is the largest rejection probability achievable at $h$ while retaining $\theta$ with probability at least $1-\alpha$ throughout its null membership region.

For closed convex $K_\theta$, let $\Pi_{K_\theta}^\Sigma(h)$ denote the unique Mahalanobis projection of $h$ onto $K_\theta$. Given a supporting pair $(a_\theta,\kappa_\theta)$, define the exposed-normal region
\begin{equation}\label{eq:EF-region}
\mathcal H_\theta^{EF}=
\left\{h\notin K_\theta:
\begin{array}{l}
 a_\theta'\Pi_{K_\theta}^\Sigma(h)=\kappa_\theta,\\
 h-\Pi_{K_\theta}^\Sigma(h)=\lambda\Sigma a_\theta\text{ for some }\lambda>0
\end{array}
\right\}.
\end{equation}

\begin{theorem}[Sharp Gaussian membership bound for convex regions]\label{thm:frontier}
Let $h\notin K_\theta$. If $K_\theta$ is nonempty, closed, and convex, then
\begin{equation}\label{eq:frontier}
\mathcal A_\theta^\star(h)=\Phi\!\left[z_{1-\alpha}-d_\Sigma(h,K_\theta)\right].
\end{equation}
Suppose in addition that $K_\theta\subseteq\{u:a_\theta'u\le\kappa_\theta\}$ for some $a_\theta\neq0$. Then for every $h\in\mathcal H_\theta^{EF}$,
\begin{equation}\label{eq:EF-stat}
\mathcal T_\theta(Y)=\frac{a_\theta'Y-\kappa_\theta}{\sqrt{a_\theta'\Sigma a_\theta}}
\end{equation}
with acceptance rule $\mathcal T_\theta(Y)\le z_{1-\alpha}$ attains \eqref{eq:frontier}. For a finite nonconvex union of closed convex repair branches, the right side of \eqref{eq:frontier} remains a lower bound, but equality need not hold unless one supporting half-space is valid over the entire union. With $K_\theta=\{-2,2\}$, $\Sigma=1$, $h=0$, and $\alpha=0.05$, for example, the bound is $0.36$ while the best honest test retains $\theta$ with probability $0.67$ (Appendix \oaref{oa:nonconvex}).
\end{theorem}

A plug-in version with uniformly consistent estimates of the supporting hyperplane and of $\Sigma$ attains the same bound uniformly on compact exposed-normal alternatives (Appendix).

I call \eqref{eq:EF-stat} the \emph{exposed-face (EF) procedure}; an exposed face is the intersection of a convex set with a supporting hyperplane. Theorem \ref{thm:frontier} is a sharp benchmark for convex Gaussian membership, not an efficiency claim for PR, whose finite-repair membership regions are generally nonconvex.

\section{Resolution-Adaptive Inference}\label{sec:ra}

Once the feasibility margins dominate statistical resolution, Theorem \ref{thm:PR} recovers the repair correspondence, and inference can adapt at the same point.

Split the sample into an independent pilot of size $m_n$ and main sample of size $N_n=n-m_n$, indexed by $\mathcal I_{1n}$ and $\mathcal I_{2n}$, with
\[
m_n\to\infty,\qquad m_n/n\to0.
\]
Let the pilot confidence level approach one, with $\bar\alpha_n\downarrow0$. Write $\underline q^{\mathrm{pil}}_{R,n}$ and $\overline q^{\mathrm{pil}}_{R,n}$ for the pilot feasibility bounds. A repair is \emph{definitely feasible} if $\overline q^{\mathrm{pil}}_{R,n}\le0$ and \emph{definitely infeasible} if $\underline q^{\mathrm{pil}}_{R,n}>0$. These are the exact implications of the pilot interval, since feasibility is the weak inequality $q_R\le0$. For a cost value $a$, define
\[
\mathscr D_n(a)=\{R:c(R)=a,\ \overline q^{\mathrm{pil}}_{R,n}\le0\}.
\]
The layer $a$ is certified when every cheaper repair is definitely infeasible, every repair in the layer is classified as either definitely feasible or definitely infeasible, and $\mathscr D_n(a)\neq\varnothing$. At most one layer can be certified, because a certified layer contains a definitely feasible repair. The resolution-adaptive PR set, denoted $C_n^{\RA}$, uses the oracle procedure for a certified layer and PR otherwise. Formally, for a deterministic repair collection $\mathscr D$, let $C_{n,\mathscr D}^{O,M}$ denote the chosen regular main-sample confidence procedure that treats $\mathscr D$ as known, and let $C_{n,\mathrm{main}}^{\PR}$ denote PR built from the main sample. Define
\[
C_n^{\RA}=\begin{cases}
C_{n,\mathscr D_n(a)}^{O,M},&\text{if layer }a\text{ is certified},\\
C_{n,\mathrm{main}}^{\PR},&\text{if certification fails}.
\end{cases}
\]
For comparison, let $C_{n,\mathscr D}^{O,F}$ denote the corresponding procedure based on the full sample and define the infeasible oracle benchmark
\[
C_n^{\mathrm{oracle}}=C_{n,\Rstar(P)}^{O,F},
\]
which knows the population-minimum repair correspondence. Write $\mathcal E_n^{\mathrm{pil}}(P)$ for the analogue of $\mathcal E_n(P)$ formed from the pilot bounds and $\eta^{\mathrm{pil}}_{m_n,\bar\alpha_n}(P)$ for the analogue of $\eta_n(P)$ in \eqref{eq:eta}.

\begin{assumption}[Resolution adaptation]\label{ass:A}
(i) \emph{Pilot.} The pilot and main samples are independent, the pilot endpoints satisfy the requirements of Assumption \ref{ass:E}, $\sup_{P\in\cP}\Pp_P\{(\mathcal E_n^{\mathrm{pil}}(P))^c\}\le\bar\alpha_n+o(1)$ with $\bar\alpha_n\downarrow0$, and $\eta^{\mathrm{pil}}_{m_n,\bar\alpha_n}$ satisfies \eqref{eq:C-uniform} with a deterministic rate $\rho^{\mathrm{pil}}_{m_n,\bar\alpha_n}\downarrow0$ in place of $\rho_n$.
(ii) \emph{Main-sample honesty.} The fallback $C_{n,\mathrm{main}}^{\PR}$ and, for every deterministic $\mathscr D$, the oracle procedure $C_{n,\mathscr D}^{O,M}$ are functions of the main sample only; the fallback has asymptotic entire-target coverage at least $1-\alpha$ uniformly over $\cP$, and $C_{n,\mathscr D}^{O,M}$ has it uniformly over $\{P\in\cP:\Rstar(P)=\mathscr D\}$ whenever that set is nonempty.
(iii) \emph{Oracle equivalence.} Along any sequence on which oracle equivalence is asserted, the full-sample and main-sample oracle procedures share an asymptotically linear representation and are locally Lipschitz in their inputs, as stated in Appendix \oaref{oa:ra-conditions}.
\end{assumption}

\begin{theorem}[Uniform honesty and oracle recovery]\label{thm:RA}
Under Assumption \ref{ass:A}(i)--(ii),
\begin{equation}\label{eq:RA-honesty}
\liminf_n\inf_{P\in\cP}\Pp_P\{\ThetaMR(P)\subseteq C_n^{\RA}\}\ge1-\alpha.
\end{equation}
If, in addition, $P_n\in\cP$ and
\begin{equation}\label{eq:RA-separation}
\Delta(P_n)/\rho^{\mathrm{pil}}_{m_n,\bar\alpha_n}\to\infty,
\end{equation}
then the pilot certifies $\Rstar(P_n)$ with probability tending to one. If Assumption \ref{ass:A}(iii) also holds along this sequence, then
\begin{equation}\label{eq:RA-oracle}
d_H(C_n^{\RA},C_n^{\mathrm{oracle}})=o_{P_n}(n^{-1/2}).
\end{equation}
\end{theorem}

In the fixed-dimensional IV class, where each oracle set is a finite union of intervals, Appendix \oaref{oa:ra-conditions} gives a primitive sufficient condition for part (iii): every population branch must have width $\kappa_n$ with $\sqrt n\,\kappa_n\to\infty$. Near branch collapse the condition fails: a branch can then be empty in the main sample and nonempty in the full sample, and the two sets differ by a fixed distance. The probability of this event is of order $\sqrt{m_n/n}$, so it vanishes, but slowly; in the Gaussian calculation of Appendix \oaref{oa:ra-conditions}, with $m_n=n/\log n$, it is still about $0.10$ at $n=25{,}600$. Theorem \ref{thm:RA} does not cover these sequences.

Because $m_n=o(n)$ and $\bar\alpha_n\downarrow0$, the pilot resolves boundaries more slowly than the full sample: with $m_n=n/\ell_n$, $\ell_n\to\infty$, a regular pilot band has $\rho^{\mathrm{pil}}_{m_n,\bar\alpha_n}\asymp c_{1-\bar\alpha_n}\sqrt{\ell_n/n}$, where $c_{1-\bar\alpha_n}$ is the simultaneous critical value, so certification requires $\sqrt n\,\Delta(P_n)/(c_{1-\bar\alpha_n}\sqrt{\ell_n})\to\infty$. This extra separation is the price of a vanishing pilot error probability. The same construction works under a finite partial order: the pilot then has to resolve only minimality status, not every feasibility sign (Appendix, Corollary \oaref{cor:partial-order-ra}).

\section{Potentially Invalid Instruments}\label{sec:iv}

For the scalar linear IV specialization, some instruments may be invalid. The bounded-validity restriction below is in the spirit of plausibly exogenous sensitivity analysis \citep{ConleyHansenRossi2012}; related work includes imperfect-instrument bounds \citep{NevoRosen2012}, invalid-instrument selection \citep{KangZhangCaiSmall2016}, and falsification-adaptive IV analysis \citep{ApfelWindmeijer2024}. The focus here is different: sampling uncertainty over which discrete repair is population-minimum and hence which repair-specific sets enter the post-relaxation target.

Fix $P\in\cP$, and write $\EE_P$ for expectation under $P$. Let $Y$ be the outcome, $X$ the endogenous regressor, $C$ a vector of included exogenous controls (including a constant), and $Z_1,\ldots,Z_J$ the candidate instruments. After partialling out the controls, write
\begin{equation}\label{eq:iv-model}
Y=\beta X+U.
\end{equation}
For candidate instrument $Z_j$, let tildes denote population residuals after projection on the included controls and set
\[
\pi_j=\EE_P[\widetilde Z_j\widetilde X],\qquad |\pi_j|\ge\underline\pi>0,
\]
with just-identified estimand
\begin{equation}\label{eq:iv-center}
\beta_j=\frac{\EE_P[\widetilde Z_j\widetilde Y]}{\EE_P[\widetilde Z_j\widetilde X]}.
\end{equation}
For each candidate validity claim, fix a prespecified tolerance $r_j\ge0$ and impose the sign-invariant bounded-validity restriction
\begin{equation}\label{eq:iv-validity}
\frac{|\EE_P[\widetilde Z_j(\widetilde Y-\beta\widetilde X)]|}{|\EE_P[\widetilde Z_j\widetilde X]|}\le r_j.
\end{equation}
Because $\EE_P[\widetilde Z_j(\widetilde Y-\beta\widetilde X)]=\pi_j(\beta_j-\beta)$, \eqref{eq:iv-validity} is equivalent to $|\beta-\beta_j|\le r_j$. Thus validity claim $j$ contributes the interval $I_j=[\beta_j-r_j,\beta_j+r_j]$.

Let the structural coefficient lie in fixed nondegenerate ex ante bounds
\[
\beta\in[\underline\beta,\overline\beta],\qquad \underline\beta<\overline\beta.
\]
Index a repair $R$ by the set of abandoned validity claims, and let $S_R$ denote the complementary set of retained instruments. Define
\begin{equation}\label{eq:iv-LU}
L_R=\max\left\{\underline\beta,\max_{j\in S_R}(\beta_j-r_j)\right\},\qquad
U_R=\min\left\{\overline\beta,\min_{j\in S_R}(\beta_j+r_j)\right\}.
\end{equation}
With the conventions $\max_{j\in\varnothing}(\beta_j-r_j)=-\infty$ and $\min_{j\in\varnothing}(\beta_j+r_j)=+\infty$, deleting every candidate validity claim yields the structural-bound repair $[\underline\beta,\overline\beta]$, so the repair menu remains feasible even when all candidate instrument restrictions conflict. Then $\Theta_R=[L_R,U_R]$ when $L_R\le U_R$, and
\begin{equation}\label{eq:iv-q}
q_R=L_R-U_R.
\end{equation}
In the primitive form \eqref{eq:primitive-rep}, each retained claim $j$ contributes the restrictions $\beta_j-r_j-\theta\le0$ and $\theta-\beta_j-r_j\le0$, and the structural bounds contribute $\underline\beta-\theta\le0$ and $\theta-\overline\beta\le0$.

\subsection{Three-instrument benchmark}\label{subsec:iv-benchmark}

The tolerance $r_j$ bounds the direct violation relative to the first-stage covariance, which makes the intervals comparable across instruments. Repair costs record which identifying claims are more costly to abandon. The benchmark uses equal costs only to normalize the ranking, not as a claim that all forms of invalidity are equally serious; Section \ref{subsec:iv-envelope} removes the dependence on costs altogether.

For the benchmark application, each abandoned validity claim has unit ex ante cost,
\begin{equation}\label{eq:unit-cost}
c(R)=|R|.
\end{equation}
When the structural bounds do not bind, a common tolerance $r$ gives exact algebra: a repair retaining one instrument has
\begin{equation}\label{eq:single-q}
q_R=-2r,
\end{equation}
a repair retaining instruments $j,k$ has $q_R=|\beta_j-\beta_k|-2r$, and the full three-instrument model has $q_\varnothing=\max_{j<k}|\beta_j-\beta_k|-2r$.
When the structural bounds do not bind, a one-instrument repair is feasible for every value of the coefficients, so there is nothing to estimate. I use these identities only when the simultaneous coefficient rectangle, enlarged by the tolerance, lies strictly inside $(\underline\beta,\overline\beta)$, and the bounded formulas \eqref{eq:iv-LU} otherwise; on the coverage event the identities then hold in population. A simultaneous band for each contrast $\beta_j-\beta_k$ gives bounds for $|\beta_j-\beta_k|-2r$, and the maximum of the three gives a sandwich for $q_\varnothing$. I call these the \emph{contrast-aware} feasibility bounds. They are sharper than $Q_R$ applied to marginal coefficient intervals and enter Assumption \ref{ass:E} directly. At $r=0$ every one-instrument repair has $q_R=0$, so the strict predecessor rule in \eqref{eq:plausible} rules out the delete-all repair. Values within rounding error of zero are treated as zero.

\subsection{Influence functions and contrast-aware calibration}

Let $\widetilde Y$, $\widetilde X$, and $\widetilde Z_j$ denote population residuals after projection on the included controls. Under iid sampling, a nonsingular second-moment matrix for the controls, finite moments sufficient for the sample projection coefficients, and the relevance bound above, the sample FWL ratio estimator is asymptotically linear with influence function
\begin{equation}\label{eq:IF}
\psi_j(W)=
\frac{\widetilde Z_j(\widetilde Y-\beta_j\widetilde X)}{\EE_P[\widetilde Z_j\widetilde X]}.
\end{equation}
The Appendix derives this representation with the residualizing projections estimated.

For three instruments the empirical PR calibration uses the six-dimensional vector
\begin{equation}\label{eq:six-vector}
V=(\beta_1,\beta_2,\beta_3,\beta_1-\beta_2,\beta_1-\beta_3,\beta_2-\beta_3)'.
\end{equation}
Its influence vector is
\[
\Psi=(\psi_1,\psi_2,\psi_3,\psi_1-\psi_2,\psi_1-\psi_3,\psi_2-\psi_3)'.
\]
The six coordinates are linearly dependent, so the covariance matrix is singular; the studentized maximum needs only marginal variances bounded away from zero.

\begin{proposition}[Uniform six-functional IV band]\label{prop:iv-uniform-band}
Fix $\epsilon>0$, $B<\infty$, $\lambda_C>0$, $\underline\pi>0$, and $\underline\sigma>0$. Let $\cP_{IV}$ be a fixed-dimensional iid class for $W=(Y,X,Z_1,Z_2,Z_3,C')'$ such that
\[
\sup_{P\in\cP_{IV}}\EE_P\|W\|^{4+\epsilon}\le B,\qquad
\inf_{P\in\cP_{IV}}\lambda_{\min}\EE_P[CC']\ge\lambda_C,
\]
\[
\inf_{P\in\cP_{IV}}\min_{j\le3}|\EE_P[\widetilde Z_j\widetilde X]|\ge\underline\pi,
\qquad
\inf_{P\in\cP_{IV}}\min_{k\le6}\operatorname{Var}_P(\Psi_k)\ge\underline\sigma^2,
\]
where $\Psi_k$ denotes coordinate $k$ of the influence vector just defined. Write $V(P)$ for the population value of \eqref{eq:six-vector}. On samples for which the control residualization and all three FWL ratios are well defined and all six plug-in marginal variances are positive, let the critical value be the conditional $(1-\alpha)$ quantile of the Gaussian-multiplier maximum; otherwise use the full-band convention in the Appendix. Then
\[
\liminf_n\inf_{P\in\cP_{IV}}
\Pp_P\{V(P)\text{ is covered coordinatewise}\}\ge1-\alpha.
\]
The primitive IV bands and the contrast-aware feasibility-index sandwiches constructed from this event therefore satisfy Assumption \ref{ass:E} uniformly on $\cP_{IV}$. The same holds when the conditional quantile is approximated by a number of multiplier draws that grows with $n$.
\end{proposition}

The Appendix handles estimated residualization, singular Gaussian limits, and quantile continuity, the last through the anti-concentration inequality of \citet{ChernozhukovChetverikovKato2015Comparison}, which needs only marginal variances bounded away from zero; no denominator is truncated away from zero. The relevance bound is substantive: if a first stage approaches zero, the ratio experiment changes, and Proposition \ref{prop:iv-uniform-band} no longer applies. Section \ref{subsec:weak-iv} removes the bound.

\subsection{Conclusions that hold over all tolerances and costs}\label{subsec:iv-envelope}

The tolerance and the cost ranking are chosen by the researcher, and a reader may want to know whether a conclusion depends on either. With a common tolerance $r_j=r$, a single simultaneous band answers this for a whole range of tolerances and for every additive cost vector at once. Let $J$ be the number of candidate validity claims ($J=3$ in the application), let $[\ell_{j,n},u_{j,n}]$ be the coordinates of the simultaneous band of Proposition \ref{prop:iv-uniform-band} that correspond to $\beta_j$, and write $C^{\PR}_{n,r,w}$ for the PR set computed with tolerance $r$ and additive cost weights $w=(w_1,\ldots,w_J)$. For $r\ge0$, define the conclusion envelope
\begin{equation}\label{eq:iv-uniform-envelope}
B_n^-(r)=\min_j\ell_{j,n}-r,
\qquad
B_n^+(r)=\max_j u_{j,n}+r.
\end{equation}
Fix a maximum tolerance $\bar r\ge0$.

\begin{corollary}[Tolerance- and cost-uniform IV conclusion envelope]\label{cor:iv-uniform-sign}
Use the contrast-aware feasibility bounds of Section \ref{subsec:iv-benchmark}. Suppose the deletion menu contains all subsets of the candidate validity claims and repair cost is additive,
\[
c_w(R)=\sum_{j\in R}w_j,\qquad w_j>0.
\]
If
\begin{equation}\label{eq:iv-uniform-nonbinding}
[B_n^-(\bar r),B_n^+(\bar r)]\subset(\underline\beta,\overline\beta),
\end{equation}
then for every $r\in[0,\bar r]$ the structural bounds are nonbinding for every coefficient vector in the simultaneous rectangle, the delete-all repair is not PR-plausible for any strictly positive additive weight vector $w$, and
\begin{equation}\label{eq:iv-uniform-containment}
C_{n,r,w}^{\PR}\subseteq[B_n^-(r),B_n^+(r)]
\qquad
\forall r\in[0,\bar r],\quad w\in(0,\infty)^J.
\end{equation}
Consequently, any threshold $\tau<B_n^-(r)$ is certified as a universal lower conclusion, $C_{n,r,w}^{\PR}\subset(\tau,\infty)$, for every such $w$; symmetrically, any $\tau>B_n^+(r)$ is certified as a universal upper conclusion. In particular, $B_n^-(\bar r)>0$ implies positivity uniformly over the whole tolerance-and-cost class. On the common primitive coverage event, the corresponding population post-relaxation targets inherit all of these conclusions simultaneously.
\end{corollary}

The envelope separates uncertainty about the repair from uncertainty about the conclusion: the repair can remain unresolved while every value supported by every plausible minimum repair lies on the same side of a threshold. The realized $B_n^-(r)$ is also a lower confidence bound, and a rounded value below it is a summary of that bound rather than a hypothesis chosen in advance. The contrast-aware bounds matter: with the canonical sandwich, a one-instrument repair cannot be shown feasible for small $r$, the delete-all repair stays plausible, and PR returns the structural bounds. The proof is in the Appendix.

\subsection{Uniform reduced-form inference without a relevance bound}\label{subsec:weak-iv}

Let $\mathcal J=\{1,\ldots,J\}$ index the candidate instruments, and let $W$ collect the observed variables $Y$, $X$, $Z_1,\ldots,Z_J$, and $C$. Define
\begin{equation}\label{eq:weak-rf}
\gamma_j=\EE_P[\widetilde Z_j\widetilde Y],\qquad
\pi_j=\EE_P[\widetilde Z_j\widetilde X],\qquad
\nu(P)=(\gamma_1,\ldots,\gamma_J,\pi_1,\ldots,\pi_J)'.
\end{equation}
For a deletion repair $R\subseteq\mathcal J$, the retained reduced-form restrictions are
\begin{equation}\label{eq:weak-primitive}
b_j(\theta,\nu)=|\gamma_j-\theta\pi_j|-r_j|\pi_j|\le0,
\qquad j\in\mathcal J\setminus R,
\end{equation}
where $r_j\ge0$. If $\pi_j\ne0$, this restriction is exactly $|\theta-\gamma_j/\pi_j|\le r_j$; no ratio is used when $\pi_j=0$. Set
\begin{equation}\label{eq:weak-repair-set}
\Theta_R^W(\nu)=\{\theta\in\Theta:b_j(\theta,\nu)\le0\ \forall j\in\mathcal J\setminus R\},
\end{equation}
with $\Theta_{\mathcal J}^W(\nu)=\Theta$ for delete-all. For repairs retaining at least one instrument, define
\begin{equation}\label{eq:weak-q}
q_R^W(\nu)=\min_{\theta\in\Theta}\max_{j\in\mathcal J\setminus R}b_j(\theta,\nu),
\end{equation}
and set $q_{\mathcal J}^W(\nu)=-1$. Compactness of $\Theta$ and continuity of \eqref{eq:weak-primitive} make every $q_R^W$ continuous and sign-equivalent to feasibility. Let $\Theta_W^{MR}(\nu)$ denote the minimum-cost union formed from the sets in \eqref{eq:weak-repair-set}, using the fixed repair costs from Section \ref{sec:population}.

To construct a simultaneous reduced-form region, let $\widehat\nu$ collect the sample residual cross-products after projection on the controls, with influence vector
\begin{equation}\label{eq:weak-if}
\varphi(W;P)=
\big(\widetilde Z_1\widetilde Y-\gamma_1,\ldots,
\widetilde Z_J\widetilde Y-\gamma_J,
\widetilde Z_1\widetilde X-\pi_1,\ldots,
\widetilde Z_J\widetilde X-\pi_J\big)'.
\end{equation}
Fix a compact rectangle $\mathcal N\subset\RR^{2J}$ containing $\nu(P)$ for every distribution in the class. Let $\mathcal K_n$ be the studentized Gaussian-multiplier rectangle for $\nu$ at level $1-\alpha$, built from the plug-in version of \eqref{eq:weak-if} and intersected with $\mathcal N$, with $\mathcal K_n=\mathcal N$ when residualization fails or the intersection is empty (Appendix \oaref{oa:weak-pr} gives the construction). Finally define
\begin{equation}\label{eq:weak-pr}
C_n^{W\!PR}=\operatorname{cl}\left\{\bigcup_{\nu\in \mathcal K_n}\Theta_W^{MR}(\nu)\right\}.
\end{equation}
The closure cannot generally be omitted: minimum-repair status requires strict infeasibility of every cheaper repair, so the graph of $\nu\mapsto\Theta_W^{MR}(\nu)$ need not be closed even though each repair-specific graph is closed.

\begin{theorem}[Uniform weak-IV post-relaxation projection]\label{thm:weak-pr}
Fix $\epsilon>0$, $B<\infty$, $\lambda_C>0$, and $\underline\sigma>0$. Let $\mathcal P_W$ be a fixed-dimensional iid class for $W=(Y,X,Z_1,\ldots,Z_J,C')'$ such that
\[
\sup_{P\in\mathcal P_W}\EE_P\|W\|^{4+\epsilon}\le B,
\qquad
\inf_{P\in\mathcal P_W}\lambda_{\min}\EE_P[CC']\ge\lambda_C,
\]
\[
\nu(P)\in\mathcal N,
\qquad
\inf_{P\in\mathcal P_W}\min_{k\le2J}\operatorname{Var}_P\{\varphi_k(W;P)\}
\ge\underline\sigma^2.
\]
No lower bound is imposed on any $|\pi_j|$. Then
\begin{equation}\label{eq:weak-pr-coverage}
\liminf_n\inf_{P\in\mathcal P_W}
\Pp_P\{\Theta_W^{MR}\{\nu(P)\}\subseteq C_n^{W\!PR}\}\ge1-\alpha.
\end{equation}
The covariance matrix of \eqref{eq:weak-if} may be singular. The set $C_n^{W\!PR}$ is a measurable random closed set. Conditional on $\mathcal K_n$, it is the inclusion-smallest closed subset of $\Theta$ containing $\Theta_W^{MR}(\nu)$ for every $\nu\in \mathcal K_n$. The same statements hold for a fixed strict partial order and for a prespecified finite family of orders after taking one closure around the corresponding union.
\end{theorem}
The proof is in the Appendix.

The theorem covers fixed zero first stages and every local-to-zero sequence in $\mathcal P_W$. It needs nondegenerate marginal score variances but not a nonsingular joint covariance matrix.

The boundary cases separate identification failure from refutation. If $\pi_j=\gamma_j=0$, restriction $j$ is satisfied by every $\theta$ and carries no identifying content. If $\pi_j=0$ but $\gamma_j\ne0$, every repair retaining $j$ is refuted. If all instruments retained by a repair have $\pi_j=\gamma_j=0$, then that repair supports all of $\Theta$. A minimum repair of this kind forces the post-relaxation target, and hence any honest entire-target confidence set, to include the structural bounds. A two-instrument example shows why the distinction matters. Let $\gamma_1=\pi_1=1$, $r=0.2$, equal costs, and $\Theta=[-10,10]$. If $(\gamma_2,\pi_2)=(0.5,0)$, the second claim is refuted, the minimum repair drops it, and the target is $[0.8,1.2]$. If $(\gamma_2,\pi_2)=(0,0)$, the second instrument is merely irrelevant, the repair that keeps both claims is feasible at cost zero, and the target is again $[0.8,1.2]$. The sets agree but the diagnoses differ: in the first case the data say that a claim must be abandoned, in the second they say nothing about it. Ratio-based inference cannot tell the two apart, because both have $\pi_2=0$.

The set \eqref{eq:weak-pr} checks the unreduced score $\gamma_j-\theta\pi_j$ and never divides by $\widehat\pi_j$. The common $\nu\in\mathcal K_n$ is essential: repair-by-repair extrema can use incompatible witnesses.

The rectangle $\mathcal K_n$ treats $\gamma_j$ and $\pi_j$ as if they could move independently, which is what makes Theorem \ref{thm:weak-pr} valid even at $\pi_j=0$. When the first stages are nonzero but possibly weak, much of that freedom is wasted, because the restriction on $\theta$ depends on $(\gamma_j,\pi_j)$ only through the score $\gamma_j-\theta\pi_j$. Inverting the score directly gives a sharper set. For $b\in\RR$, let $\widehat\omega_{ij}(b)=\widehat Z_{ji}(\widehat Y_i-b\widehat X_i)$, where hats denote sample residuals after projection on the controls, let $\widehat s_j^2(b)$ be the sample variance of $\widehat\omega_{ij}(b)$ over $i$, and for a critical value $c$ define
\begin{equation}\label{eq:score-set}
F_{n,j}=\Big\{b\in\RR:\ \sqrt n\,\Big|n^{-1}\sum_{i=1}^n\widehat\omega_{ij}(b)\Big|\le c\,\widehat s_j(b)\Big\}.
\end{equation}
This is the Anderson--Rubin set for the ratio $\gamma_j/\pi_j$; it may be an interval, a union of two half-lines, the whole line, or empty \citep{Fieller1954}. For a ratio vector $b=(b_1,\ldots,b_J)$, let $T(b)$ be the minimum-cost union of the sets $[b_j-r_j,b_j+r_j]\cap\Theta$ over the retained claims, with $\Theta$ for the delete-all repair, and define
\begin{equation}\label{eq:score-pr}
C_n^{S\!PR}=\operatorname{cl}\Big\{\bigcup_{b\in F_{n,1}\times\cdots\times F_{n,J}}T(b)\Big\}.
\end{equation}

\begin{proposition}[Score-inversion PR]\label{prop:score-pr}
Let $\mathcal P_S$ satisfy the moment and control conditions of Theorem \ref{thm:weak-pr}, with $\pi_j(P)\ne0$ for every $j$ but no lower bound on $|\pi_j(P)|$, and suppose the covariance matrix of $(\widetilde Z_j\widetilde Y,\widetilde Z_j\widetilde X)$ has smallest eigenvalue at least $\underline\sigma^2>0$ for every $j$ and $P\in\mathcal P_S$. Write $\beta_j(P)=\gamma_j/\pi_j$. If $c\ge z_{1-\alpha/(2J)}$, then
\[
\liminf_n\inf_{P\in\mathcal P_S}\Pp_P\{\ThetaMR(P)\subseteq C_n^{S\!PR}\}\ge1-\alpha.
\]
Suppose further that the tolerance is common, $r_j=r$, that cost is additive with strictly positive weights, and that $F_{n,k}\subseteq\Theta$ for some $k$. Then the delete-all repair belongs to no target in \eqref{eq:score-pr}, and
\[
C_n^{S\!PR}\subseteq\Big[\min_j\inf F_{n,j}-r,\ \max_j\sup F_{n,j}+r\Big]\cap\Theta .
\]
\end{proposition}

The proof is in Appendix \oaref{oa:score-pr}. At the true ratio the score has mean zero whatever the strength of the first stage, and the eigenvalue condition keeps the studentized score away from degeneracy even when $\beta_j(P)$ diverges along a local-to-zero sequence, so each Anderson--Rubin statistic is asymptotically standard normal uniformly over $\mathcal P_S$; Bonferroni over the $J$ instruments gives simultaneous coverage of $\beta(P)$, and the projection does the rest. The ratio vector $b$ is a common witness, so shared primitives are used exactly and Remark \ref{rem:shared-primitives} does not apply. Proposition \ref{prop:score-pr} gives up the case $\pi_j=0$, which Theorem \ref{thm:weak-pr} covers. In the application the difference is large: the rectangle of Theorem \ref{thm:weak-pr} keeps the positive sign only for tolerances below $0.17$, while the score-inversion set keeps it up to $r=0.60$ (Section \ref{subsec:weak-id-results}).

\subsection{Breakdown tolerances}\label{subsec:breakdown}

Corollary \ref{cor:iv-uniform-sign} and Proposition \ref{prop:score-pr} also say how much invalidity a conclusion can withstand, in the spirit of the breakdown points of \citet{MastenPoirier2020}. The tolerance has a direct economic reading. Write the income equation after partialling out the controls as $\widetilde Y=\beta\widetilde X+\delta_j\widetilde Z_j+e_j$ with $\EE(\widetilde Z_je_j)=0$, so that $\delta_j$ is the direct effect of instrument $j$ on the outcome holding the endogenous regressor fixed, and let $\pi_j^{\mathrm{fs}}=\EE(\widetilde Z_j\widetilde X)/\EE(\widetilde Z_j^2)$ be its first-stage coefficient. Substituting this equation into \eqref{eq:iv-validity} shows that the restriction is exactly $|\delta_j|\le r_j|\pi_j^{\mathrm{fs}}|$. Since the reduced-form coefficient of $Z_j$ is $\beta_j\pi_j^{\mathrm{fs}}$, an equivalent statement is that at most a fraction $r_j/|\beta_j|$ of the reduced-form association between $Z_j$ and the outcome is direct.

For a threshold $\tau$, a cost vector $w$, and a maximum tolerance $\bar r$, define the population breakdown tolerance
\[
r^\star_{\tau,w}(P)=\sup\Big\{s\in[0,\bar r]:\ \ThetaMR_{r,w}(P)\subset(\tau,\infty)\ \text{for every } r\in[0,s]\Big\},
\]
with $\sup\varnothing=-\infty$, where $\ThetaMR_{r,w}(P)$ is the target computed with common tolerance $r$ and additive costs $w$. It is the largest common tolerance up to which every minimum-repair target lies above $\tau$. Because the minimum repair can change with $r$, the condition is imposed for every smaller tolerance, not only at $s$.

For a lower endpoint $\lambda_n$ computed from the data, let $\widehat r_{\tau,n}(\lambda_n)=\min\{\bar r,\lambda_n-\tau\}$ if $\lambda_n>\tau$ and $\widehat r_{\tau,n}(\lambda_n)=-\infty$ otherwise.

\begin{corollary}[Lower confidence bounds for breakdown tolerances]\label{cor:breakdown}
(i) Under the conditions of Proposition \ref{prop:iv-uniform-band}, use the contrast-aware feasibility bounds with all deletion subsets in the menu, and suppose \eqref{eq:iv-uniform-nonbinding} holds with probability approaching one uniformly in $P$. Then
\[
\liminf_n\inf_{P\in\cP_{IV}}\Pp_P\Big\{\inf_{w\in(0,\infty)^J}r^\star_{\tau,w}(P)\ge\widehat r_{\tau,n}\big(\min_j\ell_{j,n}\big)\Big\}\ge1-\alpha .
\]
(ii) Under the conditions of Proposition \ref{prop:score-pr}, suppose $F_{n,k}\subseteq\Theta$ for some $k$ with probability approaching one uniformly in $P$. Then the same statement holds over $\mathcal P_S$ with $\min_j\ell_{j,n}$ replaced by $\min_j\inf F_{n,j}$.
\end{corollary}

Both bounds hold simultaneously for every positive additive cost vector because the envelopes do not depend on $w$. The proof is in Appendix \oaref{oa:breakdown}.

Translating $\widehat r_{\tau,n}$ into a direct effect on the outcome requires a value for $\pi_j^{\mathrm{fs}}$; an estimate gives an order of magnitude but not a confidence statement. The next remark carries the first-stage uncertainty through.

\begin{remark}[Breakdown in units of direct effects]\label{rem:direct-effects}
Add the $J$ first-stage coefficients $\pi_j^{\mathrm{fs}}$ to the functionals of Proposition \ref{prop:iv-uniform-band}, calibrate one simultaneous band for the enlarged vector, and let $[\underline\pi_j,\overline\pi_j]$ be its first-stage coordinates and $\lambda_n^+$ the smallest lower endpoint of its $\beta_j$ coordinates. Put $m_j=\min\{|\underline\pi_j|,|\overline\pi_j|\}$ if $0\notin[\underline\pi_j,\overline\pi_j]$ and $m_j=0$ otherwise, and
\[
\widehat D_j=\max\{\lambda_n^+,0\}\,m_j\,\mathrm{sd}(Z_j),\qquad j=1,\ldots,J ,
\]
where $\mathrm{sd}(Z_j)$ is the sample standard deviation of instrument $j$, used only as a fixed unit of measurement.
Suppose the true direct effects satisfy $|\delta_j|\,\mathrm{sd}(Z_j)<\widehat D_j$ for every $j$. On the coverage event of the enlarged band, all $J$ validity claims then hold with the common tolerance $r_\delta=\max_j|\delta_j|/|\pi_j^{\mathrm{fs}}|<\lambda_n^+$, because $|\pi_j^{\mathrm{fs}}|\ge m_j$. No repair is needed: the model with every claim relaxed by $r_\delta$ holds, and since $\beta\ge\beta_j-r_\delta$ for every $j$ and $\beta_j\ge\lambda_n^+$ on the event, $\beta\ge\lambda_n^+-r_\delta>0$. Hence, if the enlarged vector satisfies the conditions of Proposition \ref{prop:iv-uniform-band}, then with asymptotic probability at least $1-\alpha$ uniformly over $\cP_{IV}$ the effect is positive whenever every direct effect is below $\widehat D_j$ outcome units per standard deviation of its instrument. A weak first stage makes $m_j$ small, and the protection it offers is correspondingly small.
\end{remark}

Section \ref{subsec:breakdown-results} reports the bounds for the application and translates them into direct effects of the historical instruments on income.

\section{Monte Carlo Evidence}\label{sec:mc}

The simulations check four predictions of the theory. Seeds, complete tables, and Monte Carlo standard errors are in the replication package.

\emph{Hard selection at a boundary.} In 300,000 replications of the scalar example of Section \ref{sec:intro}, hard selection covers the boundary value with probability $0.4999$ (s.e.\ $0.0009$), as the one-half limit predicts, and never covers the disconnected target. Along local alternatives the exposed-face procedure tracks the Gaussian retention bound of Theorem \ref{thm:frontier} to within simulation error.

\emph{Screening by PR.} The second design has three instruments and seven instrument-retaining repairs, with coefficients and covariance calibrated to the published RST moments and the tolerance set at the boundary where two one-deletion repairs tie. The benchmark is the \emph{restricted union} that always keeps all seven branches. At $n=80$, hard selection covers the target in only $18.6\%$ of replications, while PR and the restricted union have the same average length. As $n$ grows, PR drops branches that the data rule out, and the ratio of its average length to that of the restricted union falls to $0.466$ at $n=102{,}400$.

\emph{Adaptation.} In a separated design with $r=0.20$, the pilot certifies the correct repair correspondence with probability $0.999$ at $n=102{,}400$, and the ratio of the average lengths of the adaptive and oracle sets is $1.007$. The scaled Hausdorff distance $\sqrt{N_n}\,d_H(C_n^{\RA},C_n^{\mathrm{oracle}})$ falls more slowly, from $3.2$ at $n=6{,}400$ to $1.3$ at $n=102{,}400$, so set-level equivalence is visibly slower than equivalence in length. A final experiment confirms that the strict predecessor rule excludes the delete-all repair under all seven cost rankings. Appendix \oaref{oa:shared} compares PR with the joint projection that exploits shared coefficients.

\input{generated/mc_weak_id.tex}

\section{Institutions, Validity Restrictions, and Minimum Repair}\label{sec:application}

\subsection{Empirical design}

The application revisits the institutions-and-development regressions of \citet{RodrikSubramanianTrebbi2004}. The structural equation is
\begin{equation}\label{eq:RST-model}
Y_i=\mu_0+\beta\,\mathit{RULE}_i+\mu_1\mathit{DISTEQ}_i+U_i,
\end{equation}
where $Y$ is log GDP per capita, $\mathit{RULE}$ is the rule-of-law index, and $\mathit{DISTEQ}$ is distance from the equator. There are three candidate validity claims, one for each historical instrument: log settler mortality (SM), the fraction of the population speaking English (ENG), and the fraction speaking a major European language (EUR). I use two samples. The main one is the 80-country sample in which all three instruments are observed (Section \ref{subsec:empirical-results}); combining the three on one sample is a new exercise rather than a replication of a particular RST column. The second is RST's 140-country larger sample, in which settler mortality is missing and the two language instruments are jointly rejected (Section \ref{subsec:larger-sample}).

The goal is not a new preferred estimate of the effect of institutions but to ask whether the economic conclusion survives disagreement about where identification comes from. Settler mortality is the colonial-settlement instrument of \citet{AcemogluJohnsonRobinson2001}; its data have themselves been disputed \citep{Albouy2012}. The language shares come from a different tradition and were used by RST in place of settler mortality in their larger sample, where RST reported overidentification concerns about them. The three instruments rest on different exclusion arguments, so dropping one rather than another is a substantive choice, and the 80-country sample puts the competing strategies on the same countries.

The question is whether the choice of which claim to discard changes the sign of $\beta$, or an economically relevant lower bound on it, when the only maintained principle is to relax as little as the data require. A positive PR set means that every structural value supported by every band-compatible minimum repair is positive. It does not show that the exclusion restrictions hold, and it says nothing about a specification that treats every instrument as invalid.

The source file is matched to RST's 80-country working-paper sample and checked against the published summary statistics before any inference is run. The common sample, covariance estimator, calibration, structural bounds $[\underline\beta,\overline\beta]=[-10,10]$, tolerance grid, and cost-sensitivity schemes were recorded before the observation-level analysis; Appendix \oaref{oa:analysis-record} gives the details.

Each abandoned claim has unit cost, as in \eqref{eq:unit-cost}. Equal costs encode the view that none of the three historical arguments is privileged a priori: the case for settler mortality rests on colonial disease environments and the case for the language shares on European settlement and legal transplants, and each has been disputed on its own terms. Readers who rank them differently can read off the cost-uniform results in Section \ref{subsec:empirical-results} and the unequal-cost rows of Table \ref{tab:benchmarks}. The tolerance $r$ is measured in units of the structural coefficient: by \eqref{eq:iv-validity}, it bounds the direct effect of the instrument on income relative to its first-stage effect on the rule-of-law index. The standard deviation of $\mathit{RULE}$ is about $0.85$ in the 80-country sample, so a tolerance $r$ allows a direct effect equivalent to $0.85r$ log points of income per standard deviation of the institutions measure. The reference value $r=0.42$ was fixed from published moments before the observation-level file was opened (Appendix \oaref{oa:analysis-record}); since those moments come from the same sample, the choice is not independent of the outcome, which is why every conclusion is also stated for the whole range. It is about $39\%$ of the ENG estimate and $21$--$22\%$ of the SM and EUR estimates. The tolerance is a sensitivity scale, not an estimate of the size of any exclusion violation.

\input{generated/empirical_results.tex}

\section{Conclusion}\label{sec:conclusion}

When a model is refuted, several admissible relaxations may remain, and they may support different values of the parameter of interest. PR carries this uncertainty through a single simultaneous confidence statement. With a rectangular band on repair-specific primitives it reports exactly the post-relaxation targets that the band still permits, and it remains valid when the scientific ranking is a prespecified finite family of partial orders. How far the population sits from the nearest repair boundary determines whether the repair can be learned and how closely adaptive inference can approach the oracle.

PR is narrower than a worst case over every relaxation in the menu, which keeps dominated repairs, and broader than a single repair picked from the data, which can lose part of the target near a repair boundary.

The RST data illustrate both sides. In the 80-country sample the diagnosis and the economic conclusion do not resolve together: the data cannot tell which instrument claim, if any, should be dropped, yet every PR set is positive for $r\in[0,0.60]$ and every strictly positive additive cost vector, with or without a first-stage relevance bound. In the 140-country sample the model is rejected, and the two repairs that remain give different answers about how large the effect of institutions is. PR keeps both, and the conclusion it supports is a lower bound that holds under either one.

Two extensions seem natural. The first is a menu that grows with the sample, for example when many moment inequalities can be dropped in blocks; the simultaneous band then has to be calibrated in high dimension, and Theorem \ref{thm:bandwise-complete} suggests that exploiting shared primitives becomes more important as the menu grows. The second is a ranking of repairs that is itself uncertain in a way that cannot be captured by a finite family of orders. Infinite repair menus, noncompact parameter spaces, and rankings estimated from the data remain outside the present theory.

\begingroup
\begin{spacing}{1.0}

\input{bibliography.tex}
\end{spacing}
\endgroup

\clearpage
\appendix
\renewcommand{\thesection}{S\arabic{section}}
\renewcommand{\theequation}{S.\arabic{equation}}
\renewcommand{\thetable}{S\arabic{table}}
\setcounter{equation}{0}\setcounter{table}{0}
\begin{center}{\Large\bfseries Appendix}\end{center}

The appendix contains the proofs, the calibration theory for the IV implementation, and additional empirical material. Sections, equations, and tables carry the prefix S.

\section{Primitive repair lemmas}

Throughout, $\cP$ is the distribution class, $\Pp_P$ denotes probability under $P\in\cP$, $0<\alpha<1$, $\Phi$ is the standard-normal cdf, and $z_{1-\alpha}=\Phi^{-1}(1-\alpha)$. The finite repair menu is $\cR$, with fixed costs $c(R)$ and at least one feasible repair for every $P\in\cP$. Let $(\Theta,d_\Theta)$ be compact. For $R\in\cR$, let $K_R<\infty$ and let $f=(f_1,\ldots,f_{K_R})$ denote a continuous real-valued array on $\Theta$. Define
\[
\Theta_R(P)=\{\theta\in\Theta:b_{Rj}(\theta,P)\le0,\ j=1,\ldots,K_R\},
\qquad Q_R(f)=\min_{\theta\in\Theta}\max_{1\le j\le K_R} f_j(\theta),
\]
and let $q_R(P)$ be sign-equivalent to feasibility: $q_R(P)\le0$ iff $\Theta_R(P)\neq\varnothing$. Write $c^\star(P)=\min_{R:q_R(P)\le0}c(R)$, $\Rstar(P)=\{R:q_R(P)\le0,\ c(R)=c^\star(P)\}$, and $\ThetaMR(P)=\bigcup_{R\in\Rstar(P)}\Theta_R(P)$. For $B\subseteq\Theta$, $\dist\{\theta,B\}=\inf_{b\in B}d_\Theta(\theta,b)$, with $\dist\{\theta,\varnothing\}=+\infty$. For compact $A,B$ (possibly empty), set $e(A,B)=\sup_{a\in A}\dist\{a,B\}$ with $e(\varnothing,B)=0$, and $d_H(A,B)=\max\{e(A,B),e(B,A)\}$; hence $d_H(A,\varnothing)=+\infty$ for nonempty $A$.

\begin{lemma}[Monotonicity and Lipschitz continuity]\label{lem:lipschitz}
If $f_j(\theta)\le g_j(\theta)$ for all $j,\theta$, then $Q_R(f)\le Q_R(g)$. In addition,
\[
|Q_R(f)-Q_R(g)|\le\|f-g\|_\infty.
\]
\end{lemma}
\begin{proof}
With $\delta=\|f-g\|_\infty$, $\max_j f_j(\theta)\le\max_j g_j(\theta)+\delta$ for every $\theta$; minimize and reverse $f,g$ for the Lipschitz bound. Coordinatewise ordering gives monotonicity.
\end{proof}

\begin{lemma}[Primitive continuity implies graph closedness]\label{lem:primitive-graph-app}
If $\theta\mapsto b_{Rj}(\theta,P)$ is continuous and $\max_j\|b_{Rj}(\cdot,P_n)-b_{Rj}(\cdot,P_0)\|_\infty\to0$ whenever $P_n\to P_0$, then $P\mapsto\Theta_R(P)$ has sequentially closed graph.
\end{lemma}
\begin{proof}
If $P_n\to P_0$, $\theta_n\to\theta_0$, and $\theta_n\in\Theta_R(P_n)$, then for every $j$, $|b_{Rj}(\theta_n,P_n)-b_{Rj}(\theta_0,P_0)|\le \|b_{Rj}(\cdot,P_n)-b_{Rj}(\cdot,P_0)\|_\infty+|b_{Rj}(\theta_n,P_0)-b_{Rj}(\theta_0,P_0)|\to0$; hence $b_{Rj}(\theta_0,P_0)\le0$.
\end{proof}

\begin{lemma}[Exact target membership]\label{lem:membership}
For any $P$ and $\theta\in\Theta$,
$\theta\in\ThetaMR(P)$ if and only if there exists a repair $R$ such that $b_{Rj}(\theta,P)\le0$ for every $j$ and $q_S(P)>0$ for every $S$ satisfying $c(S)<c(R)$.
\end{lemma}
\begin{proof}
If $\theta\in\ThetaMR(P)$, choose $R\in\Rstar(P)$ such that $\theta\in\Theta_R(P)$. Every cheaper repair must be infeasible, so $q_S(P)>0$. Conversely, if $R$ is feasible at $\theta$ and every cheaper repair is infeasible, then $c(R)$ is the minimum feasible cost and hence $R\in\Rstar(P)$.
\end{proof}

\section{Proof of Theorem \ref{thm:geometry}}

\subsection{Lower semicontinuity}
Suppose $P_n\to P_0$ and $c^\star(P_n)<c^\star(P_0)$ along an infinite subsequence. Because the repair menu is finite, some fixed lower-cost repair $R$ is feasible along a further subsequence. Continuity then gives $q_R(P_0)\le0$, contradicting the definition of $c^\star(P_0)$.

\subsection{Local cost and repair stability}
Take $R_0\in\Rstar(P_0)$ with $q_{R_0}(P_0)<0$. By continuity, $R_0$ remains feasible in a neighborhood of $P_0$. Every lower-cost repair has a strictly positive feasibility index at $P_0$ and remains infeasible there, so the minimum cost cannot change locally. The same argument keeps any infeasible same-cost repair out of the correspondence. If all repairs in $\Rstar(P_0)$ are strictly feasible, finiteness gives one neighborhood on which they all remain feasible, which gives local equality of the repair correspondences.

\subsection{Target-changing full-sequence separation}
Let $P_n\to P_0$ and
\[
\mathscr E=\{R:R\in\Rstar(P_n)\text{ infinitely often}\}.
\]
Suppose $\theta^\dagger\in\ThetaMR(P_0)$ but $\theta^\dagger\notin\Theta_R(P_0)$ for every $R\in\mathscr E$. Because the repair collection is finite, every repair outside $\mathscr E$ is minimum only finitely often. Hence there is $N_0$ such that $\Rstar(P_n)\subseteq\mathscr E$ for all $n\ge N_0$.

If no eventual positive separation existed, one could choose, for each $k$, an index $n_k\ge k$ such that $\dist\{\theta^\dagger,\ThetaMR(P_{n_k})\}<1/k$. Passing to a strictly increasing subsequence and choosing $\theta_k\in\ThetaMR(P_{n_k})$ with $d_\Theta(\theta_k,\theta^\dagger)<2/k$ gives $\theta_k\to\theta^\dagger$. Choose $R_k\in\Rstar(P_{n_k})$ with $\theta_k\in\Theta_{R_k}(P_{n_k})$. For large $k$, $R_k\in\mathscr E$, and finiteness yields a further subsequence with $R_k=R$. Sequential closedness of the repair graph implies $\theta^\dagger\in\Theta_R(P_0)$, a contradiction. Therefore there exist $\delta>0$ and $N$ such that
\[
\dist\{\theta^\dagger,\ThetaMR(P_n)\}\ge\delta\qquad\forall n\ge N.
\]
Since $\theta^\dagger\in\ThetaMR(P_0)$, the Hausdorff liminf is at least $\delta$.

\section{Proofs of Theorem \ref{thm:classification} and Corollary \ref{cor:carrier}}

Let $P_{n,0}$ and $P_{n,1}$ denote the two local one-observation laws, and write $P_{n,j}^n\{\cdot\}$ for probability under the corresponding $n$-sample law. For a repair estimator $\widehat\cR_n$, define
\[
e_{n,j}=P_{n,j}^n\{\widehat\cR_n\neq\Rstar(P_{n,j})\},\qquad j=0,1,
\]
and use $\|P-Q\|_{TV}=\sup_A|P(A)-Q(A)|$.

\begin{lemma}[Two-point classification]\label{lem:twopoint}
Let the two local states have different population-minimum repair correspondences. Then every repair estimator satisfies
\[
\max\{e_{n,0},e_{n,1}\}\ge\frac{1-\|P_{n,0}^n-P_{n,1}^n\|_{TV}}{2}.
\]
\end{lemma}
\begin{proof}
Construct the binary test $\varphi_n=1\{\widehat\cR_n\neq\Rstar(P_{n,0})\}$ from the repair estimator. Its type-I error is $e_{n,0}$. Because $\Rstar(P_{n,1})\neq\Rstar(P_{n,0})$, the event $\{\varphi_n=0\}$ under state one implies $\widehat\cR_n\neq\Rstar(P_{n,1})$, so its type-II error is no larger than $e_{n,1}$. For every test, the sum of type-I and type-II errors is at least $1-\|P_{n,0}^n-P_{n,1}^n\|_{TV}$. Hence $2\max\{e_{n,0},e_{n,1}\}$ is at least that quantity, which proves the bound.
\end{proof}

For the Gaussian specialization of the repair-classification theorem, the stated total-variation limit is enough; no deficiency argument is needed. For $N(0,\Sigma)$ versus $N(h,\Sigma)$, the affine supports are disjoint when $h\notin\operatorname{Range}(\Sigma)$, so the total variation distance is one. When $h\in\operatorname{Range}(\Sigma)$, restrict attention to the effective support and put $d_G^2=h'\Sigma^+h$. The likelihood ratio is then one-dimensional in the discriminating direction, and the midpoint rule has equal errors $\Phi(-d_G/2)$. Hence
\[
\|N(0,\Sigma)-N(h,\Sigma)\|_{TV}=2\Phi(d_G/2)-1.
\]
Combining this identity with the finite-sample total-variation bound proves the Gaussian conclusion whenever the local binary experiment has the total-variation limit imposed in the repair-classification theorem.

\subsection{Carrier-region coverage}
For a repair collection $\mathcal D\subseteq\cR$, write $\Theta_{\mathcal D}(P)=\cup_{R\in\mathcal D}\Theta_R(P)$. Let $P_n\in\cP$ be a sequence and $\theta_n^\dagger\in\ThetaMR(P_n)$; probabilities and stochastic orders in this subsection are under the $n$-sample law $P_n^n$. Define
\[
\mathfrak G_n=\{\mathcal D\subseteq\cR:\theta_n^\dagger\in\Theta_{\mathcal D}(P_n)\}.
\]
Let $\widehat{\mathcal D}_n$ be the hard-selected repair collection. Suppose $C_n^{\mathrm{hard}}$ and $\Theta_{\widehat{\mathcal D}_n}(P_n)$ are nonempty compact with probability tending to one,
\[
d_H\{C_n^{\mathrm{hard}},\Theta_{\widehat{\mathcal D}_n}(P_n)\}=o_{P_n}(1),
\]
and for some $\delta>0$,
\[
\inf_{\mathcal D\notin\mathfrak G_n}\dist\{\theta_n^\dagger,\Theta_{\mathcal D}(P_n)\}\ge\delta
\]
eventually over collections with nonempty population target. On the event that the Hausdorff distance is less than $\delta/2$, selection outside $\mathfrak G_n$ implies
\[
\dist\{\theta_n^\dagger,C_n^{\mathrm{hard}}\}\ge\delta/2,
\]
so $\theta_n^\dagger\notin C_n^{\mathrm{hard}}$. Therefore
\[
P_n^n\{\theta_n^\dagger\in C_n^{\mathrm{hard}}\}
\le P_n^n\{\widehat{\mathcal D}_n\in\mathfrak G_n\}+o(1).
\]
Because $\theta_n^\dagger\in\ThetaMR(P_n)$, entire-target coverage also requires coverage of $\theta_n^\dagger$ and is subject to the same upper bound. The carrier region is the set of local Gaussian selection outcomes for which the hard-selected repair collection still carries $\theta_n^\dagger$. For an isolated boundary with limiting carrier event $\{Z\le0\}$, $Z\sim N(0,1)$, its probability is $1/2$. With several unresolved decisions, the carrier-region probability replaces that constant; an independent $m$-dimensional negative orthant gives $2^{-m}$, while correlated decisions give the corresponding Gaussian orthant probability.

\section{Proof of Theorem \ref{thm:adapt-lb}}

For the local laws $P_{n,h}$, let $P_{n,h}^n\{\cdot\}$ and $E_{n,h}$ denote probability and expectation under the $n$-sample law $P_{n,h}^n$.

\begin{lemma}[LAN power envelope]\label{lem:LAN}
Suppose
\[
\log\frac{dP_{n,h}^n}{dP_{n,0}^n}=h\zeta_n-\frac12h^2I+o_{P_{n,0}^n}(1),
\qquad \zeta_n\Rightarrow N(0,I).
\]
For tests $\varphi_n\in[0,1]$ satisfying $\limsup_n E_{n,0}\varphi_n\le\alpha$,
\[
\limsup_n E_{n,h}\varphi_n\le1-\Phi(z_{1-\alpha}-|h|\sqrt I).
\]
\end{lemma}
\begin{proof}
Choose a subsequence along which the alternative rejection probability attains its limsup. Tightness of $\zeta_n$ and boundedness of $\varphi_n$ allow a further subsequence for which $(\zeta_n,\varphi_n)$ converges weakly under $P_{n,0}^n$ to $(\zeta,\varphi)$, with $\zeta\sim N(0,I)$. The LAN likelihood ratio converges jointly to $\exp\{h\zeta-h^2I/2\}$; Le Cam's third lemma therefore identifies the limit law under $P_{n,h}^n$. The limiting test has null rejection probability at most $\alpha$. Although $\varphi$ may contain randomization beyond the central sequence, conditioning it on $\zeta$ does not change either null size or alternative power because the limiting likelihood ratio is measurable with respect to $\zeta$. Thus $\varphi$ may be replaced by $\bar\varphi(\zeta)=E(\varphi\mid\zeta)$. In the Gaussian experiment $N(0,I)$ versus $N(hI,I)$ for the central sequence, Neyman--Pearson therefore implies that no level-$\alpha$ test has power exceeding $1-\Phi(z_{1-\alpha}-|h|\sqrt I)$. Because the original subsequence was arbitrary, the same upper bound holds for the full limsup.
\end{proof}

For Theorem \ref{thm:adapt-lb}, let $A_n=\{\theta^\dagger\in C_n\}$ and set $\varphi_n=1-1\{A_n\}$. Entire-target honesty at the boundary makes this a test with asymptotic size at most $\alpha$. Lemma \ref{lem:LAN} therefore gives
\[
\liminf_nP_{n,h}\{\theta^\dagger\in C_n\}\ge\Phi(z_{1-\alpha}-|h|\sqrt I).
\]
If $\dist\{\theta^\dagger,\ThetaMR(P_{n,h})\}\ge\delta$ eventually, then on $\{\theta^\dagger\in C_n\}$,
\[
e(C_n,\ThetaMR(P_{n,h}))\ge
\dist\{\theta^\dagger,\ThetaMR(P_{n,h})\}\ge\delta.
\]
Combining this implication with the preceding retention-probability bound gives the directed-excess conclusion of Theorem \ref{thm:adapt-lb}.

\subsection{An IV boundary at which a target component disappears}\label{oa:iv-lan}
Take the bounded-validity IV model of Section \ref{sec:iv} with four candidate instruments, a common tolerance $r>0$, unit deletion costs, and structural bounds wide enough not to bind. Let the just-identified coefficients be
\[
\beta_1=0,\qquad \beta_2=r,\qquad \beta_3=10r,\qquad \beta_4=12r+t .
\]
A set of retained instruments is feasible exactly when all its pairwise gaps are at most $2r$. The pair $\{1,2\}$ is always feasible, with target $[0,r]$; the pair $\{3,4\}$ is feasible exactly when $t\le0$, with target $[11r+t,11r]$; every other pair, every triple, and the full model are infeasible for $|t|<r$. The minimum cost is therefore that of dropping two claims, and
\[
\ThetaMR(t)=[0,r]\cup[11r+t,11r]\quad(t\le0),\qquad \ThetaMR(t)=[0,r]\quad(t>0).
\]
The value $\theta^\dagger=11r$ belongs to the target for every $t\le0$ and lies at distance $10r$ from it for every $t>0$, while the two repair-specific sets that stay feasible do not move; $t=0$ is a target-changing repair boundary in the sense of Definition \ref{def:target-changing}. Along the submodel in which only $\beta_4$ moves, suppose $\widehat\beta_4$ is asymptotically $N(\beta_4,\sigma^2/n)$, and parametrize $t=h/\sqrt n$. The experiment is LAN with $I=\sigma^{-2}$, so Theorem \ref{thm:adapt-lb} gives, for every $h>0$ and every procedure with asymptotic entire-target coverage $1-\alpha$ at the boundary,
\[
\liminf_nP_{n,h}\{11r\in C_n\}\ge\Phi(z_{1-\alpha}-h/\sigma),
\]
even though $11r$ is then at distance $10r$ from the target. Here the estimate of $\beta_4$ is taken to be asymptotically independent of the other three; with correlated estimates, $\sigma^{-2}$ is replaced by the $(4,4)$ element of the inverse asymptotic covariance matrix, which is the information about $t$ along this submodel.

\section{Proofs of Theorems \ref{thm:bandwise-complete} and \ref{thm:PR}}

Here $\mathcal E_n(P)$ is the common primitive-and-feasibility event, $\widehat{\mathfrak R}_n$ is its plausible-repair set, $C_{n,R}$ is the repair-specific outer set, and $C_n^{PR}=\bigcup_{R\in\widehat{\mathfrak R}_n}C_{n,R}$; these objects are defined in Section \ref{sec:pr}. Write $\underline b_{R,n}=(\underline b_{Rj,n})_j$ and $\overline b_{R,n}=(\overline b_{Rj,n})_j$.

\subsection{Deterministic inclusion}
Under the canonical construction, monotonicity in Lemma~\ref{lem:lipschitz} gives $Q_R(\underline b_{R,n})\le Q_R\{b_R(P)\}\le Q_R(\overline b_{R,n})$ on primitive-band coverage. Assumption \ref{ass:E} also allows a sharper sign-equivalent scalar sandwich to be supplied directly. In either case, on $\mathcal E_n(P)$, $\underline q_{R,n}\le q_R(P)\le\overline q_{R,n}$. For a true minimum repair $R$, $q_R(P)\le0$, so $\underline q_{R,n}\le0$. Every cheaper repair is infeasible, hence has a strictly positive population feasibility index and a strictly positive upper confidence bound. Thus every true minimum repair is retained. If $\theta\in\Theta_R(P)$, all primitive population restrictions are nonpositive, so the lower primitive endpoints are also nonpositive at $\theta$ and $\theta\in C_{n,R}$. Taking the union over the true minimum repairs gives $\ThetaMR(P)\subseteq C_n^{PR}$ on the common event.

\subsection{Bandwise completeness of the canonical rectangle}
Fix a realized primitive rectangle and canonical bounds $\underline q_R=Q_R(\underline b_R)$, $\overline q_R=Q_R(\overline b_R)$. If $f\in\mathfrak B_n^{\mathrm{rep}}$ and $\theta\in T(f)$ through a minimum repair $R$, then $\underline b_R\le f_R$ gives $\theta\in C_{n,R}$ and $Q_R(\underline b_R)\le0$. Every cheaper $S$ is infeasible under $f$, so $f_S\le\overline b_S$ gives $Q_S(\overline b_S)>0$. Hence $R$ is PR-plausible and $\theta\in C_n^{PR}$.

Conversely, if $\theta\in C_n^{PR}$ through retained $R$, set $f_R^\star=\underline b_R$, set $f_S^\star=\overline b_S$ for every cheaper $S$, and set $f_U^\star=\underline b_U$ for every remaining repair $U$. Continuity of the band endpoints makes these functions continuous and band-admissible. Then $R$ is feasible at $\theta$ and every cheaper repair is infeasible, so $R$ is minimum and $\theta\in T(f^\star)$. This proves the union identity; its minimal-set and claim-region consequences are immediate. Replacing ``cheaper'' by ``predecessor'' proves the fixed-partial-order version. The construction uses rectangular variation across repairs, which is why a separately supplied sharper scalar sandwich needs an additional joint-compatibility condition for the same equality.

\subsection{Repair recovery}
Let $\eta_n(P)$ be the maximum feasibility-endpoint error and $\Delta(P)$ the repair-resolution margin \eqref{eq:Delta}. The statistical-resolution assumption states
\[
\lim_{M\to\infty}\limsup_{n\to\infty}\sup_{P\in\mathcal P}
\Pp_P\{\eta_n(P)>M\rho_n\}=0.
\]
Take any sequence $P_n\in\mathcal P$ satisfying $\Delta(P_n)/\rho_n\to\infty$ and put $M_n=\Delta(P_n)/(3\rho_n)\to\infty$. For every fixed $M<\infty$, eventually $M_n\ge M$, so
\[
\Pp_{P_n}\{\eta_n(P_n)>\Delta(P_n)/3\}
\le \sup_{P\in\mathcal P}\Pp_P\{\eta_n(P)>M\rho_n\}.
\]
Taking $\limsup_n$ and then $M\to\infty$ gives
\[
\Pp_{P_n}\{\eta_n(P_n)\le\Delta(P_n)/3\}\to1.
\]
On this event, every true minimum repair has a strictly negative upper feasibility bound, every cheaper repair has a strictly positive lower bound, and every infeasible same-cost repair has a strictly positive lower bound. Thus exactly the true minimum repairs survive among layers up to $c^\star(P_n)$. Every higher-cost repair is ruled out by the strictly negative upper bound of a cheaper true minimum repair. Hence $\widehat{\mathfrak R}_n=\Rstar(P_n)$.

\emph{Repair-specific scales.} If the feasibility-index bounds of different repairs have different resolutions $\rho_{R,n}$, each uniformly tight on its own scale, the same argument goes through with $\Delta(P_n)/\rho_n\to\infty$ replaced by the three conditions
\begin{gather*}
\min_{R\in\Rstar(P_n)}\frac{-q_R(P_n)}{\rho_{R,n}}\to\infty,\qquad
\min_{S:\,c(S)<c^\star(P_n)}\frac{q_S(P_n)}{\rho_{S,n}}\to\infty,\\
\min_{S:\,c(S)=c^\star(P_n),\,S\notin\Rstar(P_n)}\frac{q_S(P_n)}{\rho_{S,n}}\to\infty,
\end{gather*}
with empty minima omitted: on the event that every endpoint error is below one third of its own margin, the classification argument above applies repair by repair. This version is invariant to rescaling each index separately.

\subsection{Random-set measurability}
\begin{lemma}[Measurability of PR]\label{lem:PR-measurable}
Let $(\Theta,d_\Theta)$ be a compact metric space (hence separable). Suppose every primitive lower and upper band endpoint is jointly measurable in $(\omega,\theta)$ and continuous in $\theta$, and every scalar feasibility endpoint is measurable. Then $C_n^{PR}$ is a measurable random closed set (allowing the empty set). In particular, for every nonempty deterministic compact $K\subseteq\Theta$, the event $\{K\subseteq C_n^{PR}\}$ is measurable.
\end{lemma}
\begin{proof}
For each repair let $f_R(\omega,\theta)=\max_j\underline b_{Rj,n}(\omega,\theta)$, and let $\widetilde C_{n,R}=C_{n,R}$ if $R\in\widehat{\mathfrak R}_n$ and $\widetilde C_{n,R}=\varnothing$ otherwise. It is Carath\'eodory, so $C_{n,R}(\omega)=\{\theta:f_R(\omega,\theta)\le0\}$ is closed; see \citet[Chapter~14]{RockafellarWets1998}. For a direct measurability proof, fix a countable dense $\Theta_0\subseteq\Theta$. For each $x\in\Theta$,
\[
\dist(x,C_{n,R})=\lim_{m\to\infty}\inf\{d_\Theta(x,\theta):\theta\in\Theta_0,\ f_R(\omega,\theta)<1/m\},
\]
with $\inf\varnothing=+\infty$. The right side is measurable as a limit of countable infima; compactness and continuity give the equality. Measurability of the scalar feasibility endpoints makes every indicator $1\{R\in\widehat{\mathfrak R}_n\}$ measurable. Since $\dist(x,\varnothing)=+\infty$,
\[
\dist(x,C_n^{PR})=\min_{R\in\cR}\dist(x,\widetilde C_{n,R})
\]
is measurable for every $x$, while the finite union $C_n^{PR}$ is closed. Thus $C_n^{PR}$ is a measurable random closed set. Finally, if $K_0$ is countable dense in compact $K$, closedness gives
\[
K\subseteq C_n^{PR}\quad\Longleftrightarrow\quad
\sup_{x\in K_0}\dist(x,C_n^{PR})=0,
\]
so the inclusion event is measurable.
\end{proof}

\section{Partial orders, order-robust targets, and the conditions for oracle equivalence}\label{oa:partial-order}

This section collects results that extend Section \ref{sec:pr} and Section \ref{sec:ra} . The notation $\Rstar_\prec$, $\widehat{\mathfrak R}_{n,\prec}$, $C^{\PR}_{n,\prec}$, $a_R$, $\delta_R^\prec$, $\Delta_\prec$, and $\mathcal M_\prec$ is defined in Section \ref{subsec:partial-order}.

\subsection{PR under a fixed partial order}

\begin{corollary}[PR under a finite partial order]\label{cor:partial-order}
Under the primitive repair setup and Assumption \ref{ass:E}, on $\mathcal E_n(P)$,
\[
\Rstar_\prec(P)\subseteq\widehat{\mathfrak R}_{n,\prec}
\qquad\text{and}\qquad
\Theta^{MR}_\prec(P)\subseteq C^{\PR}_{n,\prec}.
\]
Consequently,
\[
\liminf_n\inf_{P\in\cP}
\Pp_P\{\Theta^{MR}_\prec(P)\subseteq C^{\PR}_{n,\prec}\}
\ge1-\alpha.
\]
If, in addition, Assumption \ref{ass:C} holds and $P_n\in\cP$ satisfies
\[
\Delta_\prec(P_n)/\rho_n\to\infty,
\]
then
\[
\Pp_{P_n}\{\widehat{\mathfrak R}_{n,\prec}=\Rstar_\prec(P_n)\}\to1.
\]
The confidence set $C^{\PR}_{n,\prec}$ is a measurable random closed set under the same measurability conditions as $C_n^{\PR}$.
\end{corollary}

With deletion-set repairs ordered by strict inclusion, Corollary \ref{cor:partial-order} covers the union over all inclusion-minimal data-consistent relaxations. When the repair-specific sets coincide with those used in the misspecification-robust bound of \citet{LiKedagniMourifie2024}, this yields a uniformly valid confidence region for the finite-repair version of that target under the simultaneous-band conditions. The qualification matters: the argument here uses a compact finite-repair representation and does not cover all of the infinite-dimensional or noncompact models allowed in their identification analysis.

Once the repair correspondence and structural target are specified, the local impossibility results do not depend on how the ranking was represented. In particular, Theorem \ref{thm:classification} and Corollary \ref{cor:carrier} apply after replacing $\Rstar$ by $\Rstar_\prec$, while Theorem \ref{thm:adapt-lb} and the Gaussian bound of Theorem \ref{thm:frontier} operate directly on target membership and therefore apply with $\ThetaMR$ replaced by $\Theta^{MR}_\prec$. What is ranking-specific is the population cost geometry in Theorem \ref{thm:geometry} and the scalar-layer certification rule; Section \ref{subsec:partial-order-ra} gives the corresponding adaptive certification rule for a partial order.

\subsection{Robustness to a prespecified family of repair orders}\label{subsec:order-robust}

The scientific ranking of repairs may itself be unsettled. Let $\mathfrak O$ be a nonempty finite collection of strict partial orders on $\cR$, fixed before the data are observed. Define
\begin{equation}\label{eq:order-robust-target}
\Rstar_{\mathfrak O}(P)=\bigcup_{\prec\in\mathfrak O}\Rstar_\prec(P),
\qquad
\Theta^{MR}_{\mathfrak O}(P)=\bigcup_{\prec\in\mathfrak O}\Theta^{MR}_\prec(P),
\end{equation}
and
\begin{equation}\label{eq:order-robust-pr}
\widehat{\mathfrak R}_{n,\mathfrak O}=\bigcup_{\prec\in\mathfrak O}\widehat{\mathfrak R}_{n,\prec},
\qquad
C^{\PR}_{n,\mathfrak O}=\bigcup_{\prec\in\mathfrak O}C^{\PR}_{n,\prec}.
\end{equation}
This target is robust to uncertainty over the prespecified orders; it does not select an order from the data.

\begin{proposition}[Order-robust post-relaxation projection]\label{thm:order-robust}
Under the primitive repair setup and Assumption \ref{ass:E}, on $\mathcal E_n(P)$,
\[
\Rstar_{\mathfrak O}(P)\subseteq\widehat{\mathfrak R}_{n,\mathfrak O},
\qquad
\Theta^{MR}_{\mathfrak O}(P)\subseteq C^{\PR}_{n,\mathfrak O}.
\]
Consequently, $C^{\PR}_{n,\mathfrak O}$ is a measurable random closed set and has uniform asymptotic entire-target coverage at least $1-\alpha$. For a primitive array $f$, let $T_\prec(f)$ be the union of the $\prec$-minimal feasible repair targets generated by $f$. Under the canonical rectangular band,
\begin{equation}\label{eq:order-robust-completeness}
C^{\PR}_{n,\mathfrak O}=\bigcup_{f\in\mathfrak B_n^{\mathrm{rep}}}\ \bigcup_{\prec\in\mathfrak O}T_\prec(f).
\end{equation}
Thus the inclusion-minimality and claim-certification equivalence in Theorem \ref{thm:bandwise-complete} also hold for the order-robust target. Define
\begin{equation}\label{eq:order-robust-margin}
\Delta_{\mathfrak O}(P)=\min_{\prec\in\mathfrak O}\Delta_\prec(P).
\end{equation}
If Assumption \ref{ass:C} holds and $\Delta_{\mathfrak O}(P_n)/\rho_n\to\infty$, then
\[
\Pp_{P_n}\{\widehat{\mathfrak R}_{n,\prec}=\Rstar_\prec(P_n)
\text{ for every }\prec\in\mathfrak O\}\to1,
\]
and hence $\widehat{\mathfrak R}_{n,\mathfrak O}=\Rstar_{\mathfrak O}(P_n)$ with probability tending to one.
\end{proposition}

Finiteness of $\mathfrak O$ is substantive. It permits one common simultaneous event, allows the two unions in \eqref{eq:order-robust-completeness} to commute, and makes simultaneous recovery depend on the least separated candidate order. Estimated or growing order families are not covered.

\subsection{Sharpness of conclusion certificates}

\begin{corollary}[Sharpness and local resolution of conclusion certificates]\label{cor:certificate-resolution}
Let $A\subseteq\Theta$ be a fixed claim region for which $\{C_n\subseteq A\}$ is measurable, let $C\subseteq\Theta$ be nonempty and compact, and let $g:\Theta\to\RR$ be continuous. If the only information about a nonempty compact target is $T\subseteq C$, then $C\subseteq A$ is necessary and sufficient to guarantee $T\subseteq A$ for every such target, and $\min_C g,\max_C g$ are the sharp corresponding functional envelopes. Moreover, if $T_n=T(P_n)$, $s_n>0$ is deterministic, $d_H(C_n,T_n)=O_{P_n}(s_n)$, and $\kappa_A(P_n)=\dist(T_n,\Theta\setminus A)$ satisfies $\kappa_A(P_n)/s_n\to\infty$, then $\Pp_{P_n}\{C_n\subseteq A\}\to1$. Along the LAN experiment of Assumption \ref{ass:R}, if $\theta^\dagger\in T(P_{n,0})\setminus A$ and boundary entire-target coverage is at least $1-\alpha$, then for every fixed $h$,
\begin{equation}\label{eq:claim-certification-bound}
\limsup_n P_{n,h}\{C_n\subseteq A\}\le\Phi\!\left(|h|\sqrt I-z_{1-\alpha}\right).
\end{equation}
If $T(P_{n,h})\subseteq A$ eventually, this bounds the probability of correctly certifying the newly true repair-robust claim. The same statements hold for the partial-order target.
\end{corollary}

The local bound is an information limit, not a peculiarity of PR. For example, at $\alpha=.05$ and $|h|\sqrt I=1$, its upper bound is about $0.260$; it reaches one half only at $|h|\sqrt I=z_{.95}$.

\begin{remark}[Why the proof avoids a generic derivative of the feasibility map]\label{rem:no-derivative}
The canonical map $Q_R$ can be nonsmooth because minimizers move and active inequalities change, and a direct feasibility index such as the IV gap $L_R-U_R$ can be nonsmooth for the same reason. PR does not differentiate either object. For the canonical construction, monotonicity of $Q_R$ propagates the primitive band into a feasibility sandwich; for sharper implementations, Assumption \ref{ass:E} allows that sandwich to be established directly. Standard directional methods remain available when a local derivative can be estimated reliably, but they are not needed for the entire-target coverage argument.
\end{remark}

\subsection{Conditions for oracle equivalence}\label{oa:ra-conditions}

Part (iii) of Assumption \ref{ass:A} consists of the following two conditions.
\begin{enumerate}[leftmargin=*]
\item[(iii.a)] \textbf{Common asymptotic linearity.} Along any sequence $P_n$ on which oracle equivalence is asserted, all stochastic orders in (iii.a) and (iii.b) are under the $n$-sample law $P_n^n$. There are a finite-dimensional $v(P_n)$ and a common influence array $\psi_{n,i}$ such that $\widehat v_n^F=v(P_n)+n^{-1}\sum_{i=1}^n\psi_{n,i}+\mathrm{rem}_n^F$ and $\widehat v_n^M=v(P_n)+N_n^{-1}\sum_{i\in\mathcal I_{2n}}\psi_{n,i}+\mathrm{rem}_n^M$, where $\sqrt n\,\|\mathrm{rem}_n^F\|=o_{P_n}(1)$ and $\sqrt{N_n}\,\|\mathrm{rem}_n^M\|=o_{P_n}(1)$. With $S_{1n}=\sum_{i\in\mathcal I_{1n}}\psi_{n,i}$ and $S_{2n}=\sum_{i\in\mathcal I_{2n}}\psi_{n,i}$, require $S_{1n}=O_{P_n}(\sqrt{m_n})$ and $S_{2n}=O_{P_n}(\sqrt{N_n})$. Independent, mean-zero arrays with second moments bounded uniformly in $n$ are sufficient for these two tightness conditions.
\item[(iii.b)] \textbf{Oracle-set stability.} Along any sequence covered by (iii.a), for every $\mathscr D$ occurring as $\Rstar(P_n)$ there are random radii $\upsilon_n^F,\upsilon_n^M$ and one deterministic map $\Gamma_{n,\mathscr D}$ such that $C_{n,\mathscr D}^{O,F}=\Gamma_{n,\mathscr D}(\widehat v_n^F,\upsilon_n^F)$ and $C_{n,\mathscr D}^{O,M}=\Gamma_{n,\mathscr D}(\widehat v_n^M,\upsilon_n^M)$, with $|\upsilon_n^M-\upsilon_n^F|=o_{P_n}(n^{-1/2})$. There are a finite constant $L$ and deterministic neighborhoods $U_{n,\mathscr D}$ such that, with probability tending to one under $P_n^n$, both random input pairs lie in $U_{n,\mathscr D}$, their images are nonempty compact, and $d_H\{\Gamma_{n,\mathscr D}(v,\upsilon),\Gamma_{n,\mathscr D}(v',\upsilon')\}\le L\{\|v-v'\|+|\upsilon-\upsilon'|\}$ for all input pairs in $U_{n,\mathscr D}$, uniformly in $n$ and over the finitely many possible $\mathscr D$.
\end{enumerate}

In the fixed-dimensional IV class, Section \ref{oa:ivsummary} supplies the common Frisch--Waugh--Lovell (FWL) influence representation, $N_n^{-1/2}-n^{-1/2}=O(m_n/n^{3/2})=o(n^{-1/2})$, and Proposition \ref{prop:iv-uniform-band}'s covariance/quantile argument applies to both samples. Here $\Gamma_{n,\mathscr D}$ is the finite union of repair-specific interval branches, so condition (iii.b) has a primitive sufficient condition. Suppose each branch $R\in\mathscr D$ is an interval $[a_R(v,\upsilon),b_R(v,\upsilon)]$ whose endpoints are maxima and minima of finitely many $L$-Lipschitz functions, and that along $P_n$ the population branch widths $b_R-a_R$ at $(v(P_n),\upsilon)$ are at least $\kappa_n$ with $\sqrt n\,\kappa_n\to\infty$. Then every estimated branch is nonempty with probability tending to one, each branch moves by at most $L(\|v-v'\|+|\upsilon-\upsilon'|)$ in Hausdorff distance, and $d_H(\bigcup_RA_R,\bigcup_RB_R)\le\max_Rd_H(A_R,B_R)$ for nonempty sets gives condition (iii.b). Near branch collapse this argument breaks down, as the next paragraph shows.

The width condition cannot simply be dropped. Suppose a branch has population width zero and its estimated width is a sample average, so that the main-sample and full-sample widths are $S_{2n}/N_n$ and $(S_{1n}+S_{2n})/n$ with $S_{1n}$ and $S_{2n}$ the pilot and main-sample sums. With Gaussian sums the two widths have correlation $\sqrt{N_n/n}$, and the probability that one is negative and the other nonnegative is
\[
\frac1\pi\arctan\sqrt{m_n/N_n}\asymp\sqrt{m_n/n}.
\]
On that event the branch is empty in one sample and not in the other, and the two oracle sets differ by a fixed distance (or by $+\infty$ if the branch is the only one). The probability vanishes, but slowly: with $m_n=n/\log n$ it is $0.133$ at $n=400$, $0.102$ at $n=25{,}600$, and $0.095$ at $n=102{,}400$. Oracle equivalence at a collapsing branch therefore holds, if at all, only at a rate too slow to be useful, and Theorem \ref{thm:RA} excludes such sequences.

\subsection{Resolution adaptation under a finite partial order}\label{subsec:partial-order-ra}

The pilot need not resolve the feasibility sign of every repair in order to resolve partial-order minimality. Define the definitely minimal and definitely nonminimal collections
\begin{align}
\mathscr D_{n,\prec}^{\mathrm{pil}}
&=\left\{R:\overline q_{R,n}^{\mathrm{pil}}\le0,\ \underline q_{S,n}^{\mathrm{pil}}>0\ \forall S\prec R\right\},\label{eq:po-def-min}\\
\mathscr N_{n,\prec}^{\mathrm{pil}}
&=\left\{R:\underline q_{R,n}^{\mathrm{pil}}>0
\text{ or }\overline q_{S,n}^{\mathrm{pil}}\le0\text{ for some }S\prec R\right\}.\label{eq:po-def-nonmin}
\end{align}
The correspondence is \emph{partial-order certified} when
$\mathscr D_{n,\prec}^{\mathrm{pil}}\neq\varnothing$,
$\mathscr D_{n,\prec}^{\mathrm{pil}}\cap\mathscr N_{n,\prec}^{\mathrm{pil}}=\varnothing$,
and their union is $\cR$. On a valid pilot band, certification can occur only with
$\mathscr D_{n,\prec}^{\mathrm{pil}}=\Rstar_\prec(P)$.
Define
\[
\begin{aligned}
C_{n,\prec}^{\RA}
&=
\begin{cases}
C_{n,\mathscr D_{n,\prec}^{\mathrm{pil}}}^{O,M},&\text{if partial-order certification succeeds},\\
C_{n,\mathrm{main},\prec}^{\PR},&\text{otherwise},
\end{cases}\\[-1mm]
C_{n,\prec}^{\mathrm{oracle}}&=C_{n,\Rstar_\prec(P)}^{O,F}.
\end{aligned}
\]
For the next result, impose the \emph{partial-order analogue of Assumption \ref{ass:A}}: part (i) is unchanged; in part (ii), replace $\ThetaMR$, $\Rstar$, and the PR fallback by $\Theta^{MR}_\prec$, $\Rstar_\prec$, and $C_{n,\mathrm{main},\prec}^{\PR}$, respectively; and in part (iii), replace $\Rstar(P_n)$ by $\Rstar_\prec(P_n)$ throughout. These replacements are the only changes.

\begin{corollary}[Partial-order resolution adaptation]\label{cor:partial-order-ra}
Under parts (i)--(ii) of the partial-order analogue of Assumption \ref{ass:A},
\[
\liminf_n\inf_{P\in\cP}
\Pp_P\{\Theta^{MR}_\prec(P)\subseteq C_{n,\prec}^{\RA}\}
\ge1-\alpha.
\]
If, in addition, $P_n\in\cP$ and
\[
\Delta_\prec(P_n)/\rho^{\mathrm{pil}}_{m_n,\bar\alpha_n}\to\infty,
\]
then partial-order certification succeeds and returns $\Rstar_\prec(P_n)$ with probability tending to one. Under part (iii) of the partial-order analogue,
\[
d_H(C_{n,\prec}^{\RA},C_{n,\prec}^{\mathrm{oracle}})=o_{P_n}(n^{-1/2}).
\]
\end{corollary}

The pilot is designed to resolve \emph{minimality status}; it need not determine every feasibility sign. With the weak feasibility boundary treated exactly, the two status sets above are the maximal deterministic conclusions obtainable from the rectangular pilot feasibility-index box: $R$ is definitely minimal precisely when its own upper endpoint is nonpositive and every predecessor's lower endpoint is positive, and it is definitely nonminimal precisely when its own lower endpoint is positive or some predecessor's upper endpoint is nonpositive. A nonminimal repair may therefore lie on its own feasibility boundary and still be safely excluded when a scientifically preferred predecessor is known feasible. Proposition \ref{prop:partial-order-radius} shows that $\Delta_\prec$ is exactly the deterministic perturbation radius governing that distinction.

\section{Proofs for the finite partial-order extension}

This section proves Proposition \ref{prop:partial-order-radius}, Corollary \ref{cor:partial-order}, and Corollaries \ref{cor:robust-conclusions} and \ref{cor:certificate-resolution}. Let $\prec$ be a fixed strict partial order on the finite repair menu and define
\[
\Rstar_\prec(P)
=
\{R:q_R(P)\le0,\ q_S(P)>0\ \forall S\prec R\}.
\]
On the simultaneous primitive event, Lemma~\ref{lem:lipschitz} gives
\begin{equation}\label{eq:oa-q-band-partial}
\underline q_{R,n}\le q_R(P)\le\overline q_{R,n}
\qquad\forall R.
\end{equation}

For arbitrary deterministic coordinatewise feasibility intervals, the finite-sample projection rule is exact: $R$ can be $\prec$-minimal feasible for some vector in the rectangle if and only if $\underline q_R\le0$ and $\overline q_S>0$ for every $S\prec R$. The strict predecessor inequality is essential: if $\overline q_S=0$, predecessor $S$ is feasible throughout the rectangle and $R$ can never be minimal. Theorem \ref{thm:bandwise-complete} strengthens this coordinate statement for the canonical construction by projecting the entire primitive rectangle into the structural target.

\subsection{Exact perturbation geometry}

For a deterministic vector $v=(v_R)_{R\in\cR}$, write
\[
\mathcal M_\prec(v)=\{R:v_R\le0,\ v_S>0\ \forall S\prec R\},
\qquad
a_R(v)=\min_{S\prec R}v_S,
\]
with $\min\varnothing=+\infty$. Fix $R$.

If $R\in\mathcal M_\prec(v)$, then $v_R\le0$ and $a_R(v)>0$. A perturbation changes $R$ from minimal to nonminimal precisely by making either its own coordinate strictly positive or at least one predecessor coordinate nonpositive. The infimum sup-norm distances to those two events are, respectively, $-v_R$ and $a_R(v)$. Hence the distance to their union is
\[
\min\{-v_R,a_R(v)\}=\delta_R^\prec(v).
\]
If $v_R=0$, the first distance is zero but is not attained: for every $\varepsilon>0$, setting $w_R=\varepsilon$ and leaving all other coordinates fixed changes the status with $\|w-v\|_\infty=\varepsilon$. Hence $\delta_R^\prec(v)=0$, and therefore $\Delta_\prec(v)=0$ whenever any minimal repair lies on its own feasibility boundary. If $v_R<0$, the analogous infimum $-v_R$ is also generally not attained because leaving feasibility requires the strict inequality $w_R>0$.

Now suppose $R\notin\mathcal M_\prec(v)$. Three cases remain. If $v_R>0$ and $a_R(v)>0$, entering the minimality region requires only moving $v_R$ to a nonpositive value, so the infimum distance is $v_R$. If $v_R\le0$ and $a_R(v)\le0$, entering the minimality region requires moving every nonpositive predecessor coordinate strictly above zero; the infimum sup-norm distance is $-a_R(v)$. If $v_R>0$ and $a_R(v)\le0$, both changes are necessary, so the infimum distance is
\[
\max\{v_R,-a_R(v)\}.
\]
These cases give $\max\{v_R,-a_R(v)\}=\delta_R^\prec(v)$ on the nonminimal region; repairs without predecessors reduce to ordinary feasibility. This proves the repair-specific identity \eqref{eq:repair-status-radius}.

The correspondence changes if and only if the minimality status of at least one repair changes. Since the menu is finite, the distance to that finite union of status-changing sets is the minimum of the repair-specific distances. Therefore
\[
\inf\{\|w-v\|_\infty:\mathcal M_\prec(w)\ne\mathcal M_\prec(v)\}
=\min_R\delta_R^\prec(v)
=\Delta_\prec(v).
\]

Consider now arbitrary lower and upper endpoint vectors satisfying
\[
\max_R\max\{|\underline q_R-v_R|,|\overline q_R-v_R|\}<\Delta_\prec(v).
\]
If $R\in\mathcal M_\prec(v)$, then $\delta_R^\prec(v)\ge\Delta_\prec(v)$ implies $v_R\le-\Delta_\prec(v)$ and $v_S\ge\Delta_\prec(v)$ for every $S\prec R$. Hence $\underline q_R<0$ and $\overline q_S>0$ for every predecessor, so the interval rule retains $R$. If $R\notin\mathcal M_\prec(v)$, then either $v_R\ge\Delta_\prec(v)$ or some predecessor $S\prec R$ satisfies $v_S\le-\Delta_\prec(v)$. Accordingly either $\underline q_R>0$ or $\overline q_S<0$, so the interval rule excludes $R$. Thus the interval rule equals $\mathcal M_\prec(v)$. Conversely, fix $\varepsilon>\Delta_\prec(v)$. By the distance identity there exists a status-changing vector $w$ with $\|w-v\|_\infty<\varepsilon$. Under the strict interval rule, degenerate endpoints $\underline q=\overline q=w$ reproduce the population minimality correspondence exactly, including at predecessor coordinates equal to zero:
\[
\{R:w_R\le0,\ w_S>0\ \forall S\prec R\}=\mathcal M_\prec(w).
\]
Since $\mathcal M_\prec(w)\ne\mathcal M_\prec(v)$, this endpoint pair changes the classified correspondence relative to $v$. No deterministic radius strictly larger than $\Delta_\prec(v)$ can therefore guarantee equality uniformly over endpoint vectors.

It remains to verify the scalar-cost specialization. Put $S\prec R$ if and only if $c(S)<c(R)$ and let $c^\star$ be the minimum feasible cost. Every repair below $c^\star$ is infeasible, so for such a repair $R$, $\delta_R^\prec=q_R$. The same is true for an infeasible repair with cost exactly $c^\star$. For $R\in\Rstar(P)$,
\[
\delta_R^\prec(P)
=\min\left\{-q_R(P),\min_{S:c(S)<c^\star}q_S(P)\right\}.
\]
If $c(R)>c^\star$, the predecessor set contains every true minimum repair. Hence
\[
-a_R(P)
\ge \max_{S\in\Rstar(P)}\{-q_S(P)\}
\ge \min_{S\in\Rstar(P)}\{-q_S(P)\},
\]
so no higher-cost repair can reduce the global minimum below the three groups appearing in the scalar definition of $\Delta(P)$. Taking the minimum over repairs therefore gives $\Delta_\prec\{q(P)\}=\Delta(P)$.

\subsection{Coverage}

Take $R\in\Rstar_\prec(P)$. Feasibility gives $q_R(P)\le0$, so \eqref{eq:oa-q-band-partial} implies $\underline q_{R,n}\le0$. For every $S\prec R$, population minimality gives $q_S(P)>0$, and therefore $\overline q_{S,n}\ge q_S(P)>0$. Hence
\[
R\in\widehat{\mathfrak R}_{n,\prec}.
\]
Thus $\Rstar_\prec(P)\subseteq\widehat{\mathfrak R}_{n,\prec}$. If $\theta\in\Theta_R(P)$ for such an $R$, then $b_{Rj}(\theta,P)\le0$ for every $j$, while the simultaneous band gives
\[
\underline b_{Rj,n}(\theta)\le b_{Rj}(\theta,P)\le0.
\]
Hence $\theta\in C_{n,R}\subseteq C^{\mathrm{PR}}_{n,\prec}$. Taking the union over $R\in\Rstar_\prec(P)$ proves
\[
\Theta^{MR}_\prec(P)\subseteq C^{\mathrm{PR}}_{n,\prec}
\]
on $\mathcal E_n(P)$. Uniform coverage follows from Assumption \ref{ass:E}. Measurability follows from exactly the proof of Lemma~\ref{lem:PR-measurable}: only the finite measurable activity indicator is changed, from comparisons indexed by lower scalar cost to comparisons indexed by the fixed finite predecessor set $\{S:S\prec R\}$.

\subsection{Exact recovery under separation}

For any repair $R$, let
\[
a_R(P)=\min_{S\prec R}q_S(P),\qquad \min\varnothing=+\infty.
\]
Then
\[
R\in\Rstar_\prec(P)
\quad\Longleftrightarrow\quad
q_R(P)\le0\ \text{ and }\ a_R(P)>0.
\]
The margin \eqref{eq:partial-order-margin} is
\[
\delta_R^\prec(P)
=
\begin{cases}
\min\{-q_R(P),a_R(P)\},&R\in\Rstar_\prec(P),\\
\max\{q_R(P),-a_R(P)\},&R\notin\Rstar_\prec(P).
\end{cases}
\]
It is nonnegative. For a minimal repair it vanishes precisely when its own feasibility or a predecessor's infeasibility is on the classification boundary; for a nonminimal repair it vanishes only when neither its own infeasibility nor strict feasibility of a preferred repair separates its status.

On the event $\eta_n(P)\le\varepsilon$,
\begin{equation}\label{eq:oa-q-endpoint-error-partial}
|\underline q_{R,n}-q_R(P)|\le\varepsilon,
\qquad
|\overline q_{R,n}-q_R(P)|\le\varepsilon
\qquad\forall R.
\end{equation}
Suppose now that $P_n\in\mathcal P$ and $\Delta_\prec(P_n)/\rho_n\to\infty$. The same fixed-$M$ argument used for the scalar-cost recovery theorem implies
\[
\Pp_{P_n}\{\eta_n(P_n)\le\Delta_\prec(P_n)/3\}\to1.
\]
Work on this event and abbreviate $\Delta_n=\Delta_\prec(P_n)$.

If $R\in\Rstar_\prec(P_n)$, then $\delta_R^\prec(P_n)\ge\Delta_n$ forces
\[
q_R(P_n)\le-\Delta_n,
\qquad
q_S(P_n)\ge\Delta_n\quad\forall S\prec R.
\]
Therefore $\underline q_{R,n}<0$ and $\overline q_{S,n}>0$ for every predecessor $S\prec R$, so $R\in\widehat{\mathfrak R}_{n,\prec}$.

If $R\notin\Rstar_\prec(P_n)$, then
\[
\max\{q_R(P_n),-a_R(P_n)\}\ge\Delta_n.
\]
If $q_R(P_n)\ge\Delta_n$, \eqref{eq:oa-q-endpoint-error-partial} gives $\underline q_{R,n}>0$, so $R$ is excluded. Otherwise $-a_R(P_n)\ge\Delta_n$. Because the predecessor set is finite, some $S\prec R$ attains $a_R(P_n)$ and satisfies $q_S(P_n)\le-\Delta_n$. Then $\overline q_{S,n}<0$, again excluding $R$. Thus all and only the population-minimal repairs survive, proving
\[
\widehat{\mathfrak R}_{n,\prec}=\Rstar_\prec(P_n)
\]
with probability tending to one.

\subsection{Inclusion-minimal deletion repairs}

Suppose each repair $R$ is the set of assumptions deleted from the maintained model and put
\[
S\prec R\quad\Longleftrightarrow\quad S\subsetneq R.
\]
Then $\Rstar_\prec(P)$ is exactly the set of inclusion-minimal feasible deletion sets: $R$ is retained if it is feasible and no strict subset of its deletions is feasible. Therefore $\Theta^{MR}_\prec(P)$ is the union of the repair-specific identified sets across all inclusion-minimal data-consistent relaxations. Under the finite-repair primitive representation, the preceding coverage result is consequently a valid confidence-region construction for that union target.

\subsection{Repair-robust functional and conclusion certificates}

This subsection proves Corollary \ref{cor:robust-conclusions}. The argument is deterministic on the entire-target coverage event. Suppose throughout this subsection that $C_n\subseteq\Theta$ is a measurable random closed set for a nonempty target $T(P)$ and that
\[
\liminf_n\inf_{P\in\mathcal P}\Pp_P\{T(P)\subseteq C_n\}\ge 1-\alpha.
\]
Write
\[
G_n(P)=\{T(P)\subseteq C_n\}.
\]
Fix continuous functionals $g_1,\ldots,g_L:\Theta\to\mathbb R$ and define $\underline g_{\ell,n}=\inf_{\theta\in C_n}g_\ell(\theta)$ and $\overline g_{\ell,n}=\sup_{\theta\in C_n}g_\ell(\theta)$, with the extended-real conventions of Corollary \ref{cor:robust-conclusions}. On $G_n(P)$, set inclusion immediately gives
\[
\inf_{\theta\in C_n}g_\ell(\theta)
\le
\inf_{\theta\in T(P)}g_\ell(\theta)
\le
\sup_{\theta\in T(P)}g_\ell(\theta)
\le
\sup_{\theta\in C_n}g_\ell(\theta).
\]
The same event $G_n(P)$ implies these inequalities simultaneously for every $\ell\le L$; no union bound is used. Because $C_n$ is closed in compact $\Theta$, it is a random compact set, and measurability of the extrema follows from the measurable maximum theorem for continuous functions on random compact sets. With the stated extended-real conventions, the displayed implication remains valid when $C_n$ is empty because $T(P)$ is nonempty under repairability and hence $G_n(P)$ is then false. Therefore
\[
\Pp_P\left\{
\underline g_{\ell,n}\le \inf_{T(P)}g_\ell
\le \sup_{T(P)}g_\ell\le\overline g_{\ell,n}
\quad\forall\ell\le L
\right\}
\ge \Pp_P\{G_n(P)\}.
\]
Taking $\inf_P$ and $\liminf_n$ proves the simultaneous functional-envelope conclusion.

For fixed claim regions $A_1,\ldots,A_K\subseteq\Theta$, define the false-certification event
\[
F_n(P)=\left\{\exists k\le K:
C_n\subseteq A_k\ \text{and}\ T(P)\not\subseteq A_k\right\}.
\]
If $G_n(P)$ occurs and $C_n\subseteq A_k$, transitivity of set inclusion gives
$T(P)\subseteq C_n\subseteq A_k$. Hence no false certificate can occur on $G_n(P)$ and therefore
\[
F_n(P)\subseteq G_n(P)^c.
\]
It follows that
\[
\sup_{P\in\mathcal P}\Pp_P\{F_n(P)\}
\le
\sup_{P\in\mathcal P}\Pp_P\{G_n(P)^c\}
=
1-\inf_{P\in\mathcal P}\Pp_P\{G_n(P)\}.
\]
Taking $\limsup_n$ and using the displayed entire-target coverage condition yields \eqref{eq:robust-claim-fwer}. Because the event $F_n(P)$ already allows an arbitrary configuration of true and false claims, the bound is strong familywise control for the reported universal claims. The absence of an additional multiplicity penalty is not a multiple-testing shortcut: it follows because all certificates are deterministic consequences of one common entire-target coverage event.

\subsection{Proof of Corollary \ref{cor:certificate-resolution}}
The corollary has three parts: sharpness of containment, local resolution, and the LAN bound. For the first, fix a claim region $A\subseteq\Theta$, a nonempty compact set $C\subseteq\Theta$, and a continuous functional $g:\Theta\to\mathbb R$. If $C\subseteq A$, every nonempty compact $T'\subseteq C$ also lies in $A$. If $C\not\subseteq A$, any $\theta_0\in C\setminus A$ makes the singleton $T'=\{\theta_0\}$ a counterexample. Thus $C\subseteq A$ is necessary and sufficient when the only information is $T'\subseteq C$. Continuity of $g$ on compact $C$ gives minimizers and maximizers $\theta_-,\theta_+$, and the singleton targets $\{\theta_-\}$ and $\{\theta_+\}$ show that $\min_C g$ and $\max_C g$ are the sharp lower and upper bounds based only on containment.

For the second, put $T_n=T(P_n)$ and $\kappa_n=\dist(T_n,\Theta\setminus A)$. If $d_H(C_n,T_n)<\kappa_n$ and some $c\in C_n\cap(\Theta\setminus A)$, then
\[
\kappa_n\le \dist\{c,T_n\}\le d_H(C_n,T_n),
\]
a contradiction. Hence $\{d_H(C_n,T_n)<\kappa_n\}\subseteq\{C_n\subseteq A\}$. If $s_n>0$ is deterministic, $d_H(C_n,T_n)=O_{P_n}(s_n)$, and $\kappa_n/s_n\to\infty$, tightness therefore implies $\Pp_{P_n}\{C_n\subseteq A\}\to1$.

For the third, measurability of $\{C_n\subseteq A\}$ makes $\varphi_n=1\{C_n\subseteq A\}$ a valid test. At the boundary, $\theta^\dagger\in T(P_{n,0})\setminus A$, so the entire-target coverage event and $\{C_n\subseteq A\}$ are disjoint. Boundary honesty gives $\limsup_nE_{n,0}\varphi_n\le\alpha$. Lemma~\ref{lem:LAN} then yields
\[
\limsup_nP_{n,h}\{C_n\subseteq A\}
\le \Phi\!\left(|h|\sqrt I-z_{1-\alpha}\right).
\]
The argument uses only boundary target membership and therefore applies unchanged to the partial-order target.

\subsection{Order-robust PR}

For every $\prec\in\mathfrak O$, Corollary \ref{cor:partial-order} gives, on the same event $\mathcal E_n(P)$,
\[
\Rstar_\prec(P)\subseteq\widehat{\mathfrak R}_{n,\prec},
\qquad
\Theta^{MR}_\prec(P)\subseteq C^{\PR}_{n,\prec}.
\]
Taking unions over the fixed finite family $\mathfrak O$ proves the deterministic inclusions in Proposition \ref{thm:order-robust}. No probability union bound is needed because all inclusions hold on the common band event, whose probability therefore gives uniform coverage.

For canonical bandwise completeness, the fixed-order identity yields
\[
\bigcup_{\prec\in\mathfrak O}C^{\PR}_{n,\prec}
=\bigcup_{\prec\in\mathfrak O}\ \bigcup_{f\in\mathfrak B_n^{\mathrm{rep}}}T_\prec(f)
=\bigcup_{f\in\mathfrak B_n^{\mathrm{rep}}}\ \bigcup_{\prec\in\mathfrak O}T_\prec(f).
\]
The last equality only commutes two unions. Inclusion-minimality follows because any strict subset omits a target value generated by some compatible $f$ and some candidate order. Claim-certification equivalence follows by testing containment of the displayed union in the claim region.

Finally, $\Delta_{\mathfrak O}(P_n)/\rho_n\to\infty$ implies $\Delta_\prec(P_n)/\rho_n\to\infty$ for every $\prec\in\mathfrak O$. Fixed-order recovery holds for each order. Since $\mathfrak O$ is finite, the probability that any one of those equalities fails is bounded by a finite sum of probabilities tending to zero. Recovery therefore holds simultaneously, and taking unions proves recovery of $\Rstar_{\mathfrak O}(P_n)$.

\section{Proof of Theorem \ref{thm:frontier} and its feasible version}

Let $Y\sim N(h,\Sigma)$, write $P_u=N(u,\Sigma)$, and let $K=K_\theta$ be nonempty, closed, and convex. For $h\notin K$, let $h_0$ be the Mahalanobis projection of $h$ onto $K$ and set
\[
a=\frac{\Sigma^{-1}(h-h_0)}{d_\Sigma(h,K)}.
\]
Then $a'\Sigma a=1$ and the projection inequality gives $a'(u-h_0)\le0$ for every $u\in K$. The test that rejects when $a'(Y-h_0)>z_{1-\alpha}$ has size at most $\alpha$ uniformly over $K$ and acceptance probability $\Phi(z_{1-\alpha}-d_\Sigma(h,K))$ at $h$. Any honest test is in particular level $\alpha$ at the simple null $h_0$, so Neyman--Pearson gives the matching lower bound.

\subsection{Stable exposed face}
Suppose $K\subseteq\{u:a'u\le\kappa\}$ with $a\neq0$, and let $h_0=\Pi_K^\Sigma(h)$ satisfy
$a'h_0=\kappa$ and $h-h_0=\lambda(h)\Sigma a$ for some $\lambda(h)>0$. Then
\[
d_\Sigma(h,K)
=\lambda(h)\sqrt{a'\Sigma a}
=\frac{a'h-\kappa}{\sqrt{a'\Sigma a}}.
\]
Therefore the fixed statistic
\[
\mathcal T(Y)=\frac{a'Y-\kappa}{\sqrt{a'\Sigma a}}
\]
has null mean at most zero throughout $K$ and alternative mean exactly $d_\Sigma(h,K)$, so it attains the bound over the exposed-normal region.

For feasible implementation, fix $M<\infty$ and a compact $H\subset\mathcal H_\theta^{EF}$. Suppose $\widehat a-a=o_P(1)$, $\widehat\kappa-\kappa=o_P(1)$, and $\widehat\Sigma-\Sigma=o_P(1)$ uniformly under the local laws indexed by $K_{\theta,M}\cup H$, where $K_{\theta,M}=K_\theta\cap\{u:\|u\|\le M\}$. Since $a'\Sigma a>0$, the estimated denominator is then bounded away from zero with probability tending to one uniformly on this index set. Gaussian tightness gives $Y=O_P(1)$ uniformly on bounded local-drift sets, so
\[
\widehat{\mathcal T}-\mathcal T=o_P(1)
\]
uniformly over $K_{\theta,M}\cup H$. The ideal supporting-half-space rule has rejection probability at most $\alpha$ for every $u\in K_\theta$ and retention probability $\Phi(z_{1-\alpha}-d_\Sigma(h,K_\theta))$ for $h\in H$.

To justify replacing $\mathcal T$ by $\widehat{\mathcal T}$ uniformly, fix $\varepsilon>0$. For every local mean $u$ in the bounded index set,
\[
\left|P_u\{\widehat{\mathcal T}\le z_{1-\alpha}\}-P_u\{\mathcal T\le z_{1-\alpha}\}\right|
\le
P_u\{|\widehat{\mathcal T}-\mathcal T|>\varepsilon\}
+
P_u\{|\mathcal T-z_{1-\alpha}|\le\varepsilon\}.
\]
The first term is $o(1)$ uniformly by the preceding stochastic approximation. Under every local mean, $\mathcal T$ is Gaussian with unit variance, so the second term is bounded by $2\varepsilon/\sqrt{2\pi}$ uniformly. Letting first $n\to\infty$ and then $\varepsilon\downarrow0$ gives asymptotic level $\alpha$ uniformly over $K_{\theta,M}$ and uniform convergence to the bound over $H$.

\subsection{Nonconvex branches}
If $K=\cup_{m=1}^M K_m$ with closed convex components, an acceptance region $G$ honest over $K$ must be honest over each component. Applying the convex result componentwise yields
\[
P_h\{Y\in G\}\ge\Phi(z_{1-\alpha}-d_\Sigma(h,K)).
\]
Equality requires a supporting half-space associated with a nearest component to support the entire union; otherwise the nearest-branch lower bound need not be attainable.

\subsection{A two-point example}\label{oa:nonconvex}
The gap can be large. Let $Y\sim N(h,1)$, $K=\{-2,2\}$, $h=0$, and $\alpha=0.05$. Then $d_\Sigma(0,K)=2$ and the right side of \eqref{eq:frontier} is $\Phi(1.645-2)=0.361$. The best honest test can be computed exactly. The likelihood ratio of $N(0,1)$ against the equal mixture of $N(-2,1)$ and $N(2,1)$ is decreasing in $|y|$, so the test that rejects when $|Y|\le c$, with $c$ chosen so that $P_{2}\{|Y|\le c\}=0.05$, has size exactly $0.05$ at both points of $K$; the mixture is therefore least favorable, and this test is most powerful. Solving gives $c=0.425$ and power $2\Phi(c)-1=0.329$ at $h=0$, so the best attainable retention probability is $0.671$, almost twice the bound. No single supporting half-space of $K$ separates $0$ from both points, which is why equality fails.

\section{Proofs of Theorem \ref{thm:RA} and Corollary \ref{cor:partial-order-ra}}

\subsection{Safe certification}
For a cost value $a$, the rule of Section \ref{sec:ra} forms
\[
\mathscr D_n(a)=\{R:c(R)=a,\ \overline q^{\mathrm{pil}}_{R,n}\le0\}
\]
and certifies layer $a$ only when every cheaper repair has $\underline q^{\mathrm{pil}}_{S,n}>0$, every repair in the layer has either $\overline q^{\mathrm{pil}}_{R,n}\le0$ or $\underline q^{\mathrm{pil}}_{R,n}>0$, and $\mathscr D_n(a)$ is nonempty. On a valid pilot band, a repair with nonpositive upper feasibility bound is truly feasible, whereas a repair with positive lower bound is truly infeasible. Therefore, if layer $a$ is certified, all cheaper repairs are truly infeasible and the definitely feasible repairs in layer $a$ are exactly all feasible repairs in that layer. Since at least one is feasible, $a=c^\star(P)$ and $\mathscr D_n(a)=\Rstar(P)$. A boundary repair with $q_R(P)=0$ can be certified feasible only in the special but logically valid case in which its pilot upper endpoint is exactly zero; otherwise a band that straddles zero correctly prevents certification of its status. Thus the rule is maximally decisive given only the rectangular pilot feasibility-index intervals, rather than imposing an artificial strict-feasibility requirement.

\subsection{Uniform honesty}
Let $G_n(P)=\mathcal E_n^{\mathrm{pil}}(P)$ be the valid-pilot event. By the pilot-validity part of the resolution-adaptation assumption,
\[
\sup_{P\in\mathcal P}\Pp_P\{G_n(P)^c\}\le\bar\alpha_n+o(1).
\]
Fix $P$. On $G_n(P)$, safe certification either fails or returns exactly $\Rstar(P)$. If it fails, the main sample uses $C_{n,\mathrm{main}}^{PR}$; if it succeeds, it uses $C_{n,\Rstar(P)}^{O,M}$. By the main-sample-honesty condition, conditional on any deterministic repair label these procedures are functions of the independent main sample only, so conditioning on the pilot realization does not change their main-sample law under $P$.

For the uniform remainder, write
\[
\delta_n^{PR}=\sup_{P\in\mathcal P}
\left[\Pp_P\{\ThetaMR(P)\not\subseteq C_{n,\mathrm{main}}^{PR}\}-\alpha\right]_+,
\]
and for each certifiable deterministic $\mathscr D$ with nonempty state class define
\[
\delta_{n,\mathscr D}^{O}
=
\sup_{P\in\mathcal P:\Rstar(P)=\mathscr D}
\left[\Pp_P\{\ThetaMR(P)\not\subseteq C_{n,\mathscr D}^{O,M}\}-\alpha\right]_+.
\]
The main-sample-honesty condition gives $\delta_n^{PR}\to0$ and $\delta_{n,\mathscr D}^{O}\to0$ for every such $\mathscr D$. Because the repair menu is finite, only finitely many deterministic collections can be certified, so $\max_{\mathscr D}\delta_{n,\mathscr D}^{O}\to0$. Thus conditional entire-target failure on $G_n(P)$ is at most $\alpha+o(1)$ uniformly in $P$. Hence
\[
\Pp_P\{\ThetaMR(P)\not\subseteq C_n^{RA}\}
\le \Pp_P\{G_n(P)^c\}+\alpha+o(1)
\le \bar\alpha_n+\alpha+o(1).
\]
Taking the supremum over $P$ and using $\bar\alpha_n\to0$ proves the uniform-honesty conclusion.

\subsection{Pilot resolution}
Let $P_n\in\mathcal P$ and write $\rho_n^{\mathrm{pil}}=\rho^{\mathrm{pil}}_{m_n,\bar\alpha_n}$. By pilot validity and $\Delta(P_n)/\rho_n^{\mathrm{pil}}\to\infty$, the fixed-$M$ repair-recovery argument above gives
\[
\Pp_{P_n}\{\eta^{\mathrm{pil}}_{m_n,\bar\alpha_n}(P_n)\le\Delta(P_n)/3\}\to1.
\]
On this event every true minimum repair has strictly negative pilot upper feasibility bound, every cheaper repair and every excluded same-cost repair has strictly positive pilot lower feasibility bound, and at least one true minimum repair is definitely feasible. Hence every repair in the minimum-cost layer is classified correctly, all cheaper layers are ruled out, and the certification rule returns exactly $\Rstar(P_n)$. This proves pilot recovery.

\subsection{Common asymptotic linearity and oracle equivalence}
Let $\mathcal I_{1n}$ and $\mathcal I_{2n}$ index the pilot and main subsamples, of sizes $m_n$ and $N_n=n-m_n$. On the certification event, put $\mathscr D_n=\Rstar(P_n)$, so $C_n^{RA}=C_{n,\mathscr D_n}^{O,M}$ and $C_n^{\mathrm{oracle}}=C_{n,\mathscr D_n}^{O,F}$. Under common asymptotic linearity,
\[
\widehat v_n^F=v(P_n)+\frac1n\sum_{i=1}^n\psi_{n,i}+\mathrm{rem}_n^F,
\qquad
\widehat v_n^M=v(P_n)+\frac1{N_n}\sum_{i\in\mathcal I_{2n}}\psi_{n,i}+\mathrm{rem}_n^M,
\]
with $\sqrt n\,\|\mathrm{rem}_n^F\|=o_{P_n}(1)$ and $\sqrt{N_n}\,\|\mathrm{rem}_n^M\|=o_{P_n}(1)$. Write $S_{1n}=\sum_{i\in\mathcal I_{1n}}\psi_{n,i}$ and $S_{2n}=\sum_{i\in\mathcal I_{2n}}\psi_{n,i}$. Then
\[
\widehat v_n^M-\widehat v_n^F
=\left(\frac1{N_n}-\frac1n\right)S_{2n}-\frac1nS_{1n}+o_{P_n}(n^{-1/2}).
\]
The same condition gives $S_{1n}=O_{P_n}(\sqrt{m_n})$ and $S_{2n}=O_{P_n}(\sqrt{N_n})$. Since $m_n/n\to0$, the two displayed leading terms are respectively $O_{P_n}\{m_n/(n\sqrt{N_n})\}$ and $O_{P_n}(\sqrt{m_n}/n)$, both $o_{P_n}(n^{-1/2})$. Hence
\[
\|\widehat v_n^M-\widehat v_n^F\|=o_{P_n}(n^{-1/2}).
\]
By oracle-set stability, the same deterministic map $\Gamma_{n,\mathscr D_n}$ generates the main- and full-sample oracle sets. With probability tending to one, both random input pairs lie in the same neighborhood $U_{n,\mathscr D_n}$ on which the stated Hausdorff-Lipschitz bound holds. Therefore,
\[
d_H(C_{n,\mathscr D_n}^{O,M},C_{n,\mathscr D_n}^{O,F})
\le L\left(\|\widehat v_n^M-\widehat v_n^F\|+|\upsilon_n^M-\upsilon_n^F|\right)
=o_{P_n}(n^{-1/2}).
\]
The probability of noncertification tends to zero by the preceding subsection, so the same rate holds for $d_H(C_n^{RA},C_n^{\mathrm{oracle}})$. The common-map condition is essential: separate Lipschitz properties for two unrelated main- and full-sample set constructions would not by themselves imply oracle equivalence.

\subsection{Partial-order certification and oracle recovery}
On the valid pilot event, $R\in\mathscr D_{n,\prec}^{\mathrm{pil}}$ implies that $R$ is feasible and every predecessor is infeasible, hence $R\in\Rstar_\prec(P)$; while $R\in\mathscr N_{n,\prec}^{\mathrm{pil}}$ is nonminimal because either $R$ is infeasible or a predecessor is feasible. These two status conditions are exact for a Cartesian product of scalar pilot intervals: if neither holds, two vectors inside the same box can be chosen so that $R$ is minimal under one and nonminimal under the other. Thus, whenever the two status sets partition $\cR$ and the definitely minimal set is nonempty, safe certification returns exactly $\Rstar_\prec(P)$.

Uniform honesty follows from the same conditioning argument as in the scalar-cost case. On the valid pilot event, the procedure either falls back to $C_{n,\mathrm{main},\prec}^{PR}$ or uses the independent main-sample oracle-subset procedure indexed by the true deterministic collection $\Rstar_\prec(P)$. The main-sample-honesty part of the partial-order analogue bounds either failure probability by $\alpha+o(1)$ uniformly. Because $\cR$ is finite, the maximum oracle remainder vanishes, while pilot failure contributes $\bar\alpha_n+o(1)$. Hence
\[
\liminf_n\inf_{P\in\mathcal P}\Pp_P\{\Theta^{MR}_\prec(P)\subseteq C_{n,\prec}^{RA}\}\ge1-\alpha.
\]
If $\Delta_\prec(P_n)/\rho^{\mathrm{pil}}_{m_n,\bar\alpha_n}\to\infty$, the same fixed-$M$ argument used above gives $\eta^{\mathrm{pil}}_{m_n,\bar\alpha_n}(P_n)\le\Delta_\prec(P_n)/3$ with probability tending to one. The exact stability-radius proposition then classifies every repair's minimality status correctly, so certification returns $\Rstar_\prec(P_n)$. On that event put $\mathscr D_n=\Rstar_\prec(P_n)$. Part (iii) of the partial-order analogue provides the same common primitive representation and common Hausdorff-Lipschitz map $\Gamma_{n,\mathscr D_n}$ as in the scalar proof. The preceding argument therefore gives
\[
d_H(C_{n,\prec}^{RA},C_{n,\prec}^{\mathrm{oracle}})=o_{P_n}(n^{-1/2}).
\]

\section{Invalid-IV influence functions and calibration}\label{oa:ivsummary}
Let $\mathcal P_{IV}$ be the fixed-dimensional iid class of Proposition \ref{prop:iv-uniform-band}: the common $4+\epsilon$ moment envelope, control-matrix eigenvalue bound, first-stage relevance bound, and six marginal influence-variance bounds all hold uniformly on this class. Fix $P\in\mathcal P_{IV}$, let $C$ be the included-control vector, and write $\EE_P$ for expectation under $P$. After population residualization on $C$, write
\[
\beta_j=\frac{\EE_P[\widetilde Z_j\widetilde Y]}{\EE_P[\widetilde Z_j\widetilde X]},
\qquad
\psi_j(W)=\frac{\widetilde Z_j(\widetilde Y-\beta_j\widetilde X)}{\pi_j},
\quad \pi_j=\EE_P[\widetilde Z_j\widetilde X].
\]
Let $\mathcal B_n$ be the event that the sample control second-moment matrix is nonsingular and all three sample FWL denominators $\widehat\pi_j$ are nonzero; off $\mathcal B_n$, set the sample FWL ratios and plug-in influence functions to zero.

\begin{proposition}[Uniform FWL linearization]\label{prop:fwl}
On the class $\mathcal P_{IV}$ just defined, $\sup_{P\in\mathcal P_{IV}}\Pp_P(\mathcal B_n^c)\to0$, and, with the fixed extension just described, uniformly in $P$,
\[
\sqrt n(\widehat\beta_j-\beta_j)=n^{-1/2}\sum_{i=1}^n\psi_j(W_i)+o_P(1),
\qquad j=1,2,3,
\]
jointly, and
\[
\max_{j\le3}n^{-1}\sum_{i=1}^n|\widehat\psi_{ji}-\psi_{ji}|^2=o_P(1).
\]
\end{proposition}
\begin{proof}
Write $M_{AB}=\EE_P[AB]-\EE_P[AC']\EE_P[CC']^{-1}\EE_P[CB]=\EE_P[\widetilde A\widetilde B]$ and collect the finitely many raw second moments entering $M_{Z_jY}$ and $M_{Z_jX}$ in $m(P)$. The uniform $4+\epsilon$ envelope gives $\widehat m-m(P)=O_P(n^{-1/2})$ and uniform integrability. The eigenvalue and relevance bounds imply $\sup_P\Pp_P(\mathcal B_n^c)\to0$ and put $g_j(m)=M_{Z_jY}/M_{Z_jX}$ in a common neighborhood with bounded first two derivatives, so Taylor expansion is uniform. Since the influence function of $M_{AB}$ is $\widetilde A\widetilde B-M_{AB}$ and $M_{Z_jY}=\beta_jM_{Z_jX}$, the ratio derivative is $\psi_j=\widetilde Z_j(\widetilde Y-\beta_j\widetilde X)/M_{Z_jX}$. The same envelope yields the stated empirical $L^2$ convergence of $\widehat\psi_j$. The fixed extension is negligible because $\Pp_P(\mathcal B_n^c)\to0$ uniformly.
\end{proof}

Set $V=(\beta_1,\beta_2,\beta_3,\beta_1-\beta_2,\beta_1-\beta_3,\beta_2-\beta_3)'$ and $\Psi=(\psi_1,\psi_2,\psi_3,\psi_1-\psi_2,\psi_1-\psi_3,\psi_2-\psi_3)'$. Let $\widehat V$ and $\widehat\Psi_i$ be their plug-in sample analogues and define $\Sigma_V=\EE_P[\Psi\Psi']$.

\begin{proposition}[Uniform Gaussian-multiplier calibration]\label{prop:iv-calibration}
Under the conditions defining $\mathcal P_{IV}$, let $\widehat\Sigma_V=n^{-1}\sum_i\widehat\Psi_i\widehat\Psi_i'$, let $\widehat\sigma_k^2$ be its diagonal entries, and put $A_n=\mathcal B_n\cap\{\min_{k\le6}\widehat\sigma_k^2>0\}$. On $A_n$, draw iid $\xi_i\sim N(0,1)$ and let $c_{1-\alpha,n}$ be the conditional $(1-\alpha)$ quantile of $\max_{k\le6}|n^{-1/2}\sum_i\xi_i\widehat\Psi_{ik}|/\widehat\sigma_k$. Define
\[
\mathcal C_{V,n}=\left\{v:\max_{k\le6}\frac{\sqrt n|\widehat V_k-v_k|}{\widehat\sigma_k}\le c_{1-\alpha,n}\right\}\ \text{on }A_n,
\qquad \mathcal C_{V,n}=\mathbb R^6\ \text{on }A_n^c.
\]
Then $\sup_{P\in\mathcal P_{IV}}|\Pp_P\{V(P)\in\mathcal C_{V,n}\}-(1-\alpha)|\to0$. The covariance matrix may be singular; only the six marginal variance lower bounds are required.
\end{proposition}
\begin{proof}
Proposition~\ref{prop:fwl} and the contrast map give uniform asymptotic linearity and $\widehat\Sigma_V-\Sigma_V=o_P(1)$. Along any sequence $P_n\in\mathcal P_{IV}$, a subsequence of the standardized covariance matrices converges to a correlation matrix $\Upsilon$, possibly singular. Such a subsequence exists because the set of $6\times6$ correlation matrices is compact. The uniform $2+\epsilon/2$ influence-moment bound gives a uniform Lindeberg condition, so the statistic converges to $G\sim N(0,\Upsilon)$; covariance consistency gives the same conditional multiplier limit. The subsequence principle therefore yields uniform approximation, and Proposition~\ref{prop:fwl} plus the marginal-variance bound gives $\sup_P\Pp_P(A_n^c)\to0$. Applied to the twelve-dimensional vector $(G',-G')'$, whose coordinates have unit variance, the anti-concentration inequality of \citet[Theorem~3]{ChernozhukovChetverikovKato2015Comparison} bounds the L\'evy concentration function of $\max_k|G_k|$ uniformly over correlation matrices, singular ones included, so the multiplier quantiles are stable. The full-band convention on $A_n^c$ changes coverage by only $o(1)$ uniformly and uses no denominator or variance floor. The same six-coordinate event covers the centers, contrasts, primitive bands, and feasibility sandwiches used by PR. If the number of multiplier draws grows with $n$, empirical multiplier quantiles add $o_P(1)$ by conditional Glivenko--Cantelli and the same anti-concentration bound.
\end{proof}

\section{Proof of Theorem \ref{thm:weak-pr}}\label{oa:weak-pr}

The objects needed for a standalone proof are restated here. Let $\mathcal J=\{1,\ldots,J\}$ be fixed, let $\Theta$ be a compact interval, and set
\[
\gamma_j=\EE_P[\widetilde Z_j\widetilde Y],\qquad
\pi_j=\EE_P[\widetilde Z_j\widetilde X],\qquad
\nu(P)=(\gamma_1,\ldots,\gamma_J,\pi_1,\ldots,\pi_J)'.
\]
For a deletion repair $R\subseteq\mathcal J$, define
\[
b_j(\theta,\nu)=|\gamma_j-\theta\pi_j|-r_j|\pi_j|,\qquad
\Theta_R^W(\nu)=\{\theta\in\Theta:b_j(\theta,\nu)\le0\ \forall j\in\mathcal J\setminus R\},
\]
where $r_j\ge0$, with delete-all target $\Theta_{\mathcal J}^W(\nu)=\Theta$. For a repair retaining at least one instrument, let
\[
q_R^W(\nu)=\min_{\theta\in\Theta}\max_{j\in\mathcal J\setminus R}b_j(\theta,\nu),
\]
and set $q_{\mathcal J}^W(\nu)=-1$. Fixed ex ante costs generate the minimum-repair union $\Theta_W^{MR}(\nu)$. Let $\mathcal N\subset\mathbb R^{2J}$ be a compact rectangle containing every $\nu(P)$ in the distribution class $\mathcal P_W$. For fixed constants $\epsilon>0$, $B<\infty$, $\lambda_C>0$, and $\underline\sigma>0$, that class is the fixed-dimensional iid class for $W=(Y,X,Z_1,\ldots,Z_J,C')'$ satisfying $\EE_P\|W\|^{4+\epsilon}\le B$, $\lambda_{\min}\EE_P[CC']\ge\lambda_C$, and the lower bound $\underline\sigma^2$ on every marginal reduced-form influence variance, but no lower bound on $|\pi_j|$.

The rectangle $\mathcal K_n$ is constructed as follows. Let $\widehat\varphi_i$ be the plug-in analogue of the influence vector $\varphi(W_i;P)$ of \eqref{eq:weak-if}, computed from sample residuals and centered at its sample mean, and let $\widehat\sigma_k^2=n^{-1}\sum_i\widehat\varphi_{ik}^2$. On the event that sample residualization is defined and every $\widehat\sigma_k$ is positive, let $c^W_{1-\alpha,n}$ be the conditional $(1-\alpha)$ quantile of
\[
\max_{k\le2J}\Big|n^{-1/2}\sum_{i=1}^n\xi_i\widehat\varphi_{ik}/\widehat\sigma_k\Big|,
\qquad \xi_i\stackrel{iid}{\sim}N(0,1),
\]
and let $\mathcal K_n$ be the intersection of $\mathcal N$ with the rectangle $\{\nu:|\nu_k-\widehat\nu_k|\le c^W_{1-\alpha,n}\widehat\sigma_k/\sqrt n\ \forall k\}$. On the exceptional event, or when the intersection is empty, $\mathcal K_n=\mathcal N$. Define
\[
C_n^{W\!PR}=\operatorname{cl}\left\{\bigcup_{\nu\in \mathcal K_n}\Theta_W^{MR}(\nu)\right\}.
\]
The proof establishes uniform entire-target coverage, random-closed-set measurability, closed-hull minimality, and the finite-order extension in three steps. All stochastic statements below are uniform over $P\in\mathcal P_W$.

\subsection{Reduced-form linearization and simultaneous coverage}

For generic variables $A$ and $B$, write
\[
M_{AB}=\EE_P[AB]-\EE_P[AC']\EE_P[CC']^{-1}\EE_P[CB]
=\EE_P[\widetilde A\widetilde B].
\]
The vector $\nu(P)$ consists of the $2J$ functionals $M_{Z_jY}$ and $M_{Z_jX}$. Define its influence vector by
\[
\varphi(W;P)=
\big(\widetilde Z_1\widetilde Y-\gamma_1,\ldots,
\widetilde Z_J\widetilde Y-\gamma_J,
\widetilde Z_1\widetilde X-\pi_1,\ldots,
\widetilde Z_J\widetilde X-\pi_J\big)'.
\]
The sample analogues are smooth functions of finitely many raw second moments whenever the sample control Gram matrix is nonsingular. The common $4+\epsilon$ moment bound and the control-eigenvalue bound therefore give
\begin{equation}\label{eq:oa-weak-linear}
\sqrt n\{\widehat\nu-\nu(P)\}
=n^{-1/2}\sum_{i=1}^n\varphi(W_i;P)+o_P(1),
\end{equation}
uniformly in $P$. Orthogonality of the population projection residuals makes the influence function of $M_{AB}$ equal to $\widetilde A\widetilde B-M_{AB}$, which gives the displayed vector $\varphi$. The same smooth-moment expansion and the $4+\epsilon$ envelope imply
\begin{equation}\label{eq:oa-weak-l2}
\max_{k\le2J}n^{-1}\sum_{i=1}^n
|\widehat\varphi_{ik}-\varphi_k(W_i;P)|^2=o_P(1).
\end{equation}
The probability that sample residualization is undefined tends to zero uniformly.

Let $\Sigma_\nu(P)=\EE_P[\varphi(W;P)\varphi(W;P)']$ and standardize by its marginal standard deviations. Along an arbitrary sequence $P_n\in\mathcal P_W$, compactness of the set of $2J\times2J$ correlation matrices yields a subsequence converging to a correlation matrix $\Upsilon$, which may be singular. The $4+\epsilon$ envelope for $W$ supplies a uniform $2+\epsilon/2$ moment bound for $\varphi$ and hence the Lindeberg condition. Equations \eqref{eq:oa-weak-linear}--\eqref{eq:oa-weak-l2} give the same Gaussian limit $N(0,\Upsilon)$ for the statistic and its conditional multiplier analogue. The marginal-variance bound makes the exceptional event in the definition of $\mathcal K_n$ uniformly negligible. The anti-concentration bound used in the proof of Proposition \ref{prop:iv-calibration}, which holds uniformly over correlation matrices including singular $\Upsilon$, gives stability of the conditional quantile. The subsequence principle then yields
\begin{equation}\label{eq:oa-weak-region}
\liminf_n\inf_{P\in\mathcal P_W}\Pp_P\{\nu(P)\in \mathcal K_n\}\ge1-\alpha.
\end{equation}
Intersecting with $\mathcal N$ does not alter coverage because $\nu(P)\in\mathcal N$; replacing an empty intersection or an exceptional-sample rectangle by $\mathcal N$ can only increase coverage. This proves the required uniform reduced-form region without any restriction on the means $\pi_j$.

\subsection{Continuity, coverage, and zero first stages}

For every retained instrument, $(\theta,\nu)\mapsto b_j(\theta,\nu)$ is jointly continuous. Compactness of $\Theta$ and the maximum theorem imply continuity of every finite-valued $q_R^W$. If $\nu_m\to\nu$ and $\theta_m\to\theta$ with $\theta_m\in\Theta_R^W(\nu_m)$, continuity gives $\theta\in\Theta_R^W(\nu)$; hence every repair-specific correspondence has closed graph. Delete-all is constant and therefore has the same property.

On the event $\{\nu(P)\in \mathcal K_n\}$, the population target $\Theta_W^{MR}\{\nu(P)\}$ is one of the sets appearing in the raw union defining $C_n^{W\!PR}$. Thus
\[
\Theta_W^{MR}\{\nu(P)\}
\subseteq\bigcup_{\nu\in \mathcal K_n}\Theta_W^{MR}(\nu)
\subseteq C_n^{W\!PR}.
\]
Combining this deterministic inclusion with \eqref{eq:oa-weak-region} proves the uniform coverage assertion. The argument never divides by $\pi_j$ and is unchanged at $\pi_j=0$. At $(\gamma_j,\pi_j)=(0,0)$ the primitive is identically zero in $\theta$; at $\pi_j=0$ and $\gamma_j\ne0$ it is the positive constant $|\gamma_j|$. These are, respectively, loss of identification and refutation of every repair retaining instrument $j$.

\subsection{Random-set measurability and closed-hull completeness}

The fallback definition makes $\mathcal K_n$ a nonempty measurable random compact subset of $\mathcal N$. It remains to verify that $C_n^{W\!PR}$ is a measurable random closed set. Fix an open set $O\subseteq\Theta$. Closure does not change whether a set intersects $O$, so $C_n^{W\!PR}\cap O\ne\varnothing$ if and only if, for some repair $R$, there exist $\nu\in \mathcal K_n$ and $\theta\in O$ satisfying
\[
b_j(\theta,\nu)\le0\quad\forall j\in\mathcal J\setminus R,
\qquad
q_S^W(\nu)>0\quad\forall S:c(S)<c(R).
\]
The strict inequalities and the open-set restriction are equivalent to the existence of positive integers $m$ and $\ell$ such that
\[
q_S^W(\nu)\ge1/m\quad\forall S:c(S)<c(R),
\qquad
\operatorname{dist}(\theta,\Theta\setminus O)\ge1/\ell,
\]
together with the weak primitive inequalities above. For fixed $(R,m,\ell)$ these conditions define a closed subset of $\mathcal N\times\Theta$. Its intersection with the random compact set $\mathcal K_n\times\Theta$ is a random compact set, and the standard hit-measurability theorem makes its nonemptiness event measurable. A countable union over $(m,\ell)$ and a finite union over repairs proves Effros measurability. Repairs without cheaper predecessors use the vacuous-predecessor convention; when $O=\Theta$, take $\operatorname{dist}(\theta,\varnothing)=+\infty$.

Finally, $C_n^{W\!PR}$ is, by construction, the closure of the union of all minimum-repair targets generated by $\mathcal K_n$. It is therefore the inclusion-smallest closed set containing every such target. For a fixed strict partial order, replace the cheaper-repair inequalities by $q_S^W(\nu)>0$ for every predecessor $S\prec R$. A prespecified finite family of orders adds only a finite union before the final closure. This proves the remaining claims of Theorem \ref{thm:weak-pr}.

\section{Proof of Proposition \ref{prop:score-pr}}\label{oa:score-pr}

Fix $j$ and write $M_j(P)$ for the covariance matrix of $(\widetilde Z_j\widetilde Y,\widetilde Z_j\widetilde X)$, $\widehat\nu_j=(\widehat\gamma_j,\widehat\pi_j)'$ for the sample residual cross-products, and $\nu_j=(\gamma_j,\pi_j)'$. For $b\in\mathbb R$ put $e(b)=(1,-b)'$, so that $n^{-1}\sum_i\widehat\omega_{ij}(b)=e(b)'\widehat\nu_j$ and, at the true ratio $\beta_j=\gamma_j/\pi_j$, $e(\beta_j)'\nu_j=0$.

\emph{Step 1: a uniform normal limit at the true ratio.} Let $s_j^2=e(\beta_j)'M_j(P)e(\beta_j)$ and $a_j=e(\beta_j)/s_j$. The eigenvalue bound gives $s_j^2\ge\underline\sigma^2(1+\beta_j^2)$, hence $\|a_j\|\le\underline\sigma^{-1}$, whatever the size of $\beta_j$. By the linearization \eqref{eq:oa-weak-linear} in the proof of Theorem \ref{thm:weak-pr},
\[
\frac{\sqrt n\,e(\beta_j)'\widehat\nu_j}{s_j}=a_j'\sqrt n(\widehat\nu_j-\nu_j)=n^{-1/2}\sum_{i=1}^na_j'\varphi_{(j)}(W_i;P)+o_P(1),
\]
uniformly in $P$, where $\varphi_{(j)}$ collects the two coordinates of $\varphi$ that belong to instrument $j$. The summands have unit variance and, because $\|a_j\|$ is bounded and $\EE_P\|W\|^{4+\epsilon}\le B$, a uniformly bounded moment of order $2+\epsilon/2$. The Lyapunov central limit theorem along arbitrary sequences $P_n\in\mathcal P_S$ and the subsequence argument used before give a $N(0,1)$ limit uniformly over $\mathcal P_S$. The same bounds and \eqref{eq:oa-weak-l2} give $\widehat s_j(\beta_j)/s_j=\{a_j'\widehat M_ja_j\}^{1/2}\to1$ in probability uniformly, where $\widehat M_j$ is the sample covariance of the residual cross-products. Hence
\[
\sup_{P\in\mathcal P_S}\Big|\Pp_P\{\beta_j(P)\notin F_{n,j}\}-2\{1-\Phi(c)\}\Big|\to0.
\]
No lower bound on $|\pi_j|$ is used: the statistic involves $\pi_j$ only through $e(\beta_j)'\nu_j=0$ and the bounded direction $a_j$. The hats in \eqref{eq:score-set} denote sample residuals after regression on the controls, and $\widehat s_j^2(b)$ is the sample variance of the residualized products $\widehat Z_{ji}(\widehat Y_i-b\widehat X_i)$. Replacing population residuals by sample residuals does not matter at first order, because $\EE(\widetilde Z_jC)=\EE(\widetilde YC)=\EE(\widetilde XC)=0$; this is what the linearization \eqref{eq:oa-weak-linear} records, and after division by $s_j$ the remainder is $o_P(1)$ uniformly because $\|a_j\|$ is bounded. Squaring the inequality in \eqref{eq:score-set} gives $n\{e(b)'\widehat\nu_j\}^2-c^2\widehat s_j^2(b)\le0$, a quadratic inequality in $b$ with coefficients that are measurable functions of the data, so $F_{n,j}$ is a random closed set; if $\widehat s_j(b)=0$ the inequality requires $e(b)'\widehat\nu_j=0$, and at $b=\beta_j$ this happens with probability tending to zero under the eigenvalue condition.

\emph{Step 2: simultaneous coverage.} By Bonferroni and $c\ge z_{1-\alpha/(2J)}$,
\[
\limsup_n\sup_{P\in\mathcal P_S}\Pp_P\{\beta(P)\notin F_{n,1}\times\cdots\times F_{n,J}\}\le J\cdot2\{1-\Phi(z_{1-\alpha/(2J)})\}=\alpha .
\]

\emph{Step 3: projection.} When every $\pi_j\ne0$, restriction $j$ of the IV model is $|\theta-\beta_j|\le r_j$, so $\ThetaMR(P)=T\{\beta(P)\}$. On the event of Step 2, $T\{\beta(P)\}$ is one of the sets in the union \eqref{eq:score-pr}, which proves the coverage statement. Measurability follows as in Section \ref{oa:weak-pr}, with the random closed sets $F_{n,j}$ in place of the rectangle $\mathcal K_n$.

\emph{Step 4: envelope.} Suppose $r_j=r$, costs are additive with positive weights, and $F_{n,k}\subseteq\Theta$. For every $b$ in the product set, $b_k\in\Theta$, so $[b_k-r,b_k+r]\cap\Theta\ne\varnothing$ and the repair that keeps only claim $k$ is feasible. Its cost is smaller than that of deleting every claim, so the delete-all repair is not minimal at $b$. Every other repair keeps some claim $j$, and its set is contained in $[b_j-r,b_j+r]\cap\Theta\subseteq[\inf F_{n,j}-r,\sup F_{n,j}+r]\cap\Theta$. Taking unions over $b$ and the closure gives the stated containment.

\section{Proof of Corollary \ref{cor:iv-uniform-sign}}\label{oa:iv-envelope}

Fix a realization of the simultaneous bands $[\ell_{j,n},u_{j,n}]$, $j\le J$, and suppose \eqref{eq:iv-uniform-nonbinding} holds. Fix $r\in[0,\bar r]$ and $w\in(0,\infty)^J$.

\emph{Step 1: nonbinding structural bounds.} For every coefficient vector $b$ in the rectangle $\prod_j[\ell_{j,n},u_{j,n}]$ and every $j$, $b_j-r\ge B_n^-(\bar r)>\underline\beta$ and $b_j+r\le B_n^+(\bar r)<\overline\beta$. Hence the structural bounds do not bind for any vector in the rectangle, and the contrast-aware bounds use the exact identities of Section \ref{sec:iv} ($q_R=-2r$ for one retained instrument, $|\beta_j-\beta_k|-2r$ for two, and the largest pairwise gap minus $2r$ for three).

\emph{Step 2: delete-all is excluded.} Under the contrast-aware bounds, a repair that retains exactly one instrument $j$ has $\underline q_{R,n}=\overline q_{R,n}=-2r\le0$. Its cost, $\sum_{k\ne j}w_k$, is strictly smaller than the delete-all cost $\sum_kw_k$ because $w_j>0$. The strict predecessor condition in \eqref{eq:plausible} requires $\overline q_{S,n}>0$ for every cheaper $S$, so the delete-all repair is not in $\widehat{\mathfrak R}_n$.

\emph{Step 3: containment.} Every retained repair keeps a nonempty set $S_R$ of instruments. Its structural outer set is
\[
C_{n,R}=\Big[\max\big\{\underline\beta,\max_{j\in S_R}(\ell_{j,n}-r)\big\},\ \min\big\{\overline\beta,\min_{j\in S_R}(u_{j,n}+r)\big\}\Big],
\]
interpreted as empty if the lower end exceeds the upper end. For any $j\in S_R$, $C_{n,R}\subseteq[\ell_{j,n}-r,u_{j,n}+r]\subseteq[B_n^-(r),B_n^+(r)]$. Taking the union over retained repairs gives \eqref{eq:iv-uniform-containment}.

\emph{Step 4: conclusions.} If $\tau<B_n^-(r)$, then $C^{\PR}_{n,r,w}\subseteq[B_n^-(r),B_n^+(r)]\subset(\tau,\infty)$, and symmetrically for upper thresholds. Since Steps 1--3 hold for every $(r,w)$ at once and use only the realized band, the conclusions hold simultaneously. On the coverage event of Proposition \ref{prop:iv-uniform-band}, Theorem \ref{thm:PR} gives $\ThetaMR_{r,w}(P)\subseteq C^{\PR}_{n,r,w}$ for every $(r,w)$, so the population targets inherit every certified conclusion. Step 2 is where the contrast-aware bounds matter: with the canonical sandwich, the upper feasibility bound of a one-instrument repair is $(u_{j,n}-\ell_{j,n})/2-r$, which is positive for small $r$, and the delete-all repair could then be retained.

\section{Weak-identification simulations}\label{oa:mc-weak-id}

Table \ref{tab:mc-weak-id} gives the full results of the weak-identification design summarized in Section \ref{sec:mc}. Each replication computes the three confidence events from the simulated data, including a fresh multiplier critical value from 2{,}000 Gaussian draws for the regular and rectangle procedures. The lower end of each PR set is computed by brute force over a grid of $41$ points per instrument within the relevant ratio sets, which reproduces the exact common-witness projection to grid accuracy.

\input{generated/mc_weak_id_table.tex}

\section{Proof of Corollary \ref{cor:breakdown}}\label{oa:breakdown}

\emph{Part (i).} Let $E_n$ be the intersection of the coverage event of Proposition \ref{prop:iv-uniform-band} with the event that \eqref{eq:iv-uniform-nonbinding} holds; by assumption $\liminf_n\inf_P\Pp_P(E_n)\ge1-\alpha$. The band event does not depend on the tolerance or the costs, and for each $(r,w)$ the contrast-aware feasibility bounds and the outer sets are deterministic functions of the band. Hence, on $E_n$, Theorem \ref{thm:PR} gives $\ThetaMR_{r,w}(P)\subseteq C^{\PR}_{n,r,w}$ for every $r\in[0,\bar r]$ and every $w\in(0,\infty)^J$ simultaneously, and Corollary \ref{cor:iv-uniform-sign} gives $C^{\PR}_{n,r,w}\subseteq[B_n^-(r),B_n^+(r)]$. Write $\lambda_n=\min_j\ell_{j,n}$. If $\lambda_n\le\tau$ the claimed inequality is trivial. Otherwise fix $s<\min\{\bar r,\lambda_n-\tau\}$. For every $r\in[0,s]$ and every $w$, $B_n^-(r)=\lambda_n-r>\tau$, so $\ThetaMR_{r,w}(P)\subset(\tau,\infty)$, and therefore $r^\star_{\tau,w}(P)\ge s$. Letting $s$ increase to $\min\{\bar r,\lambda_n-\tau\}$ and taking the infimum over $w$ gives the inequality on $E_n$.

\emph{Part (ii).} Let $E_n'$ be the event that $\beta(P)\in F_{n,1}\times\cdots\times F_{n,J}$ and $F_{n,k}\subseteq\Theta$ for some $k$. By the proof of Proposition \ref{prop:score-pr}, $\liminf_n\inf_P\Pp_P(E_n')\ge1-\alpha$. The event does not depend on $(r,w)$, and on it the envelope part of the proposition gives $\ThetaMR_{r,w}(P)\subseteq C^{S\!PR}_{n,r,w}\subseteq[\min_j\inf F_{n,j}-r,\infty)$ for every common tolerance $r$ and every $w$. The rest of the argument is as in part (i).

\section{Sample construction and the analysis record}\label{oa:analysis-record}

The data version matters. The 2002 working paper reports an 80-country extended-AJR sample, whereas the published 2004 article uses 79 countries after sample revisions \citep{RodrikSubramanianTrebbi2002NBER,RodrikSubramanianTrebbi2004}. The source file is matched to the 80-country profile, and its summary statistics are checked against the published values, before any inference is run.

Before running the observation-level analysis, I recorded the common sample, the HC0 heteroskedasticity-robust covariance estimator, the six-functional multiplier calibration and its seed, the structural bounds $[\underline\beta,\overline\beta]=[-10,10]$, the tolerance grid, the benchmark ranking, and seven cost-sensitivity schemes. The reference tolerance $r=0.42$ had already been chosen from published moments for the same sample (Section \ref{subsec:pre-analysis}), so this record is not an independent preregistration.

\subsection{Reconstruction from published moments}\label{subsec:pre-analysis}

Before opening the observation-level file, I reconstructed the just-identified estimates from the second moments reported for the 80-country working-paper sample:
\[
(\beta_{\mathrm{ENG}},\beta_{\mathrm{SM}},\beta_{\mathrm{EUR}})=(1.073,1.879,1.973),
\qquad
(r_{12},r_{23},r_{13})=(0.403,0.047,0.450),
\]
where instruments are indexed $1=\mathrm{ENG}$, $2=\mathrm{SM}$, $3=\mathrm{EUR}$ and $r_{jk}=|\beta_j-\beta_k|/2$ is the common tolerance at which the intervals centered on instruments $j$ and $k$ first intersect. As $r$ increases, the point-estimate minimum repair therefore moves from two deletions, to a single one-deletion repair, to two tied one-deletion repairs, and finally to no deletion. I fixed $r=0.42$ because it lies between the $0.403$ and $0.450$ thresholds. Since these moments come from the sample later analyzed at the observation level, this is not a holdout calibration, and the conclusions below are stated for the whole range $r\in[0,0.60]$.

\subsection{The 140-country sample in detail}

Table~\ref{tab:larger-sample} collects the 140-country results of Section \ref{subsec:larger-sample} and reports how many repairs PR retains at each level.

\input{generated/larger_sample.tex}

\input{generated/shared_primitive_appendix.tex}

\input{generated/empirical_appendix.tex}
\input{generated/fieller_robustness_appendix.tex}

\end{document}

%% file: generated/mc_weak_id.tex
\emph{Weak identification.} The last design mimics the application with a first stage that may be weak. Three independent instruments have ratios equal to the three estimates in Table~\ref{tab:rst-centers}; the first-stage $F$ statistics are about $24$ and $15$ for the ENG and EUR analogues and $9.9\lambda^2$ for the SM analogue, with $\lambda\in\{1,0.5,0.25,0.1\}$, $n\in\{80,400\}$, and 2,000 replications per design (Appendix~\oaref{oa:mc-weak-id}). I record whether the confidence event behind each procedure holds, which implies coverage of the whole target for every tolerance and cost vector. For the regular six-functional band this probability is $0.865$--$0.910$ across designs and $0.893$ already at $\lambda=1$, where $F\approx10$ as for SM in the data. For the score inversion of Proposition~\ref{prop:score-pr} it is $0.940$--$0.962$. The reduced-form rectangle of Theorem~\ref{thm:weak-pr} gives $0.941$--$0.955$ at $n=400$ but only $0.890$--$0.907$ at $n=80$, where the cross-products it studentizes are still far from normal. The robust procedures pay in informativeness. At $\lambda=1$ and $n=80$ the score-inversion PR set at $r=0.42$ is positive in 36\% of replications, against 84\% for the regular band; at $\lambda=0.1$ the figures are 2\% and 23\%. When the SM first stage is close to zero, its Anderson--Rubin set is unbounded, and the data cannot rule out ratio configurations in which SM is the only claim retained; the honest set then reaches the structural bound. Part of the regular band's apparent informativeness in those designs comes from undercoverage. With equicorrelated instruments (correlation $0.5$ or $0.8$) the pattern is the same: the score inversion covers $0.936$--$0.963$, the regular band $0.904$--$0.948$, and the rectangle $0.886$--$0.898$ at $n=80$.

%% file: generated/empirical_results.tex
\subsection{Results for the 80-country sample}\label{subsec:empirical-results}

The common sample consists of the $n=80$ countries with complete observations in the 2002 working-paper file posted by the authors. Summary statistics and first-stage regressions match the published extended-AJR sample, and every prespecified data check passes (Appendix \oaref{oa:empirical-results}). Table~\ref{tab:rst-centers} reports the three just-identified estimates.

\begin{table}[t]
\centering
\caption{Just-identified estimates on the common 80-country sample}
\label{tab:rst-centers}
\begin{tabular}{llrrr}
\toprule
Instrument & Role in RST & Estimate & HC0 s.e. & Robust $F=t^2$\\
\midrule
SM  & 80-country specification & 1.877 & 0.418 & 9.9\\
ENG & Larger-sample alternative & 1.071 & 0.136 & 29.7\\
EUR & Larger-sample alternative & 1.974 & 0.463 & 13.2\\
\bottomrule
\end{tabular}
\begin{minipage}{0.94\textwidth}
\vspace{0.25em}
\footnotesize\emph{Notes:} Each row is a just-identified IV regression of log GDP per capita on the rule-of-law index with a constant and distance from the equator as controls. The second column describes how RST used the instrument; using all three on one sample is what creates the repair problem. The robust $F$ statistic is the squared HC0 $t$ statistic of the excluded instrument in the first stage.
\end{minipage}
\end{table}

The estimates range from 1.071 (ENG) to 1.974 (EUR). Half the pairwise gaps, which are the common tolerances at which two instrument-specific intervals first meet, are 0.048 for SM and EUR, 0.403 for ENG and SM, and 0.451 for ENG and EUR, close to the values reconstructed from the published moments before the file was opened (Appendix \oaref{oa:analysis-record}).

The $95\%$ multiplier critical value for the six studentized coordinates is $2.508$. The implied simultaneous bands for the contrasts are $[-1.845,0.234]$ for ENG$-$SM, $[-2.021,0.216]$ for ENG$-$EUR, and $[-1.413,1.219]$ for SM$-$EUR. All three contain zero, so the simultaneous contrast bands do not reject the joint validity restrictions at $5\%$, even at $r=0$. In this sample the model is not refuted; what the data cannot do is say whether, and how, it should be repaired. Section~\ref{subsec:larger-sample} turns to a sample in which the restrictions are rejected.

At $r=0$ the point estimates call for two deletions and PR is $[0.731,3.134]$; at $r=0.20$ they call for one deletion and PR is $[0.531,3.334]$; at the reference value $r=0.42$ they still call for one deletion and PR is $[0.311,3.554]$; at $r=0.50$ no deletion is needed and PR is $[0.231,3.634]$. At every tolerance in the grid the simultaneous band leaves all seven repairs that keep at least one instrument plausible, spread over cost layers $0$, $1$, and $2$, so PR coincides with the restricted union over those seven repairs: with $n=80$ the band screens out none of them. The literal union over the whole menu would include the delete-all repair and equal the structural bounds $[-10,10]$. PR excludes that repair because a one-instrument repair is algebraically feasible and strictly cheaper under every positive additive cost vector. What the data deliver is a common sign: the whole PR set lies above zero, under the equal-cost ranking and under each of the seven prespecified alternatives (Appendix \oaref{oa:empirical-results}). The statement is uniform over the instrument-deletion repairs, tolerances, and cost vectors considered here. It does not say that the historical instruments satisfy their exclusion restrictions, and it does not say that $\beta$ is positive under unrestricted violations.

Corollary~\ref{cor:iv-uniform-sign} turns these grid results into a statement about the continuum. For every common tolerance $r\in[0,0.60]$ and every strictly positive additive cost vector, the PR set lies in $[0.731-r,\ 3.134+r]$. The lower envelope is $0.311$ at $r=0.42$ and $0.131$ at $r=0.60$; rounded down, $\beta>0.31$ and $\beta>0.13$. These are summaries of a realized lower confidence bound, not thresholds fixed in advance. The sign conclusion is not sensitive to simulation error in the critical value: at $r=0.60$ it would fail only if the critical value exceeded 2.97.

\subsection{Weak identification}\label{subsec:weak-id-results}

The results so far rely on a relevance bound, and the SM first stage ($F=9.9$) is too weak to take that bound for granted. Proposition~\ref{prop:score-pr} removes it. With $c=2.508$, the critical value used above, which exceeds the Bonferroni value $z_{1-0.05/6}=2.394$, the Anderson--Rubin sets \eqref{eq:score-set} are $[0.677,1.663]$ for ENG, $[1.180,11.133]$ for SM, and $[1.210,5.607]$ for EUR. All three are bounded intervals, and the ENG set lies inside the structural bounds, so the envelope part of the proposition applies: for every $r$ and every positive additive cost vector, the score-inversion PR set lies in $[0.677-r,\ 10]$. The upper end is the structural bound because the SM set extends beyond it; with a weak instrument the data say little about how large the effect could be. The lower end is attained by the exact common-witness projection and stays positive for every tolerance up to $0.60$, where it equals $0.077$. The positive sign therefore does not depend on the relevance bound.

The rectangle of Theorem~\ref{thm:weak-pr} is valid in still greater generality, since it also allows $\pi_j=0$, but here it is much less informative. Its critical value is $2.538$, and the ratios $\gamma_j/\pi_j$ compatible with the rectangle range over $[0.17,6.56]$ for ENG, $[0.38,38.52]$ for SM, and $[0.69,10.06]$ for EUR. The resulting PR set is positive only for $r<0.17$. The reason is that the rectangle lets the reduced-form coefficient and the first stage move independently, whereas their estimation errors are strongly correlated; the score inversion uses that correlation. Figure~\ref{fig:diagnosis-sign}(a) plots the three lower bounds against the tolerance.

\subsection{A rejected model: the 140-country sample}\label{subsec:larger-sample}

Settler mortality is available for only 80 countries. For their larger sample RST rely on the two language instruments, and in that sample the maintained model is rejected by conventional tests. I use the $n=140$ countries with complete data on income, rule of law, distance from the equator, the two language shares, and the integration variables of RST's larger-sample regressions; the integration variables only define the sample. Settler mortality is defined only for former colonies, so the larger sample adds countries that were never colonized as well as former colonies without mortality data; the two samples differ in composition, and I do not compare magnitudes across them. This exercise was added during revision and was not part of the original analysis record.

\begin{figure}[t]
\centering
\includegraphics[width=.96\textwidth]{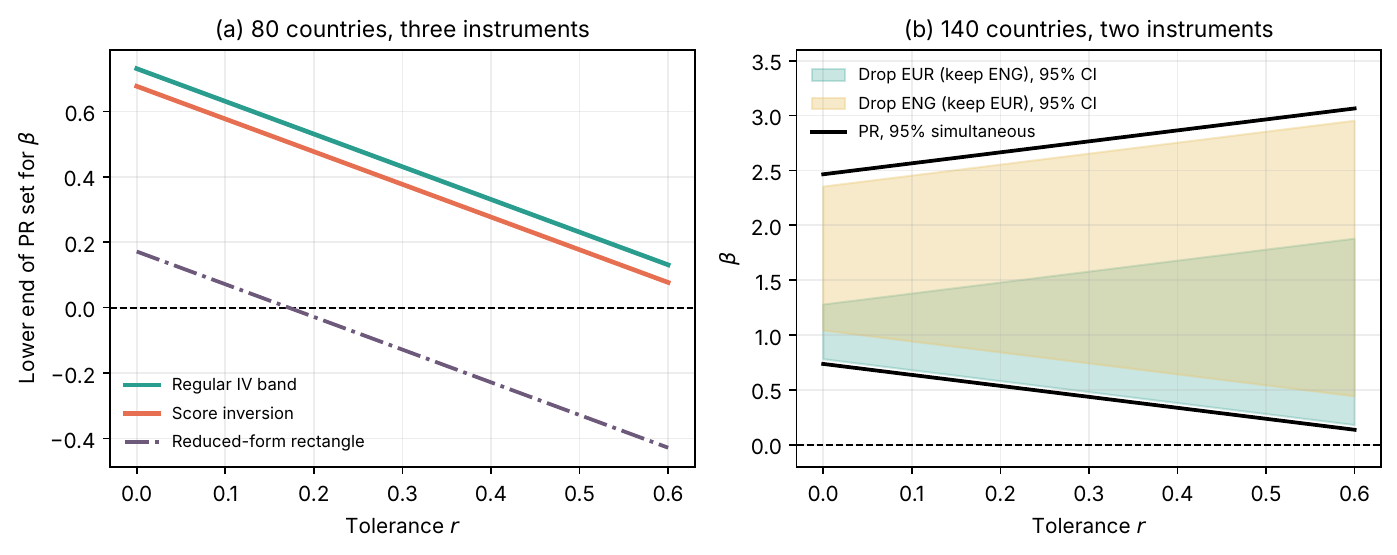}
\caption{Lower and upper ends of confidence sets as the tolerance $r$ varies. (a) The 80-country sample: lower end of PR computed from the regular IV band (Corollary~\ref{cor:iv-uniform-sign}), from score inversion without a relevance bound (Proposition~\ref{prop:score-pr}), and from the reduced-form rectangle (Theorem~\ref{thm:weak-pr}). (b) The 140-country sample: the $95\%$ interval reported after dropping one language instrument, widened by $r$, for each of the two choices, and the PR set from the $95\%$ simultaneous band.}
\label{fig:diagnosis-sign}
\end{figure}

The two just-identified estimates are $1.028$ for ENG (s.e.\ $0.127$, robust $F=35.1$) and $1.697$ for EUR (s.e.\ $0.334$, $F=17.4$). Both instruments are strong, so I use the regular implementation of Section~\ref{sec:iv}, with the three functionals $\beta_{\mathrm{ENG}}$, $\beta_{\mathrm{EUR}}$, and their difference in place of the six of Proposition~\ref{prop:iv-uniform-band}. Hansen's overidentification statistic is $J=7.08$, with $p=0.008$. A researcher who follows standard practice would conclude that at least one language instrument is invalid and report results with one of them dropped. Figure~\ref{fig:diagnosis-sign}(b) and the lower panel of Table~\ref{tab:benchmarks} show why the choice matters. Keeping only ENG gives the $95\%$ interval $[0.78,1.28]$; keeping only EUR gives $[1.04,2.35]$. They overlap only on $[1.04,1.28]$, and the 2SLS interval under the rejected model, $[0.96,1.58]$, excludes part of each.

PR does not choose. With the $95\%$ simultaneous band (critical value $2.296$), the contrast band $[-1.415,0.077]$ still contains zero: the $J$ test examines the contrast alone, whereas PR must cover the two coefficients and the contrast together, so its band is wider and the maintained model cannot be excluded. PR is $[0.74,2.46]$. With the $90\%$ band (critical value $2.005$) the contrast band is $[-1.320,-0.017]$, the maintained model is excluded for every $r<0.009$, and at $r=0$ the two one-deletion repairs are the only plausible ones; PR is their union, $[0.77,2.37]$. Neither level was fixed in advance, so I report both. At either level the lower end, $0.74$ or $0.77$, comes from the ENG branch and the upper end from the EUR branch, so a lower bound of about $0.75$ holds whichever claim is dropped. Dropping one instrument rather than the other moves the upper bound by about one unit of $\beta$. With $r=0.42$, PR at $95\%$ is $[0.32,2.88]$.

\subsection{Comparison with simpler procedures}\label{subsec:benchmarks}

Table~\ref{tab:benchmarks} puts PR next to the alternatives a referee would ask about, with every PR row computed from the same $95\%$ simultaneous band unless marked. Three comparisons stand out.

First, hard selection is narrower, but only because it ignores the repair problem. In the 80-country sample at $r=0.42$, a researcher who keeps the repair favored by the point estimates (drop ENG, keep SM and EUR) and reports conventional intervals obtains $[0.65,3.12]$; PR gives $[0.31,3.55]$. In the 140-country sample the two hard choices give $[0.78,1.28]$ and $[1.04,2.35]$, and each excludes values supported by the other. Second, the union over every repair in the menu is uninformative, $[-10,10]$, because it keeps the repair that deletes every claim; the strict predecessor rule removes it without any statistical screening. Third, imposing that all repairs share the same coefficients, which Remark~\ref{rem:shared-primitives} shows can shrink PR in principle, changes nothing here: the exact shared-coefficient projection of the same band equals PR at both tolerances, to three decimals.

The last rows show that the cost ranking is a substantive input. In the 140-country sample at $r=0$, PR with equal costs is $[0.74,2.46]$. If dropping the EUR claim is regarded as twice as costly as dropping the ENG claim, the repair that keeps only ENG is no longer minimal and PR becomes $[0.93,2.46]$; with the opposite ranking it is $[0.74,1.32]$. The lower bound moves from $0.74$ to $0.93$ and the upper bound from $2.46$ to $1.32$. The sign does not depend on the ranking; the magnitude does. A researcher who has a substantive reason to trust one language instrument more than the other should state it, because it matters.

The simplest serious competitor is the union of the one-instrument intervals from the same simultaneous band, $[\min_j\ell_{j,n}-r,\ \max_ju_{j,n}+r]$, which a careful researcher might report after deciding that any single instrument could be the valid one. With equal costs it coincides with PR in both samples (Table~\ref{tab:benchmarks}): with 80 countries the band screens out none of the seven instrument-retaining repairs, and the one-instrument branches contain all the others. In these data, then, PR does not deliver a narrower set; it delivers the justification for this one. It is the set that covers every minimum-repair target; it explains why the delete-all repair is excluded, since any single instrument within the tolerance restores feasibility at lower cost; and it carries the sign over every positive cost vector at once. The two sets separate as soon as the data certify that a cheaper repair is feasible. Hold the 80-country estimates and their correlations fixed and shrink the standard errors as if the sample had $n$ countries. At $r=0.42$ the band first certifies that SM and EUR can be kept together at $n\approx255$. The one-instrument repairs then stop being minimal, and the lower end of PR jumps to $0.87$, against $0.46$ for the union; at $n=640$ the two are $1.09$ and $0.53$. This is a power calculation at fixed estimates, not evidence about the effect, but it shows where the minimum-repair rule pays: the lower end of the union is set by the ENG-only branch, and that branch drops out once two other instruments are known to agree. The unequal-cost rows of the 140-country sample give an instance in the actual data.

\begin{table}[t]
\centering
\caption{What different procedures report}
\label{tab:benchmarks}
\small
\begin{tabular}{lcc}
\toprule
 & $r=0$ & $r=0.42$\\
\midrule
\multicolumn{3}{l}{\emph{80 countries, three instruments}}\\
SM only, conventional 95\% interval & $[1.06,2.70]$ & $[0.64,3.12]$\\
Point-estimate minimum repair, conventional & not unique & $[0.65,3.12]$\\
Union over every repair & $[-10,10]$ & $[-10,10]$\\
Union of one-instrument intervals, same band & $[0.73,3.13]$ & $[0.31,3.55]$\\
PR, regular band & $[0.73,3.13]$ & $[0.31,3.55]$\\
PR, shared-coefficient projection & $[0.73,3.13]$ & $[0.31,3.55]$\\
PR, score inversion (no relevance bound) & $[0.68,10.00]$ & $[0.26,10.00]$\\
\addlinespace
\multicolumn{3}{l}{\emph{140 countries, two instruments}}\\
2SLS under the rejected model & $[0.96,1.58]$ & ---\\
Drop EUR (keep ENG), conventional & $[0.78,1.28]$ & $[0.36,1.70]$\\
Drop ENG (keep EUR), conventional & $[1.04,2.35]$ & $[0.62,2.77]$\\
Union over every repair & $[-10,10]$ & $[-10,10]$\\
Union of one-instrument intervals, same band & $[0.74,2.46]$ & $[0.32,2.88]$\\
PR, equal costs & $[0.74,2.46]$ & $[0.32,2.88]$\\
PR, equal costs, 90\% band & $[0.77,2.37]$ & $[0.35,2.79]$\\
PR, dropping EUR costs twice as much & $[0.93,2.46]$ & $[0.51,2.88]$\\
PR, dropping ENG costs twice as much & $[0.74,1.32]$ & $[0.32,1.74]$\\
\bottomrule
\end{tabular}
\par
\begin{minipage}{0.94\textwidth}
\vspace{0.25em}
\footnotesize\emph{Notes:} Conventional rows report marginal $95\%$ intervals for the retained instruments, intersected across retained instruments and widened by $r$; they treat the chosen repair as known. PR rows use the $95\%$ simultaneous bands, except the row marked $90\%$, of Section~\ref{subsec:empirical-results} and Section~\ref{subsec:larger-sample}. The shared-coefficient projection takes the exact minimum-repair projection over coefficient vectors that lie in the band and satisfy all three contrast bands. At $r=0$ in the 80-country sample the point estimates make all three one-instrument repairs minimal, so there is no unique hard choice. All sets are intersected with $[-10,10]$.
\end{minipage}
\end{table}

\subsection{Economic magnitudes}\label{subsec:magnitudes}

The coefficient $\beta$ measures how log income responds to the rule-of-law index, a composite of survey and expert assessments, using historical variation to isolate exogenous differences. A positive $\beta$ says that the institutional quality captured by the index raises income on average; it does not say which reform would move the index or by how much. With that reading in mind, a convenient yardstick is the interquartile range. In the 80-country sample, moving from the 25th to the 75th percentile of the rule-of-law index is a change of 1.02 index points, while the interquartile gap in log income is 1.57. At $r=0$, the PR lower bound implies that such a move raises log income by at least $0.75$, about 48\% of the income gap. At $r=0.42$ the bound is $0.32$ (20\% of the gap), and at $r=0.60$ it is $0.13$ (9\%). In standard-deviation units ($\mathrm{sd}(\mathit{RULE})=0.85$), the bound at $r=0.60$ is $0.111$ log point. For comparison, the SM point estimate alone would attribute 122\% of the interquartile income gap to institutions and the ENG estimate 70\%.

In the 140-country sample, the interquartile range of the index is 1.43 and the interquartile gap in log income is 2.11. The lower end of PR at $r=0$ implies at least $1.05$ log points, about 50\% of the gap, and this bound does not depend on which language instrument is trusted. The upper end does. With only ENG, the upper confidence limit attributes at most 87\% of the gap to institutions; with only EUR, the limit is 160\%, more than the entire gap.

\subsection{How much invalidity can the sign withstand?}\label{subsec:breakdown-results}

Table~\ref{tab:breakdown} reports the lower confidence bounds of Corollary~\ref{cor:breakdown} for the sign of $\beta$ ($\tau=0$), together with their translation into direct effects. In the 80-country sample the regular band gives $\widehat r_{0,n}=0.731$: every minimum-repair target is positive for every common tolerance up to this value and every positive additive cost vector. By the discussion before Corollary~\ref{cor:breakdown}, and evaluating first-stage coefficients and $\beta_j$ at their estimates, this tolerance allows up to 68\% of the reduced-form association between English-language share and income to be direct, and up to 39\% and 37\% for settler mortality and the European-language share, all at once. In levels, the sign survives direct effects of 0.26, 0.23, and 0.20 log points of income per standard deviation of ENG, SM, and EUR. For comparison, the total reduced-form association of a one-standard-deviation change in each instrument with log income is 0.39, 0.60, and 0.53. Without a relevance bound the score-inversion bound is 0.677, only slightly smaller. The reduced-form rectangle gives 0.172, which corresponds to direct effects of about 0.06 log points per standard deviation; this is the price of allowing exactly zero first stages. In the 140-country sample the bound is 0.736 at the $95\%$ level and 0.773 at $90\%$, so at the point estimates the sign survives direct effects of about 0.18 log points per standard deviation of the English-language share even though the two language instruments are jointly rejected.

These translations hold the first-stage coefficients at their estimates. The last two rows of Table~\ref{tab:breakdown} carry their uncertainty through, as in Remark~\ref{rem:direct-effects}: the band is recalibrated for the enlarged vector (critical value $2.633$ with nine functionals in the 80-country sample, $2.475$ with five in the larger sample), and each first stage is replaced by the end of its band closest to zero. With probability at least $95\%$, $\beta$ is positive whenever the direct effects per standard deviation are below 0.13 for ENG, 0.04 for SM, and 0.05 for EUR in the 80-country sample, and below 0.10 and 0.06 for ENG and EUR in the larger sample. The guaranteed protection is much smaller than the point translations for SM and EUR. The SM first-stage band reaches $-0.043$, so the data cannot exclude a nearly irrelevant SM instrument, and a tolerance stated relative to a tiny first stage permits only a tiny direct effect. The English-language share is the only instrument whose first stage is precise enough to translate the tolerance into a substantial direct effect, about 0.13 log point per standard deviation in the 80-country sample and 0.10 in the larger one.

\begin{table}[t]
\centering
\caption{Lower confidence bounds on the breakdown tolerance for $\beta>0$}
\label{tab:breakdown}
\small\setlength{\tabcolsep}{4pt}
\begin{tabular}{llcccc}
\toprule
 & & & \multicolumn{3}{c}{Direct effect per s.d.}\\
\cmidrule(lr){4-6}
Sample & Procedure & $\widehat r_{0,n}$ & ENG & SM & EUR\\
\midrule
80 & Regular band (Cor.~\ref{cor:iv-uniform-sign}) & 0.731 & 0.26 & 0.23 & 0.20\\
 & Score inversion (Prop.~\ref{prop:score-pr}) & 0.677 & 0.24 & 0.22 & 0.18\\
 & Reduced-form rectangle (Thm.~\ref{thm:weak-pr}) & 0.172 & 0.06 & 0.05 & 0.05\\
140 & Regular band, 95\% & 0.736 & 0.18 & --- & 0.16\\
 & Regular band, 90\% & 0.773 & 0.19 & --- & 0.17\\
\addlinespace
\multicolumn{6}{l}{\emph{With first-stage uncertainty (Remark~\ref{rem:direct-effects})}}\\
80 & Regular band with first stages & 0.714 & 0.13 & 0.04 & 0.05\\
140 & Regular band with first stages & 0.714 & 0.10 & --- & 0.06\\
\bottomrule
\end{tabular}
\begin{minipage}{0.94\textwidth}
\vspace{0.25em}
\footnotesize\emph{Notes:} $\widehat r_{0,n}$ is the lower confidence bound of Corollary~\ref{cor:breakdown} with $\tau=0$; it holds simultaneously for every strictly positive additive deletion-cost vector. The last three columns report $\widehat r_{0,n}\,|\pi_j^{\mathrm{fs}}|\,\mathrm{sd}(Z_j)$, the largest direct effect on log income of a one-standard-deviation change in instrument $j$ that is compatible with tolerance $\widehat r_{0,n}$, using estimated first-stage coefficients from the same sample. In the first five rows these columns translate the confidence bound into economic units at the point estimates and are not confidence statements. The last two rows are confidence statements (Remark~\ref{rem:direct-effects}): they use a band that also covers the first-stage coefficients and replace each coefficient by the end of its band closest to zero. In these rows the third column is $\lambda_n^+$, the smallest lower end of the $\beta_j$ coordinates of the enlarged band. Rows for the 80-country sample use $95\%$ simultaneous bands.
\end{minipage}
\end{table}

Taken together, the two samples support a lower bound and not much more. Institutions matter for income under every repair that the data cannot rule out, and the magnitude of that minimum effect is economically large when the instruments are held close to exact validity. Statements that institutions explain all of the cross-country income gap, or that a specific instrument identifies the effect, depend on which historical identifying argument one is willing to drop. These conclusions are conditional on the repair menu and on relaxing as little as the data require; if every historical claim could be abandoned at once, the answer would be the uninformative structural bounds.

%% file: generated/mc_weak_id_table.tex
\begin{table}[ht]
\centering
\caption{Weak-identification simulations}
\label{tab:mc-weak-id}
\small\setlength{\tabcolsep}{3.5pt}
\begin{tabular}{rrr ccc ccc}
\toprule
 & & & \multicolumn{3}{c}{Confidence event holds} & \multicolumn{3}{c}{PR set at $r=0.42$ positive}\\
\cmidrule(lr){4-6}\cmidrule(lr){7-9}
$n$ & $\lambda$ & $F_{\mathrm{SM}}$ & Regular & Score & Rect. & Regular & Score & Rect.\\
\midrule
80 & 1.00 & 9.9 & 0.893 & 0.952 & 0.904 & 0.842 & 0.362 & 0.001\\
80 & 0.50 & 2.5 & 0.887 & 0.940 & 0.890 & 0.546 & 0.101 & 0.000\\
80 & 0.25 & 0.6 & 0.885 & 0.947 & 0.907 & 0.331 & 0.023 & 0.000\\
80 & 0.10 & 0.1 & 0.865 & 0.946 & 0.906 & 0.230 & 0.015 & 0.000\\
400 & 1.00 & 9.9 & 0.910 & 0.962 & 0.951 & 0.896 & 0.573 & 0.027\\
400 & 0.50 & 2.5 & 0.890 & 0.957 & 0.955 & 0.591 & 0.162 & 0.000\\
400 & 0.25 & 0.6 & 0.876 & 0.948 & 0.946 & 0.334 & 0.040 & 0.000\\
400 & 0.10 & 0.1 & 0.878 & 0.959 & 0.941 & 0.220 & 0.013 & 0.000\\
\bottomrule
\end{tabular}
\par
\begin{minipage}{0.96\textwidth}\vspace{0.25em}\footnotesize
Notes: 2,000 replications per row; Monte Carlo standard errors are at most $0.011$ for every entry. Design: $Z\sim N(0,I_3)$, $X=\sum_jp_jZ_j+v$, $Y=1.07X+\sum_jd_jZ_j+0.5u$, $\mathrm{corr}(u,v)=0.5$, with $d_j$ chosen so that $\beta_j=(1.07,1.88,1.97)$ and $p_j$ chosen so that the single-instrument first-stage $F$ statistics are about $(24,9.9\lambda^2,15)$. Regular: six-functional multiplier band (Proposition~\ref{prop:iv-uniform-band}). Score: Anderson--Rubin sets with $c=z_{1-0.05/6}$ (Proposition~\ref{prop:score-pr}). Rectangle: Theorem~\ref{thm:weak-pr}. PR sets use equal costs, structural bounds $[-10,10]$, and the exact common-witness projection for the score and rectangle procedures.
\end{minipage}
\end{table}

\begin{table}[ht]
\centering
\caption{Weak-identification simulations with correlated instruments}
\label{tab:mc-weak-id-corr}
\small\setlength{\tabcolsep}{3.5pt}
\begin{tabular}{rrr ccc ccc}
\toprule
 & & & \multicolumn{3}{c}{Confidence event holds} & \multicolumn{3}{c}{PR set at $r=0.42$ positive}\\
\cmidrule(lr){4-6}\cmidrule(lr){7-9}
$\rho$ & $n$ & $\lambda$ & Regular & Score & Rect. & Regular & Score & Rect.\\
\midrule
0.5 & 80 & 1.00 & 0.908 & 0.951 & 0.892 & 0.966 & 0.769 & 0.100\\
0.5 & 80 & 0.25 & 0.914 & 0.936 & 0.886 & 0.943 & 0.768 & 0.053\\
0.5 & 400 & 1.00 & 0.948 & 0.955 & 0.945 & 1.000 & 0.993 & 0.577\\
0.5 & 400 & 0.25 & 0.930 & 0.954 & 0.947 & 0.997 & 0.966 & 0.428\\
0.8 & 80 & 1.00 & 0.909 & 0.948 & 0.894 & 0.982 & 0.634 & 0.141\\
0.8 & 80 & 0.25 & 0.904 & 0.950 & 0.898 & 0.968 & 0.671 & 0.116\\
0.8 & 400 & 1.00 & 0.946 & 0.963 & 0.935 & 1.000 & 0.999 & 0.803\\
0.8 & 400 & 0.25 & 0.940 & 0.950 & 0.948 & 1.000 & 0.990 & 0.627\\
\bottomrule
\end{tabular}
\par
\begin{minipage}{0.96\textwidth}\vspace{0.25em}\footnotesize
Notes: 1,000 replications per row; Monte Carlo standard errors are at most $0.016$. Same design as Table~\ref{tab:mc-weak-id} except that $Z\sim N(0,S)$ with $S=(1-\rho)I_3+\rho\mathbf 1\mathbf 1'$. The first-stage vector $p$ is unchanged, so the covariances $\pi=Sp$ and the effective strength of each instrument change with $\rho$; the direct effects $d=S^{-1}(\beta-\beta_1)\pi$ keep the instrument-specific ratios at $(1.07,1.88,1.97)$. Correlation strengthens the covariance $\pi_{\mathrm{SM}}$ of the weak instrument, which is why $\lambda$ matters less here than in Table~\ref{tab:mc-weak-id}.
\end{minipage}
\end{table}

%% file: generated/larger_sample.tex
\begin{table}[t]
\centering
\caption{The 140-country larger sample: two language instruments}
\label{tab:larger-sample}
\begin{tabular}{lcc}
\toprule
 & $r=0$ & $r=0.42$\\
\midrule
\multicolumn{3}{l}{\emph{Inference that takes one specification as given}}\\
2SLS, both instruments (95\% CI) & $[0.96,\,1.58]$ & ---\\
ENG only, i.e.\ EUR claim dropped (95\% CI) & $[0.78,\,1.28]$ & $[0.36,\,1.70]$\\
EUR only, i.e.\ ENG claim dropped (95\% CI) & $[1.04,\,2.35]$ & $[0.62,\,2.77]$\\
\addlinespace
\multicolumn{3}{l}{\emph{Post-Relaxation Projection}}\\
PR, 95\% simultaneous band & $[0.74,\,2.46]$ & $[0.32,\,2.88]$\\
\quad repairs retained & 3 of 3 & 3 of 3\\
PR, 90\% simultaneous band & $[0.77,\,2.37]$ & $[0.35,\,2.79]$\\
\quad repairs retained & 2 of 3 & 3 of 3\\
\bottomrule
\end{tabular}
\begin{minipage}{0.94\textwidth}
\vspace{0.25em}
\footnotesize\emph{Notes:} $n=140$. Log GDP per capita on the rule-of-law index with a constant and distance from the equator. Just-identified estimates: ENG $1.028$ (s.e.\ $0.127$, robust $F=35.1$), EUR $1.697$ (s.e.\ $0.334$, $F=17.4$). Hansen's $J=7.08$ ($p=0.008$). The one-instrument rows report marginal intervals widened by $r$, the inference a researcher would report after choosing that repair. Repairs: keep both claims, drop ENG, drop EUR; dropping both is never minimal. All sets lie inside the structural bounds $[-10,10]$.
\end{minipage}
\end{table}

%% file: generated/shared_primitive_appendix.tex
\section{Shared primitives: rectangular versus joint projection}\label{oa:shared}

Remark \ref{rem:shared-primitives} shows that canonical PR can be strictly larger than the union of targets compatible with a band whose coordinates enter several repairs. This section measures how large the difference is in repeated samples. There are two just-identified coefficients $\beta_1,\beta_2$, estimated independently with normal errors of standard deviations $0.25$ and $0.015$, so the first instrument is much less precise than the second, as in the remark. The menu contains the repair that keeps both instruments (cost $0$) and the two repairs that keep one (cost $1$); $r=0.1$ and $\Theta=[-1,2]$. The band is the Bonferroni rectangle with critical value $2.241$. Canonical PR treats the two coordinates as separate primitives of each repair. The joint projection collects $\Theta^{MR}(b)$ over all points $b$ of the band; on the band event it contains the target, and it is contained in canonical PR. Both sets are computed exactly on a grid of mesh $10^{-3}$, with 2{,}000 replications per design.

\begin{table}[ht]
\centering\small
\caption{Rectangular PR and joint projection with a shared coefficient}\label{tab:shared}
\begin{tabular}{lccccccc}
\toprule
& & & \multicolumn{2}{c}{Coverage} & \multicolumn{2}{c}{Mean length} & Joint \\
Design & $\beta_1$ & $\beta_2$ & PR & Joint & PR & Joint & smaller \\
\midrule
R0 feasible & 0.100 & 0.125 & 1.000 & 1.000 & 1.320 & 1.314 & 0.082\\
R0 on its boundary & 0.325 & 0.125 & 1.000 & 1.000 & 1.333 & 1.323 & 0.132\\
R0 infeasible & 0.500 & 0.125 & 0.970 & 0.968 & 1.356 & 1.343 & 0.190\\
strong rejection & 1.000 & 0.125 & 0.955 & 0.954 & 1.521 & 1.518 & 0.052\\
\bottomrule
\end{tabular}
\par\medskip
\begin{minipage}{0.92\textwidth}\footnotesize
\emph{Notes:} ``Joint smaller'' is the fraction of replications in which the joint projection is shorter than PR by more than two grid steps. Monte Carlo standard errors for the coverage rates are at most $0.005$.
\end{minipage}
\end{table}

Both procedures attain the nominal level. The joint projection is strictly smaller in at most 19\% of replications, and its mean length is never below 99\% of that of PR. The gain appears only when the imprecise coefficient can be small enough to make the full-information repair feasible while the precise one pins the other end of the band, the configuration of the remark. Under the rectangular information set PR is sharp (Theorem \ref{thm:bandwise-complete}); this design suggests that what is lost by ignoring shared primitives is small, while the joint projection has a cost that grows with the number of repairs.

%% file: generated/empirical_appendix.tex
\section{Observation-level empirical validation and covariance}\label{oa:empirical-results}

The analysis file contains the six variables used here, extracted without modification from the spreadsheet posted by the authors. The common complete sample has $n=80$ observations and matches the 80-country extended-AJR sample of the 2002 working paper.

All prespecified data checks pass. Table \ref{tab:oa-validation} lists the eight largest absolute differences between sample statistics and the rounded published values.

\begin{table}[ht]\centering\caption{Largest deviations from published sample statistics}\label{tab:oa-validation}\begin{tabular}{llrrr}\toprule Check & Variable & Sample & Published & Difference\\\midrule

Mean & DISTEQ & 15.2250 & 15.2300 & -0.0050\\

First-stage $t$ & SM\_classical\_t & -3.6341 & -3.6300 & -0.0041\\

Std.\ dev. & ENGFRAC & 0.2860 & 0.2900 & -0.0040\\

Mean & LCGDP95 & 8.0130 & 8.0100 & 0.0030\\

Std.\ dev. & DISTEQ & 11.1571 & 11.1600 & -0.0029\\

IV estimate & SM & 1.8766 & 1.8794 & -0.0028\\

Std.\ dev. & LOGEM4 & 1.2179 & 1.2200 & -0.0021\\

Std.\ dev. & EURFRAC & 0.4119 & 0.4100 & 0.0019\\

\bottomrule\end{tabular}\end{table}

The HC0 estimate of the covariance matrix of the three ratio influence functions, ordered ENG, SM, EUR (the asymptotic covariance of $\sqrt n$ times the estimation error), is

\[\widehat\Omega=\begin{pmatrix}

1.4713 & 0.8514 & 1.3450\\

0.8514 & 13.9767 & 4.5447\\

1.3450 & 4.5447 & 17.1201

\end{pmatrix}.\]

The contrast-aware feasibility bounds use the following pairwise contrasts:

\begin{center}\begin{tabular}{lrrr}\toprule Contrast & Estimate & HC0 s.e. & $t$ from zero\\\midrule

ENG-SM & -0.806 & 0.415 & -1.94\\

ENG-EUR & -0.903 & 0.446 & -2.02\\

SM-EUR & -0.097 & 0.524 & -0.18\\

\bottomrule\end{tabular}\end{center}

The simultaneous critical value, 2.508, is the $95\%$ quantile of the maximum of the six studentized coordinates over 200,000 Gaussian multiplier draws with a fixed seed. Splitting the draws into 40 blocks gives a Monte Carlo standard error of about $0.004$ for this quantile, far below the margin that would change any conclusion in the paper.

\subsection{Prespecified deletion-cost sensitivity}

Equal costs are the primary specification. Before the observation-level analysis I also recorded the seven distinct positive cost vectors with entries in $\{1,2\}$, up to scale. Table \ref{tab:oa-costs} lists the repairs that PR keeps at $r=0.42$ under each of them.

\begin{table}[ht]\centering\caption{Repairs retained by PR at $r=0.42$ under alternative deletion costs}\label{tab:oa-costs}\begingroup\footnotesize\setlength{\tabcolsep}{4pt}\begin{tabular}{@{}cl>{\raggedright\arraybackslash}p{6.4cm}r@{}}\toprule $(c_{\mathrm{ENG}},c_{\mathrm{SM}},c_{\mathrm{EUR}})$ & Cost layers & Retained repairs (claims deleted) & No.\\\midrule

(1,1,1) & 0, 1, 2 & none; ENG; SM; EUR; ENG+SM; ENG+EUR; SM+EUR & 7\\

(2,1,1) & 0, 1, 2 & none; ENG; SM; EUR; SM+EUR & 5\\

(1,2,1) & 0, 1, 2 & none; ENG; SM; EUR; ENG+EUR & 5\\

(1,1,2) & 0, 1, 2 & none; ENG; SM; EUR; ENG+SM & 5\\

(1,2,2) & 0, 1, 2, 3 & none; ENG; SM; EUR; ENG+SM; ENG+EUR & 6\\

(2,1,2) & 0, 1, 2, 3 & none; ENG; SM; EUR; ENG+SM; SM+EUR & 6\\

(2,2,1) & 0, 1, 2, 3 & none; ENG; SM; EUR; ENG+EUR; SM+EUR & 6\\

\bottomrule\end{tabular}\endgroup\end{table}

Across the seven rankings, the lower endpoint of the PR set at $r=0.42$ ranges from 0.311 to 0.393 and the upper endpoint from 3.345 to 3.554, so every ranking gives the same sign (positive).

The same holds for all 63 combinations of reporting tolerance and cost ranking; the smallest lower endpoint among them is 0.231. All of these statements follow from the same six-functional band, so they hold simultaneously and need no separate calibration.

Corollary \ref{cor:iv-uniform-sign} extends the table to a continuum. The lower endpoints of the simultaneous coefficient bands are ENG=0.731, SM=0.828, EUR=0.813; the largest upper endpoint is 3.134. After expansion by the maximum prespecified tolerance $\bar r=0.60$, the envelope is $[0.131,3.734]$, which lies strictly inside $[-10,10]$. The sign is therefore positive for every $r\in[0,0.60]$ and every strictly positive additive deletion-cost vector.

The replication package contains the complete tolerance grid for all seven cost vectors. The cost vectors were fixed before the observation-level estimates were computed.

%% file: generated/fieller_robustness_appendix.tex
\section{Score-inversion sets for the 80-country sample}\label{oa:fieller-robustness}

This section gives the details behind Section \ref{subsec:weak-id-results}. The calculation was added after the main observation-level analysis and was not part of the original analysis record.

Let $\widehat a_j=n^{-1}\sum_i\widetilde Z_{ji}\widetilde Y_i$ and $\widehat p_j=n^{-1}\sum_i\widetilde Z_{ji}\widetilde X_i$, with residuals taken after partialling out a constant and DISTEQ. A value $b$ is retained for instrument $j$ when
\begin{equation}\label{eq:fieller-score}
\left|\widehat a_j-b\widehat p_j\right|
\le
\frac{c}{\sqrt n}\,\widehat s_j(b),
\end{equation}
where $\widehat s_j^2(b)$ is the sample variance of $\widetilde Z_{ji}(\widetilde Y_i-b\widetilde X_i)$. This is the set \eqref{eq:score-set} of Proposition \ref{prop:score-pr}: the ratio inversion of \citet{Fieller1954}, which tests a structural value through the unreduced moment, as in \citet{AndersonRubin1949}, rather than by dividing by an estimated first stage.

I use $c=2.508$, the six-functional multiplier critical value of the main analysis. Proposition \ref{prop:score-pr} requires only $c\ge z_{1-\alpha/(2J)}=2.394$ for $J=3$ and $\alpha=0.05$, so this choice is conservative and no new critical value is computed. Squaring \eqref{eq:fieller-score} gives a quadratic inequality in $b$, which is solved without assuming that the solution is bounded; intervals, pairs of half-lines, the whole line, and the empty set are all possible. In the RST sample all three solutions are bounded intervals (Table \ref{tab:fieller-robustness}).
\begin{table}[t]\centering
\caption{Score-inversion sets, instrument by instrument}\label{tab:fieller-robustness}
\begin{tabular}{lrrc}\toprule
Instrument & $\widehat\beta_j$ & $\widehat p_j$ & 95\% score-inversion set\\\midrule
ENG & 1.071 & 0.093 & [0.677,1.663] \\
SM & 1.877 & -0.291 & [1.180,11.133] \\
EUR & 1.974 & 0.102 & [1.210,5.607] \\
\bottomrule\end{tabular}
\begin{flushleft}\footnotesize
Notes: All three inversions use $c=2.508$. Residuals are taken after partialling out a constant and distance from the equator.
\end{flushleft}
\end{table}

The ENG set lies inside the structural bounds $[-10,10]$, so the envelope part of Proposition \ref{prop:score-pr} applies: for every common tolerance $r$ and every strictly positive additive cost vector, the score-inversion PR set is contained in $[0.677-r,\,10]$. The smallest lower endpoint, 0.677, exceeds the largest tolerance considered, $\bar r=0.60$, so the set is positive throughout. The upper end equals the structural bound because the SM set extends beyond $10$.